\documentclass[a4paper,11pt]{article}
\pdfoutput=1 
\usepackage[symbol]{footmisc} 
\usepackage{jheppub} 
	
\usepackage{color, colortbl}
\usepackage[T1]{fontenc}
\usepackage[utf8]{inputenc}
\usepackage[T1]{fontenc}
\usepackage{epsfig,latexsym}
\usepackage{amsmath}
\usepackage[dvipsnames]{xcolor}
\usepackage{adjustbox}
\usepackage{tensor}
\usepackage{graphicx}
\usepackage{verbatim}
\usepackage{mathrsfs}
\usepackage{amssymb}
\usepackage{multirow}
\usepackage{epsfig}
\usepackage{color,colordvi}
\usepackage{appendix}
\usepackage{slashed}
\usepackage{array}
\usepackage{cancel}
\usepackage{float}
\usepackage[font=footnotesize, justification=centering]{caption}
\usepackage{epsf}
\usepackage{amsmath}
\usepackage{physics}
\usepackage{MnSymbol}
\usepackage{rotating}
\usepackage{multirow}
\usepackage[normalem]{ulem}
\usepackage{tcolorbox}
\usepackage{bbold}
\DeclareMathAlphabet{\mathpzc}{OT1}{pzc}{m}{it}

 \csname
@addtoreset\endcsname{equation}{section}

\newcolumntype{x}[1]{>{\centering\arraybackslash\hspace{0pt}}p{#1}}

\newcommand{\beq}{\begin{equation}}
\newcommand{\eeq}{\end{equation}}
\renewcommand{\[}{\left[}
\renewcommand{\]}{\right]}
\renewcommand{\(}{\left(}
\renewcommand{\)}{\right)}
\def\eq#1{{Eq.\,(\ref{#1})}}

\newcommand{\be}{\begin{eqnarray}}
\newcommand{\ee}{\end{eqnarray}}
\newcommand{\bea}{\begin{eqnarray}}
\newcommand{\eea}{\end{eqnarray}}
\newcommand{\bi}{\begin{itemize}}
\newcommand{\ei}{\end{itemize}}
\newcommand{\ben}{\begin{enumerate}}
\newcommand{\een}{\end{enumerate}}
\def\bes{\begin{equation*}}
\def\ees{\end{equation*}}
\def\bead{\begin{aligned}}
\def\eead{\end{aligned}}
\def\bmat{\left(\begin{matrix}}
\def\emat{\end{matrix}\right)}

\def\Re{\text{Re}}
\def\Im{\text{Im}}
\def\diag{\text{diag}}

\def\cI{{\cal I}}

\def\cL{{\cal L}}

\def\cO{{\cal O}}
\def\cR{{\cal R}}
\def\cT{{\cal T}}
\def\CKM{\text{CKM}}

\title{The shift-invariant orders of an ALP}

\author[a]{Quentin Bonnefoy,}
\author[a,b]{Christophe Grojean,}
\author[a,b]{Jonathan Kley}

\affiliation[a]{Deutsches Elektronen-Synchrotron DESY, Notkestr. 85, 22607 Hamburg, Germany}
\affiliation[b]{Institut für Physik, Humboldt-Universität zu Berlin, 12489 Berlin, Germany}

\emailAdd{quentin.bonnefoy@desy.de, christophe.grojean@desy.de, jonathan.kley@desy.de}

\abstract{It is generally believed that global symmetries, in particular axion shift symmetries, can only be approximate. This motivates us to quantify the breaking of the shift invariance that characterizes the couplings of an axion-like particle (ALP), and to identify proper order parameters associated to this breaking. Focusing on the flavorful effective Yukawa couplings to Standard Model fermions, we work out explicit conditions for them to maintain an exact axion shift symmetry. Those conditions are given in terms of Jarlskog-like flavor-invariants and can be directly evaluated from the values of the different Yukawa couplings. Therefore, they represent order parameters for the breaking of the axion shift symmetry. We illustrate this construction by matching the axion EFT to UV models, and by showing that the renormalization group running closes on those shift-breaking flavor-invariants, as it should on any complete set of order parameters. Furthermore, the study of the invariants' CP-parities indicate that all but one are CP-odd, hence the assumption of CP conservation suffices to cancel all but one sources of shift-breaking in the theory. We also investigate similar conditions in the low-energy EFT below the electroweak scale, and comment on relations inherited from a UV completion which realizes the electroweak symmetry linearly. Finally, we discuss the order parameter associated to the non-perturbative shift-breaking induced by the axion-gluons coupling, which is also flavorful.}

\begin{document} 
\begin{flushright}
DESY-22-096\\
HU-EP-22/22
\end{flushright}
\maketitle
\flushbottom

\section{Introduction}\label{section:intro}

Axions, which we take here to be any sort of pseudo-Nambu--Goldstone bosons (pNGBs), are leading candidates for physics beyond the Standard Model (SM). Indeed, they are predicted by several well-motivated extensions of the SM, and help solving many issues at once, first of which the strong-CP~\cite{Peccei:1977hh,Peccei:1977ur,Weinberg:1977ma,Wilczek:1977pj,Kim:1979if,Shifman:1979if,Dine:1981rt,Zhitnitsky:1980tq} and dark matter problems~\cite{Preskill:1982cy,Abbott:1982af,Dine:1982ah} (see~\cite{DiLuzio:2020wdo} for a recent review of axion physics). Their pNGB nature is rooted in the presence of an approximate shift symmetry of the axion field, which we also refer to as a Peccei--Quinn (PQ) symmetry. The symmetry allows, for instance, QCD axions to receive their mass mostly from QCD or fuzzy axion dark matter to be ultra-light. These features are inherited from the parametrically-close presence of a shift symmetric point, and are sharper when we approach it. However, there are several reasons to study the surroundings of that point and allow for some amount of shift breaking. First, quantum gravity objects to exact global symmetries and is expected to generate irreducible corrections to axion potentials and interactions~\cite{Hawking:1987mz,Giddings:1988cx,Banks:2010zn}. The need to suppress these gravitational shift-breaking contributions is present in any pNGB model, and goes under the name of axion quality problem. Second, there are cases where shift-breaking is a key aspect of model building: for instance, a slight amount of shift-breaking is responsible for the scanning of the Higgs mass and the resolution of the hierarchy problem in relaxion models~\cite{Graham:2015cka}. Therefore, considering shift-breaking ALP interactions seems to be necessary to make contact with theory and phenomenology.

 Consequently, it is important to clearly pinpoint the presence of physical shift-symmetry-breaking couplings, as well as to quantify their magnitude. The way to do this depends on the precise framework used to describe the axion couplings. For instance, one could study a specific UV model containing a pseudoscalar. Instead, we work here at the level of effective couplings to Standard Model (SM) fields, because effective field theories are the appropriate tools to encode axion interactions with SM particles, in a way which systematically captures and connects all contributions of an axion to high-energy observables.
 Indeed, being pNGBs, axions are generically light, hence they can be produced and contribute to processes at all energy scales of interest for high-energy physics. In addition, they arise in very diverse UV models, and can couple to all particles of the SM in all the ways compatible with their pNGB nature. Therefore, in a bottom-up approach, their couplings are essentially free parameters, up to the constraints imposed by the pNGB shift symmetry, which is precisely what an EFT approach encodes. For these reasons, axion EFTs have been systematically studied since the early days of axion physics~\cite{Georgi:1986df,Srednicki:1985xd}, and are for instance used in the context of flavor physics~\cite{Izaguirre:2016dfi,Bjorkeroth:2018dzu,Gavela:2019wzg,Bauer:2019gfk,Albrecht:2019zul,MartinCamalich:2020dfe,Endo:2020mev,Ishida:2020oxl,Calibbi:2020jvd,Carmona:2021seb,Chakraborty:2021wda,Davoudiasl:2021haa,Bauer:2021mvw,Coloma:2022hlv,Gori:2020xvq} or LHC observables~\cite{Mimasu:2014nea,Jaeckel:2015jla,Brivio:2017ije,Bauer:2017nlg,Bauer:2017ris,Alonso-Alvarez:2018irt,Ebadi:2019gij,Dobrich:2019dxc,Coelho:2020saz,Haghighat:2020nuh,Bonnefoy:2020gyh, Florez:2021zoo,Wang:2021uyb,dEnterria:2021ljz,Li:2021ygc,Goncalves:2021pdc,Alves:2021puo,Bonilla:2022pxu,Carmona:2022jid,Kling:2022ehv}.

In this paper, we carry on the systematic study of the structural properties of axion EFTs, with a focus on the breaking of axion shift-invariance due to the axion couplings to SM fermions. We work in a non-redundant operator basis which captures the most generic leading-order couplings of a light pseudoscalar $a$ to SM fermions, namely
\beq
\label{eq:SMEFTaxion}
\mathcal{L} = \mathcal{L}_{\text{SM}} + \frac{1}{2} \( \partial_{\mu} a \right) \( \partial^{\mu} a \right) 
-\frac{a}{f} \left( \bar{Q} \tilde{Y}_u \tilde{H} u + \bar{Q} \tilde{Y}_d H d + \bar{L} \tilde{Y}_e H e + \text{h.c.}  \right) + \mathcal{O}\left(\frac{1}{f^2}\right) \ ,
\eeq
where $f$ is the axion decay constant (we henceforth take $f\gg v$, the electroweak scale), $\tilde Y_{u,d,e}$ are \textit{generic complex} matrices in flavor space, $\tilde H\equiv i\sigma^2H^*$ and $\mathcal{L}_{\text{SM}}$ contains the SM couplings, whose fermionic sector reads
\beq
\label{eq:SMlag}
\mathcal{L}_{\text{SM}}\supset \sum_{\psi \in \text{SM}} i\bar\psi \slashed D \psi-\left( \bar{Q} Y_u \tilde{H} u + \bar{Q} Y_d H d + \bar{L} Y_e H e + \text{h.c.}  \right) \ .
\eeq
Our main goal is then to revisit the conditions for these couplings to be interpreted as the shift-invariant couplings of an axion, and to quantify the deviations from such conditions.

The common answer to the first part of the question is that one should be able to capture those interactions using the following Lagrangian (see~\cite{Chala:2020wvs,Bonilla:2021ufe} for a discussion of redundant operators),
\beq
\label{eq:SMEFTa}
\mathcal{L} = \mathcal{L}_{\text{SM}} + \frac{1}{2} \left( \partial_{\mu} a \right) \left( \partial^{\mu} a \right) + \frac{\partial_{\mu}a}{f} \sum_{\psi \in \text{SM}}\bar{\psi} c_\psi \gamma^{\mu} \psi + \mathcal{O}\left(\frac{1}{f^2}\right) \ ,
\eeq
where the sum runs over all Weyl fermion multiplets of the SM and the $c_\psi$ are \textit{hermitian} matrices in flavor space. The Lagrangian of \eq{eq:SMEFTa} makes the axion shift symmetry $a\to a +\epsilon f$ manifest. Then, one can map the couplings of \eq{eq:SMEFTa} onto those of \eq{eq:SMEFTaxion} via field redefinitions~\cite{Bauer:2020jbp,Chala:2020wvs}, in order to describe shift-invariant couplings using \eq{eq:SMEFTaxion}. For this mapping procedure to be possible, constraints must hold on the $\tilde Y$ couplings of \eq{eq:SMEFTaxion}~\cite{Bauer:2020jbp,Chala:2020wvs}. These constraints can also be understood only in terms of the operator basis of \eq{eq:SMEFTaxion}, where the shift invariance is never manifest: they allow one to absorb an axion shift via appropriate field redefinitions~\cite{Chala:2020wvs}. 

Unfortunately, these conditions are \textit{implicit}: given a set of couplings, one has to check whether a set of equations can be solved (we will review this approach in more details later on). In addition, they do not allow to differentiate between approximate and badly broken shift symmetries, nor to identify a power counting parameter which suppresses the breaking. Instead, we will present explicit conditions on the Wilson coefficients of \eq{eq:SMEFTaxion}, which can be directly evaluated given a set of couplings and immediately yield an answer. Therefore, such conditions define quantities which vanish iff the axion shift symmetry is preserved and whose size quantifies how badly it is broken, hence those quantities are order parameters of the breaking of the axion shift symmetry.

This is very similar in spirit to finding the Jarlskog invariant for CP-violation in the SM~\cite{Jarlskog:1985ht,Jarlskog:1985cw}, or in the SM effective field theory (SMEFT)~\cite{Bonnefoy:2021tbt}, instead of scanning possible field redefinitions which absorb unphysical complex Lagrangian parameters. It may therefore not come as a surprise that our conditions are expressed in terms of flavor-invariants, namely combinations of Lagrangian parameters which are left unchanged under flavor field redefinitions. This allows us to encode the physical collective effects associated to the presence or absence of the axion shift symmetry.

Beyond explicit axion couplings to fermions, the CP-even axion-gauge bosons couplings are also flavorful when the PQ and the gauge symmetries have mixed anomalies. They do not break the shift symmetry at the perturbative level, but the gluon coupling does so at the non-perturbative level, as is crucial in QCD axion solutions to the strong CP problem. Therefore, we also study the order parameter for this non-perturbative breaking. 

The organisation of the paper is as follows. In section~\ref{section:shiftSymmetry}, we present flavor-invariant order parameters for the breaking of the axion shift-symmetry: we first identify conditions in a given flavor basis, which we then rephrase in a flavor-invariant language. We then discuss illustrative examples and properties in section~\ref{section:details}: we compute the invariants associated to specific UV models (Section~\ref{UVmodels}), we study the CP-parities of the invariants (Section~\ref{section:CPV}), and we repeat their derivation in the low-energy below the electroweak scale (Section~\ref{section:IREFT}), emphasizing the connection to UV completions which realize non-linearly the electroweak symmetry. In section~\ref{section:RGEs}, we study the renormalization group (RG) running of the invariants. First, we show in section~\ref{section:RGrunning} that the linear space which they generate is RG-closed, as it should for any complete set of order parameters. We perform a similar analysis below the electroweak scale in section~\ref{IRRG}, and show that matching conditions to a UV theory which realizes the electroweak symmetry linearly are conserved by the RG flow at leading order. Therefore, they can be used when analyzing low-energy observables, such as electric dipole moments (EDM). We end in section~\ref{section:SMEFTRGrunning} by exhibiting sum-rules on the axion-induced RG running in the Standard Model Effective Field Theory (SMEFT), which can shed light on the axion properties from observing the SMEFT RG runing. Last, section~\ref{section:thetaQCD} discusses the invariant associated to the non-perturbative shift-breaking induced by the axion-gluon coupling, which receives a contribution from flavored couplings when navigating between the bases of Eqs.~\eqref{eq:SMEFTaxion}-\eqref{eq:SMEFTa}, as well as its RG running. Finally, we conclude in section~\ref{conclusions}. Some appendices complete the paper. Appendix~\ref{appendix:matrixRelations} presents matrix relations which we used to construct the invariants. Appendix~\ref{appendix:degenerateCases} discusses the fate of the axion shift-symmetry at remarkable points of the parameter space of the SM, e.g. where fermion masses are degenerate or the CKM matrix possesses texture zeros, which require the use of flavor-invariants non-linear in the axion couplings, discussed in appendix~\ref{appendix:nonLinearDegenerate}. Eventually, appendix~\ref{appendix:higherOrderRunningCg} gives more details on the RGEs used in section~\ref{section:thetaQCD}.

\section{Flavor-invariant order parameters for the breaking of an axion shift symmetry}\label{section:shiftSymmetry}

In this section, we ask the following questions: given a set of couplings $\tilde Y_{u,d,e}$ in the operator basis of \eq{eq:SMEFTaxion}, under which conditions do they describe the couplings of a shift-symmetric axion? And if they do not, which are the order parameters of shift-symmetry breaking?

A partial answer has been long known: when the shift symmetry of the axion is exact, it is possible to express the axion-fermion interactions using the Lagrangian of \eq{eq:SMEFTa}. Integrating by parts and using the fermionic equations of motion of the SM, one finds that the Lagrangian in \eq{eq:SMEFTaxion} can arise from a Lagrangian of the form given in \eq{eq:SMEFTa} when\footnote{Strictly speaking, one should consider that there could exist a flavor transformation which sends $\tilde Y_{u,d,e}$ to that in \eq{eq:ShiftSymCoup} without changing the Yukawa couplings (that happens when those are invariant under a subset flavor transformations, such as the usual baryon and three lepton numbers). However, explicitly taking this into account  
is unnecessary, since it amounts to redefining $c_{L,e,Q,u,d}$.
In addition, although we do not write the bosonic couplings of the axion to the Higgs or the gauge fields explicitly, we assume that the basis of Eqs.~\eqref{eq:SMEFTaxion}-\eqref{eq:SMEFTa} do not contain redundant operators such as $\partial_\mu a \(iH^\dagger D^\mu H+h.c.\)$, which merely amounts to shifting the value of the fermionic Wilson coefficients. Finally, shift symmetry correlates the $\cO(1/f^2)$ couplings involving two axion fields and those at $\cO(1/f)$, and similarly for higher-point couplings. In this paper, we only focus on the constraints at $\cO(1/f)$.}
\beq
\label{eq:ShiftSymCoup}
\exists c_{Q,u,d,L,e} \text{ hermitian, such that: } \tilde{Y}_{u,d} = i (Y_{u,d} c_{u,d} - c_Q Y_{u,d}) \ , \quad \tilde{Y}_e = i (Y_e c_e - c_L Y_e) \ .
\eeq
However, these conditions are implicit, as they require one to scan over all hermitian matrices $c_{Q,u,d,L,e}$. As a consequence, for a given set of $\tilde Y$ couplings, one cannot straightforwardly identify  which entries of the matrices $\tilde Y$ violate the axion shift symmetry and should be sent to zero so as to recover the symmetry. 

Therefore, we find it useful to deal with a set of algebraic conditions on $\tilde Y_{u,d,e}$ which amount to the identities in \eq{eq:ShiftSymCoup} but do not refer to implicit matrices $c_\psi$. In this section, we thus identify explicit, independent, necessary and sufficient conditions on the entries of $\tilde Y_{u,d,e}$ in \eq{eq:SMEFTaxion} for the axion shift symmetry to hold. We focus on the case where all quark and lepton masses are non-vanishing and non-degenerate in flavor space, and where there are no texture zeros in the CKM matrix, which is the experimentally relevant case -- we comment on degenerate cases in appendix~\ref{appendix:degenerateCases}. Throughout this paper, we neglect neutrino masses.

When needed, we work in a particular flavor basis, which we take to be
\beq
	Y_u=\diag(y_u,y_c,y_t) \ , \quad Y_d=V_\CKM \cdot \diag(y_d,y_s,y_b) \ , \quad Y_e=\diag(y_e,y_\mu,y_\tau) \ .
\label{upBasis}
\eeq
The mass basis can then be reached by performing $d_L\to V_\CKM d_L$. Nevertheless, our conditions can be phrased using flavor-invariant algebraic expressions of $Y_{u,d,e}$ and $\tilde Y_{u,d,e}$ which vanish when the axion shift-symmetry is exact: we indeed exhibit below a set of flavor-invariants, which are reminiscent of the flavor-invariants built in SMEFT to capture the breaking of CP~\cite{Bonnefoy:2021tbt}, and which are in one-to-one correspondence with the independent conditions in \eq{eq:ShiftSymCoup}. Such a formulation pinpoints unambiguously which couplings respect/violate the axion shift-symmetry in \eq{eq:SMEFTaxion}.

Before going into details, let us also remind that the matching between the bases of \eq{eq:SMEFTa} and \eq{eq:SMEFTaxion} also involve relative shifts between the axion couplings to gauge fields~\cite{Bauer:2020jbp}, induced by mixed anomalies with the gauge group of the SM. These shifts are flavorful and are naturally expressed using flavor-invariants. The case of the gluon coupling is particularly interesting, since it participates in the breaking of the axion shift symmetry at the non-perturbative level. We focus here on perturbative breaking induced by the direct couplings to fermions, but we come back to the case of gluons in section~\ref{section:thetaQCD}.

\subsection{Counting of parameters with or without a shift-symmetry}\label{section:counting}

To identify all conditions on $\tilde{Y}_{u,d,e}$, let us first count how many  we expect, by comparing the free parameters in the operator bases of \eq{eq:SMEFTa} and \eq{eq:SMEFTaxion}. We can classify couplings according to their behavior under the CP transformation which transforms any fermion field $\psi$ according to\footnote{When masses are non-degenerate, one can always rephase the fields and reach a mass basis where the action of CP does not contain additional phases.} $\psi\to \gamma^0C\bar\psi^T$, where $C$ is the (antisymmetric) charge conjugation matrix such that $\gamma^\mu C=-C(\gamma^\mu)^T$. Since the axion is a pseudoscalar, real $\tilde Y_\psi$ in \eq{eq:SMEFTaxion} and imaginary $c_\psi$ in \eq{eq:SMEFTa} are CP-odd in the mass basis. When the axion shift symmetry is broken, we need to use the non-shift symmetric EFT of \eq{eq:SMEFTaxion}, in which the couplings $\tilde Y$ are arbitrary $3\times 3$ complex matrices. However, the presence of the lepton family numbers $U(1)_{L_i}$ symmetries in the SM can be used to modify at will two phases among those of $\tilde Y_{e,i\neq j}$, which are therefore unphysical. Instead, the independent rephasing-invariant quantities are $\tilde Y_{e,ii}$, $\arg(\tilde Y_{e,ij}\tilde Y_{e,ji})$ ($i< j$), $\abs{\tilde Y_{e,ij}}$ and $\arg(\tilde Y_{e,12}\tilde Y_{e,23}\tilde Y_{e,31})$. They amount to $16$ independent quantities, $7$ CP-odd and $9$ CP-even. In the quark sector, all parameters are physical and one finds $2 \times 9=18$ CP-even and $2 \times 9=18$ CP-odd couplings in the quark sector. When the shift-symmetry is exact, we can start in the explicitly shift-invariant basis of \eq{eq:SMEFTa}, where there are $2$ hermitian matrices $c_{L,e}$ in the lepton sector and $3$ hermitian matrices $c_{Q,u,d}$ in the quark sector parametrizing all couplings to fermions. The lepton number rephasings can be again used to remove two phases\footnote{The rephasing-invariants now read $c_{L/e,ii},\abs{c_{L/e,ij}}$ ($i< j$), $\arg(c_{e,ij}c_{L,ji})$ ($i<j$) and $\arg(c_{L,12}c_{L,23}c_{L,31})$.}. Furthermore, there exists a freedom in the derivative basis, associated to the addition of the operator $\partial_\mu a J^\mu$, for any conserved fermionic current of the SM $J^\mu$~\cite{Bonilla:2021ufe}. This operator does not induce any physical effect, as it can be removed (at $\cO(1/f)$) thanks to an axion-dependent flavor transformation. Given that there is an exact baryon number $U(1)_B$ symmetry in the quark sector and the $U(1)_{L_i}$ symmetries in the lepton sector, one can remove one diagonal entry of either $c_Q,c_u$ and three out of $c_{L,e}$. This leads us to count $9$ CP-even and $4$ CP-odd couplings in the lepton sector, as well as $17$ CP-even and $9$ CP-odd couplings in the quark sector. Hence, we expect $3$ CP-odd relations in the lepton sector together with $9$ CP-odd and $1$ CP-even relation in the quark sector that characterize the presence of a shift symmetry in the basis of \eq{eq:SMEFTaxion}. We summarize the different countings in Table~\ref{tableRanksNonDegenerate}.
\begin{table}[h!]
\small
\centering
\resizebox{\columnwidth}{!}{%
\begin{tabular}{c|c|c|c|c|c|c}
&\multicolumn{2}{c|}{Shift-symmetric Wilson coefficients $c_{Q,u,d,L,e}$}&\multicolumn{2}{c|}{Generic Wilson coefficients $\tilde Y_{u,d,e}$}&\multicolumn{2}{c}{Number of constraints}\\
\hline
&CP-even&CP-odd&CP-even&CP-odd&CP-even&CP-odd\\
\hline
Quark sector&$17$&$9$&$18$&$18$&$1$&$9$\\
Lepton sector&$9$&$4$&$9$&$7$&$0$&$3$
\end{tabular}
}
\caption{Number of physical coefficients at dimension-five in the EFTs of \eq{eq:SMEFTa} and \eq{eq:SMEFTaxion} (see the text for details), and numbers of constraints that $\tilde Y_{u,d,e}$ need to verify to respect an exact shift invariance.}
\label{tableRanksNonDegenerate}
\end{table}

\subsection{Flavor invariants in the lepton sector}

Let us now derive those relations. We start with the lepton case, where the constraints are simpler. As already mentioned, there exists a field redefinition which maps \eq{eq:SMEFTa} to \eq{eq:SMEFTaxion}, with~\cite{Bauer:2020jbp,Chala:2020wvs}
\beq
\label{eq:matchingLeptons}
\tilde{Y}_e = i (Y_e c_e - c_L Y_e)
\eeq
in particular. For non-vanishing lepton masses, $Y_e$ is invertible and one can solve for $c_e$,
\beq
c_e=-iY_e^{-1}\(\tilde Y_e^{\vphantom{-1}}+ic_L^{\vphantom{-1}} Y_e^{\vphantom{-1}}\) \ .
\eeq
Imposing that $c_e$ should be hermitian leads to constraints, here expressed in a flavor basis where $Y_e$ is diagonal and real,
\beq
\label{eq:conditionsLeptonsMassBasis}
\exists c_{L} \text{ hermitian s.t. } \frac{\tilde Y_{e,ij}}{y_{e,i}}+\frac{\tilde Y^*_{e,ji}}{y_{e,j}}+ic_{L,ij}\[\frac{y_{e,j}}{y_{e,i}}-\frac{y_{e,i}}{y_{e,j}}\]=0 \ \forall i,j \ .
\eeq
When $i=j$, $c_L$ disappears from the expression and one finds constraints on $\tilde Y_e$, namely that $\tilde Y_{e,ii}$ is purely imaginary. The constraints where $i<j$ and $i>j$ are complex conjugates of one another, therefore one can focus e.g. on those where $i<j$, and they can all be solved by a suitable choice of $c_L$,
\beq
\label{eq:solutionCL}
c_{L,ii}=0 \ , \quad c_{L,ij,i<j}=i \frac{y_{e,j}\tilde{Y}_{e,ij}^{\mathstrut} + y_{e,i} \tilde{Y}_{e,ji}^{*}}{y_{e,j}^2-y_{e,i}^2} \ .
\eeq
This defines a hermitian $c_L$, bringing no further constraints. Therefore, there are only $3$ conditions on $\tilde Y_e$ in order that it describes a shift-symmetric axion, consistently with our counting at the beginning of this section. Although derived in a specific flavor basis, the constraints can be expressed in a flavor-invariant way. Flavor-invariant means that they are left unchanged by the flavor symmetry, whose spurious action on the Lagrangian parameters of \eq{eq:SMlag} and \eq{eq:SMEFTaxion} are given in Table~\ref{tab:ytrasmforma}.
\begin{table}[H]
	\centering
	\begin{tabular}{c|c|c|c|c|c}
		& $SU(3)_Q$ & $SU(3)_u$ & $SU(3)_d$ & $SU(3)_L$ & $SU(3)_e$\\\hline
		$Y_u,\tilde Y_u$ &$\mathbf{3}$ & $ \mathbf{\bar{3}}$ & $\mathbf{1}$ &$\mathbf{1}$ &$\mathbf{1}$  \\[0.1cm]
		$Y_d,\tilde Y_d$ &$\mathbf{3}$ & $ \mathbf{1}$ & $\mathbf{\bar{3}}$ &$\mathbf{1}$ &$\mathbf{1}$  \\[0.1cm]
		$Y_e,\tilde Y_e$ &$\mathbf{1}$ & $ \mathbf{1}$ & $\mathbf{1}$ &$\mathbf{3}$ &$\mathbf{\bar{3}}$
	\end{tabular}
	\caption{Flavor transformation properties of the Yukawa matrices treated as spurions}
	\label{tab:ytrasmforma}
\end{table}
In a flavor-invariant language, the constraints on $\tilde Y_e$ read
\beq
\label{eq:diagonalConditionsLeptons}
\Re\Tr(X_e^{0,1,2}\tilde{Y}_e^{\mathstrut}Y_e^{\dagger}) = 0 \ ,
\eeq
where we define $X_e\equiv Y_e^{\mathstrut}Y_e^\dagger$. Later on, we also repeatedly use $X_{u,d}^{\mathstrut}\equiv Y_{u,d}^{\mathstrut}Y_{u,d}^\dagger$. This flavor-invariant expression is important for our purpose, as it identifies the flavor-invariant, hence physical, order parameters of shift-symmetry breaking in the lepton sector.

\subsection{Flavor invariants in the quark sector}\label{section:quarkInvariants}

In the quark sector, the presence of the doublet $Q$ imposes that we treat up- and down-quarks simultaneously. The pair of couplings $\tilde Y_{u,d}$ describes a shift-symmetric axion when
\beq
\label{eq:matchingQuarks}
\exists c_{Q,u,d} \text{ hermitian s.t. } \tilde{Y}_{u,d} = i (Y_{u,d} c_{u,d} - c_Q Y_{u,d}) \ .
\eeq
Similarly to what we did above for the leptons, we can solve for $c_{u,d}$ when no mass vanishes, and the fact that $c_{u,d}$ are hermitian brings the following constraints, expressed in the flavor basis of \eq{upBasis},
\beq
\exists c_Q \text{ hermitian s.t. } \bmat\frac{\tilde Y_{u,ij}}{y_{u,i}}+\frac{\tilde Y^*_{u,ji}}{y_{u,j}}+ic_{Q,ij}\[\frac{y_{u,j}}{y_{u,i}}-\frac{y_{u,i}}{y_{u,j}}\]\\
 \frac{V_{\text{CKM},ki}^{*}\tilde{Y}_{d,kj}^{\mathstrut}}{y_{d,i}} + \frac{\tilde{Y}_{d,ki}^{*}V_{\text{CKM},kj}^{\mathstrut}}{y_{d,j}} + i c_{Q,kl} V_{\text{CKM},ki}^{*}V_{\text{CKM},lj}^{\mathstrut} \left( \frac{y_{d,j}}{y_{d,i}} - \frac{y_{d,i}}{y_{d,j}} \right)\emat=0 \ \forall i,j \ ,
\eeq
where the sum over $k,l$ is implicit. The $i=j$ equations imply constraints identical to those found for the leptons,  
\beq
\label{eq:diagonalConditionsQuarks}
\Re\Tr(X_{u,d}^{0,1,2}\tilde{Y}_{u,d}^{\mathstrut}Y_{u,d}^{\dagger}) = 0 \ .
\eeq
However, the presence of $c_Q$ in both equations implies further conditions. They can be seen from first solving for the off-diagonal entries of $c_Q$ using the equations involving $\tilde Y_u$,
\begin{equation}
\label{eq:cQFromU}
c_{Q,ij,i<j} = i \frac{y_{u,j}\tilde{Y}_{u,ij}^{\mathstrut} + y_{u,i} \tilde{Y}_{u,ji}^{*}}{y_{u,j}^2-y_{u,i}^2} \ ,
\end{equation}
which can be inserted in the equations for $\tilde Y_d$ to obtain
\beq
\label{eq:constraintCombined}
\bead
&\frac{V_{\text{CKM},ki}^{*}\tilde{Y}_{d,kj}^{\vphantom{*}}}{y_{d,i}} + \frac{\tilde{Y}_{d,ki}^{*}V_{\text{CKM},kj}^{\mathstrut}}{y_{d,j}} \\
&+i\sum_k\left[c_{Q,kk}V_{\text{CKM},ki}^{*}V_{\text{CKM},kj}^{\vphantom{*}}+i\sum_{l\neq k}\frac{y_{u,l}\tilde{Y}_{u,kl}^{\mathstrut} + y_{u,k} \tilde{Y}_{u,lk}^{*}}{y_{u,l}^2-y_{u,k}^2} V_{\text{CKM},ki}^{*}V_{\text{CKM},lj}^{\vphantom{*}}\right] \left( \frac{y_{d,j}}{y_{d,i}} - \frac{y_{d,i}}{y_{d,j}} \right) = 0
\eead
\eeq
for $i<j$. For a generic CKM matrix, these three complex equations depend on two free real parameters, given by the differences\footnote{The r.h.s. of \eq{eq:matchingQuarks} is invariant under $c_{Q,u,d}\to c_{Q,u,d}+\alpha \mathbb{1}$, so that only differences between the diagonal entries of $c_Q$ can contribute. In \eq{eq:constraintCombined}, the invariance under $c_Q\to c_Q+\alpha \mathbb{1}$ is ensured by CKM unitarity.  Here $\mathbb{1}$ corresponds to the only matrix which commutes with both $Y_u$ and $Y_d$ in the case where the quark masses and the CKM entries are non-degenerate. Instead, the r.h.s. of \eq{eq:matchingLeptons} is invariant under $c_{L,e}\to c_{L,e}+\alpha M_e$, where $M_e$ is any matrix which commutes which $Y_e$. When the lepton masses are non-degenerate, we have $M_e=\diag\(m_{e,i}\in \mathbb{R}\)$ in the flavor basis where $Y_e$ is diagonal, which explains why we could choose $c_{L,kk}=0$ in \eq{eq:solutionCL}.} $c_{Q,kk}-c_{Q,ll}$, and they yield four independent genuine constraints on $\tilde Y_{u,d}$. We would like to emphasize that these four conditions are {\it collective} effects, namely they only make sense when both the up- and down-type Yukawa couplings are present. Together with the conditions in \eq{eq:diagonalConditionsQuarks}, we therefore find $10$ conditions on the entries of $\tilde Y_{u,d}$ (consistently with our earlier counting), $4$ of which entangle up- and down-sectors.

To express the four last quark relations in terms of flavor-invariants, it is helpful to write the previous relations in a matrix (i.e. flavor-covariant) form. Starting again with the implicit relation for the shift-symmetric axion Yukawa couplings
\beq
\tilde{Y}_{u,d} = i (Y_{u,d} c_{u,d} - c_Q Y_{u,d}) \, ,
\eeq
one can solve this equation for $c_{u,d}$ assuming non-vanishing quark masses
\beq
c_{u,d} = -i Y_{u,d}^{-1} \left( \tilde{Y}_{u,d} + i c_Q Y_{u,d} \right).
\eeq
When the quark Yukawas $Y_{u,d}$ are full rank matrices, the vanishing of the anti-hermitian part $(c_{u,d}^{(ah)} \sim c_{u,d}^{\mathstrut} - c_{u,d}^{\dagger})$ of $c_{u,d}$ implies the following commutator relation
\beq
\label{eq:CommRel}
\[c_Q, X_x \] = i \( \tilde{Y}_x^{\mathstrut} Y_x^{\dagger} + Y_x^{\mathstrut} \tilde{Y}_x^{\dagger} \),
\eeq
with $X_x = Y_x Y_x^{\dagger}$ and $x=u,d$. We can then find flavor-invariant constraints by exploiting well-known commutator relations. For instance, we can reproduce the constraints in \eq{eq:diagonalConditionsQuarks} by using the fact that for any two matrices $A,B$
\beq
\label{twoMatIdentity}
\Tr\(A^n \[A,B\]\) = 0 \quad \forall n \in \mathbb{Z},
\eeq
which implies
\beq
-i \Tr\(X_x^n \[c_Q,X_x \] \) = \Tr\(X_x^n\( \tilde{Y}_x^{\mathstrut} Y_x^{\dagger} + Y_x^{\mathstrut} \tilde{Y}_x^{\dagger} \) \) = 0
\eeq
For $x=u,d,e$ and $n=0,1,2$, these equations correspond to the diagonal constraints we have found above. Additional commutator identities displayed in appendix~\ref{appendix:commutatorRelations} allow us to derive extra conditions. 

\subsection{Complete set of linear invariants}\label{section:invariantsPresentation}

Eventually, we consider the following set of flavor-invariants, linear in $\tilde Y_{u,d,e}$,
\begin{tcolorbox}
\beq
\label{eq:QuarkInv}
\begin{split}
 I_u^{(1)} = \Re\Tr\left( \tilde{Y}_u^{\mathstrut} Y_u^{\dagger} \right), & \qquad I_u^{(2)} = \Re\Tr\left( X_u^{\mathstrut} \tilde{Y}_u^{\mathstrut} Y_u^{\dagger} \right), \qquad I_u^{(3)} = \Re\Tr\left( X_u^2 \tilde{Y}_u^{\mathstrut} Y_u^{\dagger} \right), \\
 I_d^{(1)} = \Re\Tr\left( \tilde{Y}_d^{\mathstrut} Y_d^{\dagger} \right), & \qquad I_d^{(2)} = \Re\Tr\left( X_d^{\mathstrut} \tilde{Y}_d^{\mathstrut} Y_d^{\dagger} \right), \qquad I_d^{(3)} = \Re\Tr\left( X_d^{2\mathstrut} \tilde{Y}_d^{\mathstrut} Y_d^{\dagger} \right), \\
& I_{ud}^{(1)} = \Re\Tr\left(X_d^{\mathstrut} \tilde{Y}_{u\vphantom{d}}^{\mathstrut} Y_{u\vphantom{d}}^{\dagger} + X_{\vphantom{d}u}^{\mathstrut} \tilde{Y}_d^{\phantom{\dagger}} Y_d^{\dagger} \right), \\
I_{ud,u}^{(2)} & = \Re\Tr\left( X_{\vphantom{d}u}^{2\mathstrut} \tilde{Y}_d^{\mathstrut} Y_d^{\dagger} + \{X_{u\vphantom{d}}^{\mathstrut},X_d^{\mathstrut}\} \tilde{Y}_{u\vphantom{d}}^{\phantom{\dagger}} Y_{u\vphantom{d}}^{\dagger} \right), \\
I_{ud,d}^{(2)} & = \Re\Tr\left( X_d^{2\mathstrut} \tilde{Y}_{u\vphantom{d}}^{\mathstrut} Y_{u\vphantom{d}}^{\dagger} + \{X_{u\vphantom{d}}^{\mathstrut},X_d^{\mathstrut}\} \tilde{Y}_d^{\mathstrut} Y_d^{\dagger} \right), \\
I_{ud}^{(3)} = & \Re\Tr\left(X_d^{\mathstrut} X_{u\vphantom{d}}^{\mathstrut} X_d^{\mathstrut} \tilde{Y}_{u\vphantom{d}}^{\mathstrut} Y_{u\vphantom{d}}^{\dagger} + X_{\vphantom{d}u}^{\mathstrut} X_d^{\mathstrut} X_{\vphantom{d}u}^{\mathstrut} \tilde{Y}_d^{\mathstrut} Y_d^{\dagger} \right) \\
I_{ud}^{(4)} = & \Im\Tr\left( \left[ X_{u\vphantom{d}}^{\mathstrut}, X_d^{\mathstrut}\right]^2 \left( \left[ X_d^{\mathstrut},\tilde{Y}_{u\vphantom{d}}^{\mathstrut} Y_{u\vphantom{d}}^{\dagger}\right] - \left[ X_{\vphantom{d}u}^{\mathstrut},\tilde{Y}_d^{\mathstrut} Y_d^{\dagger}\right]\right)\right)
\end{split}
\eeq
\end{tcolorbox}
for the quarks and
\begin{tcolorbox}
\beq
\label{eq:LeptonInv}
I_e^{(1)} = \Re\Tr\left( \tilde{Y}_e^{\mathstrut} Y_e^{\dagger} \right), \qquad I_e^{(2)} = \Re\Tr\left( X_e^{\mathstrut} \tilde{Y}_e^{\mathstrut} Y_e^{\dagger} \right), \qquad I_e^{(3)} = \Re\Tr\left( X_e^{2\mathstrut} \tilde{Y}_e^{\mathstrut} Y_e^{\dagger} \right)
\eeq
\end{tcolorbox}
for the leptons. Those invariants have to vanish for the EFT, in the Yukawa basis of \eq{eq:SMEFTaxion}, to be shift-invariant. Their vanishing also provides a sufficient condition. This is shown by taking advantage of their linearity in $\tilde Y_{u,d,e}$, which allows us to use simple linear algebra: we compute the rank of the {\it transfer matrix} $\cT_{Aa}$ which relates the set of invariants $\{I_A\}$ to the entries $\{c_a\}$ of $\tilde Y_{u,d,e}$ in a given flavor basis, arranged in a vector:
\beq
I_A=\cT_{Aa}c_a \ ,
\eeq
where $\cT$ only depends on the four-dimensional Yukawas $Y_{u,d,e,}$, due to the linearity of the invariants in $c_a$. Therefore, its rank, i.e. the number of conditions associated to the set of equalities $I_A=0 \ \forall A$, can be directly computed. It is found to be $13$, namely $10$ in the quark sector and $3$ in the lepton sector, which agrees with the number of conditions from shift-invariance. Therefore, the invariants in Eqs.~\eqref{eq:QuarkInv}-\eqref{eq:LeptonInv} vanish if and only if $\tilde Y_{u,d,e}$ describe the couplings of a shift-symmetric axion. We stress that they are algebraic and explicit: given values for $\tilde Y_{u,d,e}$, evaluating those invariants suffices to discriminate between shift-invariant or shift-breaking couplings.

The set for the quark sector is not minimal as it contains 11 invariants but only captures 10 conditions. This comes from the fact that our invariants can be arranged into a vanishing linear combination\footnote{The coefficients of an appropriate combination are themselves combinations of products of traces formed by $X_{u,d}$. See section~\ref{section:RGrunning} for more details.}, hence one invariant can be eliminated in favor of the 10 others. Consequently, it is possible to find a subset of 10 non-redundant invariants which preserves maximal rank for its transfer matrix. This is achieved for instance by the following set,
\beq
\label{eq:MinSet}
I_{u\vphantom{d}}^{(1)},I_{u\vphantom{d}}^{(2)},I_d^{(1)},I_d^{(2)}, I_{u\vphantom{d}}^{(3)}+I_d^{(3)},I_{ud}^{(1)},I_{ud,u}^{(2)},I_{ud,d}^{(2)},I_{ud}^{(3)},I_{ud}^{(4)} \ .
\eeq
In the following we will still work with the redundant set as it is easier to show that the set is closed under RG flow by projecting onto a minimal set after performing the RG evolution.

Let us end this section by stressing that our conditions apply beyond the non-redundant operator basis of \eq{eq:SMEFTaxion}, and would also capture the breaking of shift-symmetry in a redundant operator basis which would mix the couplings $\tilde Y$ of \eq{eq:SMEFTaxion} and $c$ of \eq{eq:SMEFTa}. Indeed, even in such a redundant formulation, whether {\it all} couplings can be rewritten in the basis of \eq{eq:SMEFTa} still amounts to the condition in \eq{eq:ShiftSymCoup}. This is consistent with the linearity of our invariants: let us split
\beq
\tilde Y=\tilde Y^{(\text{PQ})}+\tilde Y^{(\cancel{\text{PQ}})} \ ,
\eeq
where the couplings induced by $\tilde Y^{(\text{PQ})}$ respect a PQ symmetry and can therefore be written as in \eq{eq:SMEFTa}. Our invariants vanish on $\tilde Y^{(\text{PQ})}$, and, thanks to their linearity, 
\beq
I_A\(\tilde Y\)=I_A\(\tilde Y^{(\cancel{\text{PQ}})}\) \ .
\eeq
They therefore capture the sources of PQ breaking in the theory, irrespective of any shift-invariant couplings which are additionnally present.

\section{Examples and properties}\label{section:details}

In this section, we illustrate the use of our invariants, highlight some of their properties, and comment on their connection to other symmetries than the axion shift symmetry. More precisely, we confirm in section~\ref{UVmodels} that our invariants capture the sources of shift-symmetry-breaking, as well as their collective nature, when the axion EFT is matched to UV models. We then connect in section~\ref{section:CPV} our invariants to CP-odd invariants used in the study of CP-violation, and we finally repeat in section~\ref{section:IREFT} the analysis in the low-energy EFT below the electroweak scale. The absence of weak interactions, which arrange left-handed (LH) up- and down-quarks in a doublet, implies that the IR conditions are looser than those which hold in the UV completion.

\subsection{Matching to UV models}\label{UVmodels}

To illustrate the use of our invariants, we evaluate them when the ALP EFT is matched onto different UV models, which confirms that the invariants capture the sources of PQ-breaking and its collective nature. We focus on perturbations of models which possess an exact PQ symmetry. 

\subsubsection{Axiflavon/flaxion model}\label{flaxionSection}

Let us start with the axiflavon/flaxion model~\cite{Davidson:1981zd,Calibbi:2016hwq,Ema:2016ops} in which the Froggatt--Nielsen and Peccei--Quinn mechanisms are realized through the same spontaneously broken $U(1)$. One introduces a complex scalar $\phi$, called the flavon, with the following effective interactions with the Standard Model fields\footnote{This effective Lagrangian can be UV-completed in a theory of vectorlike fermions of mass $M$ which couple to the SM fermions and $\phi$~\cite{Froggatt:1978nt,Bonnefoy:2019lsn}.}
\beq
\label{eq:axiflavon}
- \mathcal{L} = \alpha_{ij}^d \( \frac{\phi}{M} \)^{q_{Q_i}-q_{d_j}} \bar{Q}_i H d_j + \alpha_{ij}^u \( \frac{\phi}{M} \)^{q_{Q_i}-q_{u_j}} \bar{Q}_i \tilde{H} u_j + \alpha_{ij}^e \( \frac{\phi}{M} \)^{q_{L_i}-q_{e_j}} \bar{L}_i H e_j + \text{h.c.}
\eeq
with $M$ the cut-off of the model and the $q_i\in \mathbb{R}$ are the charges of the SM fields under the $U(1)$ where a charge of $+1$  is assigned to the flavon and the Higgs is taken to be neutral under the $U(1)$. The symmetry is broken by the VEV $\langle \phi \rangle = f$ of the complex scalar which also determines the hierarchy of the SM Yukawa couplings. We can then parametrize the flavon as $\phi = \frac{1}{\sqrt{2}} \( f + s + i a \)$ and identify the field $a$ as the axion of the theory. Expanding the above Lagrangian in the unbroken electroweak phase we obtain for the interactions of the axion with the SM particles
\beq
\label{eq:axiflavonPQinv}
- \mathcal{L} = \frac{ia}{f} \( Y_{ij}^d \( q_{Q_i}-q_{d_j} \) \bar{Q}_i H d_j + Y_{ij}^u \( q_{Q_i}-q_{u_j} \) \bar{Q}_i \tilde{H} u_j + Y_{ij}^e \( q_{L_i}-q_{e_j} \) \bar{L}_i H e_j \) + \text{h.c.}
\eeq
where $Y_{ij}^x \equiv \alpha_{ij}^x \(\frac{f}{\sqrt{2}M}\)^{q_{x_{L,i}}-q_{x_{R,j}}}$ are also the SM Yukawa couplings. We can simply read off the axion EFT couplings from this Lagrangian by comparing with Eq.~\eqref{eq:SMEFTaxion}
\beq
\tilde{Y}_{u,ij} = i Y_{ij}^u \( q_{Q_i}-q_{u_j} \), \quad \tilde{Y}_{d,ij} = i Y_{ij}^d \( q_{Q_i}-q_{d_j} \), \quad \tilde{Y}_{e,ij} = i Y_{ij}^e \( q_{L_i}-q_{e_j} \).
\eeq
The Lagrangian in \eq{eq:axiflavon} is constructed to be Peccei-Quinn invariant, hence all couplings in \eq{eq:axiflavonPQinv} must correspond to a shift-symmetric axion\footnote{Beyond the precise model discussed in this section, \eq{eq:axiflavonPQinv} describes any set of shift-symmetric axion couplings in the flavor basis which diagonalizes the PQ symmetry, i.e. where it acts as a phase shift on each flavor independently. This basis is that which diagonalizes all couplings $c$ in \eq{eq:SMEFTa}, which read $c_{\psi,ij}=-q_{\psi_i}\delta_{ij}$ for each fermion field $\psi$.}. This is consistent with the fact that our invariants $I$ vanish when evaluated on the above couplings, for instance
\beq
I_u^{(1)}= -\text{Im}\Tr\(\[\text{diag}\(q_Q\)Y^u-Y^u\text{diag}\(q_u\)\]Y^u{}^\dagger\)=0 \ ,
\eeq
and similarly for all the other invariants, due to the cyclicity of the trace and the fact that the imaginary part of the trace of a hermitian matrix vanishes. The invariants become non-zero when one introduces a generic Peccei-Quinn breaking term
\beq
\label{eq:axiflavonPQbreak}
- \mathcal{L}_{\cancel{PQ}} = \epsilon \frac{ia}{f} \( \beta_{ij}^d \bar{Q}_i H d_j + \beta_{ij}^u \bar{Q}_i \tilde{H} u_j + \beta_{ij}^e \bar{L}_i H e_j \) + \text{h.c.}
\eeq
which we give a generically different power counting $\epsilon$ than the Peccei-Quinn invariant Lagrangian (these couplings can originate from terms as in \eq{eq:axiflavon}, but with $q_{x_{L,i}}-q_{x_{R,j}} \to n^x_{ij}$ and $\alpha_x \to \epsilon \alpha'_x$). We can match this Lagrangian at tree level to the ALP EFT as defined in \eq{eq:SMEFTaxion}, yielding
\beq
\label{eq:smallPQbreakinggeneric}
\tilde{Y}_{u,ij} = i Y_{ij}^u \( q_{Q_i}-q_{u_j} \) + i\epsilon \beta_{ij}^u, \quad \tilde{Y}_{d,ij} = i Y_{ij}^d \( q_{Q_i}-q_{d_j} \) + i\epsilon \beta_{ij}^d, \quad \tilde{Y}_{e,ij} = i Y_{ij}^e \( q_{L_i}-q_{e_j} \) + i \epsilon \beta_{ij}^e.
\eeq
Plugging this into our invariants $I$ gives
\beq
\label{eq:evalInv}
\{I \}_{I \in \text{minimal set}} = \epsilon f_I(Y_{ij},\beta_{ij},q_i)
\eeq
where the $f_I$ are complicated polynomials of the parameters of the theory (the dependence in $\epsilon$, $\beta_{ij}$ and $q_i$ is linear, due to the linearity of our invariants). Taking the shift-symmetric limit $\epsilon \to 0$ makes all invariants vanish as expected. 

We can further confirm that our invariants act as order parameters of the ALP shift-symmetry and illustrate their features by considering specific realizations of the PQ-breaking term. For instance, let us add to \eq{eq:axiflavon} the term
\beq
\label{eq:axiflavonPQbreak2}
- \mathcal{L}_{\cancel{PQ}} = \delta_{i1}\delta_{j1} \alpha'\(\frac{\phi}{M}\)^{ q'_{Q_i}- q'_{u_j}} \bar{Q}_i \tilde{H} u_j+ \text{h.c.} \ .
\eeq
This shifts the Yukawa and axion couplings with respect to those of the axiflavon model as follows:
\beq
Y_{u,ij}\to Y_{u,ij}+y'\delta_{i1}\delta_{j1} \ , \quad  \tilde Y_{u,ij}\to \tilde Y_{u,ij}+i(q'_{Q_i}-q'_{u_j})y'\delta_{i1}\delta_{j1} \ ,
\eeq
with $y'\equiv \(\frac{f}{\sqrt{2}M}\)^{ q'_{Q_1}- q'_{u_1}} \alpha'$, hence
\beq
\beta_{u,ij}= (q'_{Q_i}-q'_{u_j}-[q_{Q_i}-q_{u_j}])y'\delta_{i1}\delta_{j1} \ ,
\eeq
in the language of \eq{eq:axiflavonPQbreak}. Then, one finds that all our invariants are proportional to the one quantity which violates the PQ symmetry, namely $q_{Q_1}-q_{u_1}-\[ q'_{Q_1}- q'_{u_1}\]$. For instance,
\beq
I_u^{(1)}=\(q_{Q_1}-q_{u_1}-\[ q'_{Q_1}- q'_{u_1}\]\)\text{Im}\( y' Y_{11}^u{}^*\) \ .
\eeq
Another illuminating example arises when one considers the couplings in \eq{eq:axiflavon}, and changes $q_{Q_1}\to q'_{Q_1}$ in the up-quark coupling only. In this case, the quantity $ q'_{Q_1}- q_{Q_1}$ violates the PQ symmetry, but it is only resolved by invariants which are sensitive to the collective nature of PQ breaking, namely those which simultaneously involve $\tilde Y_u$ and $\tilde Y_d$. Indeed, the change $q_{Q_1}\to  q'_{Q_1}$ is a mere relabelling from the perspective of the up-quarks alone, but it breaks PQ when the down-quarks are taken into account. Consistently, we have
\beq
I_u^{(1)}=-\text{Im}\Tr\(\[\text{diag}\( q'_Q\)Y^u-Y^u\text{diag}\(q_u\)\]Y^u{}^\dagger\)=0 \ ,
\eeq
where $ q'_{Q_j}\equiv q_{Q_j}+\delta_{j1}\( q'_{Q_1}-q_{Q_1}\)$, whereas for instance
\beq
I_{ud}^{(1)}=\frac{1}{2i}\(q_{Q_1}-q'_{Q_1}\)\[X_u,X_d\]_{11} \ .
\eeq

\subsubsection{Two-Higgs-doublet model}

Another class of UV models that can embed an axion are two-Higgs-doublet models (2HDM) (see e.g.~\cite{Branco:2011iw} for a review). For definiteness, we consider a 2HDM of type II with the following PQ-preserving Lagrangian in the quark sector
\beq
\label{2HDMlag}
- \mathcal{L} =  \bar{Q} Y_u^{(1)} \tilde H_1 u + \bar{Q} Y_d^{(2)} H_2 d + \text{h.c.} \ . 
\eeq
The scalar potential is chosen to be invariant under a global $U(1)$ PQ symmetry. This fixes the PQ charges $q_{H_i}$ of the Higgses (up to a global normalization). The non-vanishing difference $q_{H_1}-q_{H_2}$ allows us to introduce PQ-breaking in the Yukawa sector, as we will see below.

After integrating out the massive Higgses as well as removing the gauge Goldstone bosons, one can describe the axion couplings as well as the fermion mass terms by the replacement
\beq
H_i=e^{iq_{H_i}\frac{a}{f}}\frac{v_i}{v}H \ , \text{ with } H=\bmat 0 \\ \frac{v}{\sqrt 2} \emat \ , \ v^2\equiv v_1^2+v_2^2 \ .
\eeq
The axion decay constant $f$ can be much larger than $v_{1,2}$, e.g. in DFSZ-like models~\cite{Dine:1981rt,Zhitnitsky:1980tq}, in which case other scalar fields appear in the scalar potential. This replacement reproduces the appropriate PQ transformations, $H_i\xrightarrow[]{\text{PQ}}e^{i\alpha_\text{PQ}q_{H_i}}H_i$ for $\alpha_\text{PQ}$ the transformation parameter, since $a\xrightarrow[]{\text{PQ}} a+2\pi\alpha_\text{PQ}f$. Without additional Lagrangian terms, we find as expected that our invariants vanish, which can be straightforwardly checked from the following matching expressions,
\beq
Y_u=\frac{v_1}{v}Y_u^{(1)} \ , \quad Y_d=\frac{v_2}{v}Y_d^{(2)} \ , \quad \tilde Y_u=-iq_{H_1}Y_u \ , \quad \tilde Y_d=iq_{H_2}Y_d \ .
\eeq
One can amend the Lagrangian so as to break the PQ symmetry, highlighting different aspects of our invariants. Starting with
\beq
- \mathcal{L}_{\cancel{PQ}} =  Y^{(2)}_{u,ij}\bar{Q}_i \tilde H_2 u_j + \text{h.c.} \text{ with } Y^{(2)}_{u,ij}=\delta_{i1}\delta_{j1}Y_{u,11}^{(2)} \ ,
\eeq
one shifts the Yukawa and axion couplings as follows,
\beq
Y_u=\frac{v_1}{v}Y_u^{(1)}+\frac{v_2}{v}Y_u^{(2)} \ , \quad \tilde Y_u=-iq_{H_1}\frac{v_1}{v}Y_u^{(1)}-iq_{H_2}\frac{v_2}{v}Y_u^{(2)} \ .
\eeq
One then finds that the up-sector-only invariants are proportional to real/imaginary parts of $(q_{H_1}-q_{H_2})Y_{u,11}^{(2)}$, as expected given the different ways to obtain an exact PQ symmetry in this sector (for a generic $Y_u^{(1)}$), e.g.
\beq
I_u^{(1)}=(q_{H_1}-q_{H_2})\frac{v_1v_2}{v^2}\text{Im}\Tr\(Y^{(2)}_uY_u^{(1)}{}^\dagger\)=-(q_{H_1}-q_{H_2})\frac{v_1v_2}{v^2}\text{Im}\(Y_{u,11}^{(2)}Y_{u,11}^{(1)}{}^*\) \ .
\eeq
As in the Froggatt--Nielsen case, one can observe collective effects at play: let us further assume that $Y_{u,1j}^{(1)}=0$, which is such that the up-quark couplings do not violate PQ, until the down-quarks are taken into account. Indeed, we find
\beq
I_u^{(1)}=0 \ ,
\eeq
whereas for instance
\beq
I_{ud}^{(1)}=-\frac{1}{2i}(q_{H_1} - q_{H_2})\frac{v_1v_2}{v^2}\[X_u, X_d\]_{11} \ .
\eeq

\subsubsection{Weakly-broken PQ symmetry}

Let us close by making a general statement for any model with an approximate PQ symmetry, characterized by a small breaking parameter $\epsilon$: the invariants of Eqs.~\eqref{eq:QuarkInv}-\eqref{eq:LeptonInv} are all $\epsilon$-suppressed. This follows from the linearity of our invariants emphasized at the end of section~\ref{section:invariantsPresentation}. Indeed, in models with a weakly-broken PQ symmetry, the Yukawa couplings are split into
\beq
\tilde Y=\tilde Y^{(\text{PQ})}+\tilde Y^{(\cancel{\text{PQ}})} \ ,
\eeq
where $\tilde Y^{(\cancel{\text{PQ}})}=\cO(\epsilon)$ and $\tilde Y^{(\text{PQ})}$ respects an exact axion shift symmetry, i.e. our invariants vanish when evaluated on $\tilde Y^{(\text{PQ})}$. Due to the linearity of the invariants, they are suppressed by $\epsilon$ as claimed.

\subsection{Connection to CPV}\label{section:CPV}

The possibility to reintroduce CP violation through the axion, which is constructed to solve the strong CP problem, has mostly been disregarded in the literature and has only gained more attention recently~\cite{DiLuzio:2020oah,Dekens:2022gha}. There is a close interplay between leading order CP violation and shift symmetry in the ALP Lagrangian. Adapting the results of~\cite{Bonnefoy:2021tbt}, we find the following necessary and sufficient conditions for CP to be conserved in the quark sector of the Yukawa basis of the ALP EFT
\beq
\begin{split}
J_4 & = L_{0000}(\tilde{Y}_x Y_x^{\dagger}) = L_{1000}(\tilde{Y}_x^{\mathstrut} Y_x^{\dagger}) = L_{0100}(\tilde{Y}_x^{\mathstrut} Y_x^{\dagger}) \\
& = L_{1100}(\tilde{Y}_x^{\mathstrut} Y_x^{\dagger}) = L_{0110}(\tilde{Y}_x^{\mathstrut} Y_x^{\dagger}) = L_{2200}(\tilde{Y}_x^{\mathstrut} Y_x^{\dagger}) \\
& = L_{0220}(\tilde{Y}_x^{\mathstrut} Y_x^{\dagger}) = L_{1220}(\tilde{Y}_x^{\mathstrut} Y_x^{\dagger}) = L_{0122}(\tilde{Y}_x^{\mathstrut} Y_x^{\dagger}) = 0
\end{split}
\eeq
with $L_{abcd}(\tilde{C}) \equiv \Re\Tr\( X_u^aX_d^bX_u^cX_d^d \tilde{C}\)$, $x=u,d$ and $J_4 = \Im\Tr([X_u,X_d]^3)$ is the Jarlskog invariant~\cite{Jarlskog:1985ht,Jarlskog:1985cw} which is the single object that captures all CP violation in the SM.
If this is compared with our set of shift symmetry invariants in \eq{eq:QuarkInv}, we find that all invariants but $I_{ud}^{(4)}$ can be expressed as combinations of the CP-odd invariants. For instance, $I_u^{(1)}=L_{0000}(\tilde{Y}_{u\vphantom{d}}^{\dagger}Y_{u\vphantom{d}}^{\mathstrut})$ and $I_{ud}^{(1)}=L_{0100}(\tilde{Y}_{u\vphantom{d}}^{\dagger}Y_{u\vphantom{d}}^{\mathstrut})+L_{1000}(\tilde{Y}_d^{\dagger}Y_d^{\mathstrut})$.
Therefore, most sources of leading order shift-breaking in the ALP EFT also source CP violation, hence CP conservation almost implies axion shift symmetry. This connection is only spoilt by $I_{ud}^{(4)}$, namely the one CP-even shift-symmetric invariant of our set that has to be included in order to obtain a full rank transfer matrix. Furthermore, the connection holds exactly in the lepton sector of the EFT.

In the degenerate cases, where the flavor symmetry of the SM is enlarged with respect to $U(1)_B\times U(1)_{L_i}$, CP conservation implies shift invariance at the level of the coefficients which can interfere with the dimension-four coefficients, i.e. at the level of observables computed at $\cO(1/f)$. It is however not sufficient for observables computed beyond that order. See appendix~\ref{appendix:degenerateCases} for more details.

Conversely, an exact shift symmetry also correlates sources of CP violation in the axion EFT. E.g., requiring that $I_{ud}^{(1)}$ vanishes implies that $L_{0100}(\tilde{Y}_{u\vphantom{d}}^{\dagger}Y_{u\vphantom{d}}^{\mathstrut})=-L_{1000}(\tilde{Y}_d^{\dagger}Y_d^{\mathstrut})$. These correlations, that emerge from the collectiveness of shift breaking, have an impact on CP violating observables like EDMs and allow us to relate the contributions of up and down quarks to those observables. If a sufficient amount of data from CP violating observables is available to constrain all parameters in the quark sector, these correlations would allow us to distinguish a shift symmetric ALP, for which the correlations are present, from a non-shift symmetric ALP. We will study the implications of axion shift invariance on EDMs in Sec.~\ref{IRRG}.

\subsection{Shift invariance below the electroweak scale or for a non-linearly realized electroweak symmetry}\label{section:IREFT}

As we saw previously, the conditions to be shift-symmetric are affected by gauge interactions. Indeed, the presence of electroweak interactions generated entangled conditions in the quark sector. Therefore, it is interesting to run the same analysis in the low-energy EFT, below the electroweak scale\footnote{Although we present the analysis for the low-energy EFT, it also applies to processes which involve the top-quarks, as long as no electroweak couplings contribute. In that case, all matrices remain $3\times 3$ complex matrices in flavor space.}
 (but above the QCD scale). 
 
At these low-energies, the gauge interactions reduce to those of electromagnetism and QCD, and the mass terms in the dimension-four Lagrangian to
\beq
\label{IRmasses}
\cL\supset -\bar u_L m_u u_R-\bar d_L m_d d_R-\bar e_L m_e e_R + h.c. \ ,
\eeq
where $m_{u,d,e}$ are ($2\times 2,3\times 3$ and $3\times 3$) complex matrices. The shift-symmetric dimension-five couplings to the axion are identical to those of \eq{eq:SMEFTa}, except that now $\psi\in\{(u,d,e)_{L,R}\}$, and the generic ones read
\beq
\label{eq:SMEFTaxionLowE}
\cL\supset -\frac{a}{f}\(\bar u_L \tilde m_u u_R+\bar d_L \tilde m_d d_R+\bar e_L \tilde m_e e_R + h.c.\) \ ,
\eeq
similarly to \eq{eq:SMEFTaxion}, with the notable difference that the up- and down-quarks sectors are decoupled. It then follows immediately from our previous analysis that the conditions for shift-invariance are similar to those of the lepton sector in the UV,
\beq
\label{IRshiftconditions}
I_x^{(i+1,IR)}\equiv\Tr(X_x^{i=0,1,...,N_x-1}\tilde m_x^{\mathstrut} m_x^{\dagger})=0 \ ,
\eeq
where $x=u,d,e$, $N_u=2,N_{d,e}=3$ and here $X_x\equiv m_xm_x^\dagger$.

The number of constraints in the IR reduces with respect to that in the UV: there are no longer conditions connecting the up- and down-sectors. That one gets strictly less conditions in the IR should not come as a surprise: we derived UV conditions under the assumptions that the axion couples to the degrees of freedom of the SM, which realize linearly the electroweak $SU(2)_L\times U(1)_Y$ symmetry (this was made transparent in \eq{eq:SMEFTa} by the use of a Higgs doublet $H$). However, the most general UV resolution of $SU(2)_L\times U(1)_Y$ may need to be phrased using the language of non-linear realizations of symmetries~\cite{Coleman:1969sm,Callan:1969sn}, which can be applied to the EW symmetry~\cite{Feruglio:1992wf} and its extension to axion couplings~\cite{Brivio:2017ije}. In this approach, one makes explicit use of the Goldstone boson multiplet $\pi^a$, which generates the longitudinal components of massive W and Z bosons, and which are conveniently packaged into a matrix $U$,
\beq
U=e^{i\pi^a\sigma^a/v} \ ,
\eeq
where $\sigma^a$ are the Pauli matrices and $v$ is the EW vev. $U$ has convenient transformations under $SU(2)_L\times U(1)_Y$,
\beq
U\to e^{i\(\alpha_Y+\alpha^a\sigma^a/2\)}U \ ,
\eeq
and the physical Higgs scalar $h$ is independently added to the theory as a gauge singlet. The usual linear realizations would be recovered by defining
\beq
H=U\bmat 0 \\ \frac{v+h}{\sqrt 2}\emat \ ,
\eeq
and by using $H$ only to write couplings. 

When $U$ and $h$ are treated independently, as it is the case for non-linear realizations, one can supplement the Lagrangian of \eq{eq:SMEFTa} by additional shift-invariant fermionic operators at dimension five (see~\cite{Brivio:2017ije} for a complete treatment),
\beq
\frac{\partial_{\mu}a}{f} \sum_{\psi=Q,L} \bar{\psi}\tilde c_\psi T \gamma^{\mu} \psi \ , \quad T\equiv U\sigma_3 U^\dagger \ .
\eeq
By working in unitary gauge where $U=\mathbb{1}$, it is clear that these operators allow one to decorrelate the couplings of the different components of an $SU(2)_L$ doublet. Instead, the axion-fermion couplings of the generic basis of \eq{eq:SMEFTaxion} map to
\beq
\frac{a}{f} \(\bar{Q}_L U\[K_Q+\sigma_3\tilde K_Q\]\bmat u_R\\ d_R \emat+\bar{L}_L U\[K_L+\sigma_3\tilde K_L\]\bmat 0\\ e_R \emat\) \ .
\eeq
and the number of building blocks is unchanged with respect to \eq{eq:SMEFTaxion}, which is seen by identifying
\beq
\tilde Y_{u,d}=K_Q\pm\tilde K_Q \ , \quad \tilde Y_e=K_L-\tilde K_L \ .
\eeq
Therefore, the conditions to be shift-invariant in a non-linear realization of the EW symmetry correspond to three copies (for $u,d,e$) of the lepton conditions of \eq{eq:diagonalConditionsLeptons}. 

This explains why a pure IR study of the shift-invariance properties of the Lagrangian of \eq{eq:SMEFTaxionLowE} cannot reproduce more than three copies of lepton-like conditions. Nevertheless, assuming a matching to a linear phase of the EW symmetry and an exact axion shift symmetry, we will show in section~\ref{IRRG} that more conditions remain valid at leading order under the RG flow. 

\section{Renormalization group evolution}\label{section:RGEs}

In previous sections, we presented flavor-invariant order parameters for the breaking of the axion shift symmetry. As any complete set of order parameters, it should be closed under the RG flow which preserves symmetry\footnote{The same statement applies for instance to flavor-invariant order parameters for CP violation in the SM. See Ref.~\cite{Feldmann:2015nia} for the RG-running of flavor-invariants in the quark sector of the SM, and Refs.~\cite{Yu:2020gre,Wang:2021wdq} in the lepton sector with Majorana neutrino masses.}. This is what we show in section~\ref{section:RGrunning}. In section~\ref{IRRG}, we descend to the IR EFT below the electroweak scale and find that the relations inherited from the UV under tree-level matching are maintained by the RG running below the electroweak scale, although they do not strictly follow from shift symmetry in the IR. We also revisit EDM bounds on CP-violating axion couplings under the assumption of an approximate shift symmetry. Finally, in section~\ref{section:SMEFTRGrunning} we illustrate the use of our invariants by working out sum-rules on the axion-induced RG running of SMEFT operators at dimension-six.

\subsection{Renormalization group running above the electroweak scale}\label{section:RGrunning}

To verify the completeness of our set of invariants, we can calculate their RG evolution under which the set should be closed. Using the RGEs of the components\footnote{Reference~\cite{Chala:2020wvs} restricts to CP-even ALP couplings, which translates into a real condition on $\tilde Y_{u,d,e}$. However, the RGEs presented in~\cite{Chala:2020wvs} can be directly upgraded to account for generic $\tilde Y$ (upon performing the replacement $a_{s\psi\phi}^T\to (a_{s\psi\phi}-ia_{\widetilde{s\psi\phi}})^\dagger$ for any fermion $\psi$, in the notation of this reference). This is due to the fact that a given diagram only depends on either $\tilde Y$ or $\tilde Y^\dagger$, according to whether the axion couples to a $\bar L R$ or $\bar R L$ vertex, where $L/R$ refers to left- and right-handed fermions respectively. Therefore, the presence of a transpose on $\tilde Y^T$, which comes from the assumption that $\tilde Y$ is real, is sufficient to pinpoint the diagrams which couple to $\tilde Y^\dagger$ and to extract their coefficient. As a consistency check of this replacement, the resulting RGEs display the appropriate flavor covariance and are consistent with the results of~\cite{Bauer:2020jbp}. In the present paper, we use this upgraded version of the RGEs.}~\cite{Chala:2020wvs,Bauer:2020jbp} of the invariants yields for the lepton invariants
\beq
\label{eq:LeptonRG}
\begin{split}
& \dot{I}_e^{(1)} = 2 \gamma_e^{\mathstrut} I_e^{(1)} + 6 I_e^{(2)} + 2 \Tr(X_e) \left( I_e^{(1)} + 3 (I_d^{(1)} - I_u^{(1)}) \right), \\
& \dot{I}_e^{(2)} = 4 \gamma_e^{\mathstrut} I_e^{(2)} + 9 I_e^{(3)} + 2 \Tr(X_e^2) \left( I_e^{(1)} + 3 (I_d^{(1)} - I_u^{(1)}) \right), \\
& \dot{I}_e^{(3)} = 6 \gamma_e^{\mathstrut} I_e^{(3)} + 12 I_e^{(4)} + 2 \Tr(X_e^3) \left( I_e^{(1)} + 3 (I_d^{(1)} - I_u^{(1)}) \right)
\end{split}
\eeq
where $\dot{I} = 16 \pi^2 \mu \frac{dI}{d\mu}$ and $\gamma_e = - \frac{15}{4}g_1^2 - \frac{9}{4} g_2^2 + \Tr\left( X_e + 3(X_u+X_d) \right)$. Furthermore, $I_e^{(4)} = \Re\Tr\left( X_e^3 \tilde{Y}_e^{\vphantom{3}} Y_e^{\vphantom{3}\smash{\dagger}} \right)$ is not independent from the invariants in \eq{eq:LeptonInv}, since due to the Cayley-Hamilton theorem any $n \times n$ matrix has to satisfy its characteristic equation and the $n$th power of the matrix can be can be expressed in terms of lower powers and traces of lower powers of the matrix. For a $3 \times 3$ matrix $A$, the Cayley-Hamilton theorem has the form~\cite{Jenkins:2009dy}
\beq
\label{eq:CHthm}
A^3 = A^2 \Tr A - \frac{1}{2} A \( (\Tr A)^2 - \Tr A^2 \) + \frac{1}{6} \mathbb{1} \( (\Tr A)^3 - 3 \Tr A^2 \Tr A + 2 \Tr A^3 \)
\eeq
which allows us to reexpress $I_e^{(4)}$ as follows
\beq
\label{eq:Ie4}
I_e^{(4)} = \Tr(X_e) I_e^{(3)} - \frac{1}{2} \left( (\Tr X_e)^2 - \Tr X_e^2 \right) I_e^{(2)} + \frac{1}{6} \left( (\Tr X_e)^3 - 3 \ \Tr X_e^2 \Tr X_e + 2 \Tr X_e^3 \right) I_e^{(1)}.
\eeq
Therefore, the set in \eq{eq:LeptonRG} does indeed form a closed set of differential equations and hence the set of lepton invariants in Eq.\eqref{eq:LeptonInv} is complete.\\
For the quark sector we find the following set of RGEs
\beq
\label{eq:QuarkRG}
\begin{split}
& \dot{I}_u^{(1)} = 2 \gamma_u^{\mathstrut} I_u^{(1)} + 6 I_u^{(2)} - 3 I_{ud}^{(1)} - 2 \Tr(X_u) \left( I_e^{(1)} + 3 (I_d^{(1)} - I_u^{(1)}) \right), \\
&\dot{I}_u^{(2)} =  4 \gamma_u^{\mathstrut} I_u^{(2)} + 9 I_u^{(3)} - 3 I_{ud,u}^{(2)} - 2 \Tr(X_u^2) \left( I_e^{(1)} + 3 (I_d^{(1)} - I_u^{(1)}) \right), \\
&\dot{I}_{u}^{(3)} = 6 \gamma_u^{\mathstrut} I_u^{(3)} + 12 I_u^{(4)} - 3 I_u^{'} - 2 \Tr(X_u^3) \left( I_e^{(1)} + 3 (I_d^{(1)} - I_u^{(1)}) \right), \\
& \dot{I}_d^{(1)} = 2 \gamma_d^{\vphantom{(}} I_d^{(1)} + 6 I_d^{(2)} - 3 I_{ud}^{(1)} + 2 \Tr(X_d) \left( I_e^{(1)} + 3 (I_d^{(1)} - I_u^{(1)}) \right), \\
&\dot{I}_d^{(2)} =  4 \gamma_d^{\vphantom{(}} I_d^{(2)} + 9 I_d^{(3)} - 3 I_{ud,d}^{(2)} + 2 \Tr(X_d^2) \left( I_e^{(1)} + 3 (I_d^{(1)} - I_u^{(1)}) \right), \\
&\dot{I}_{d}^{(3)} = 6 \gamma_d^{\vphantom{(}} I_d^{(3)} + 12 I_d^{(4)} - 3 I_d^{'} + 2 \Tr(X_d^3) \left( I_e^{(1)} + 3 (I_d^{(1)} - I_u^{(1)}) \right), \\
&\dot{I}_{ud}^{(1)} = 2(\gamma_{u\vphantom{d}}^{\mathstrut}+\gamma_d^{\mathstrut}) I_{ud}^{(1)}, \\
&\dot{I}_{ud,u}^{(2)} = (4\gamma_{u\vphantom{d}}^{\mathstrut}+2\gamma_d^{\mathstrut}) I_{ud,u}^{(2)} + 3 I_u^{'} - 6 I_{ud}^{(3)} - 2\Tr(X_u X_d X_u) \left( I_e^{(1)} + 3 (I_d^{(1)} - I_u^{(1)}) \right), \\
&\dot{I}_{ud,d}^{(2)} = (4\gamma_d^{\mathstrut}+2\gamma_{u\vphantom{d}}^{\mathstrut}) I_{ud,d}^{(2)} + 3 I_d^{'} - 6 I_{ud}^{(3)} + 2\Tr(X_d X_u X_d) \left( I_e^{(1)} + 3 (I_d^{(1)} - I_u^{(1)}) \right), \\
&\dot{I}_{ud}^{(3)} = 4(\gamma_{u\vphantom{d}}^{\mathstrut}+\gamma_d^{\mathstrut}) I_{ud}^{(3)}, \\
&\dot{I}_{ud}^{(4)} = 6 \left(\gamma_{u\vphantom{d}}^{\mathstrut}+\gamma_d^{\mathstrut} + \frac{1}{2} \Tr(X_u + X_d) \right) I_{ud}^{(4)} - \Im\Tr([X_u,X_d]^3)(I_u^{(1)}+I_d^{(1)}). \\
\end{split}
\eeq
where we have defined $\gamma_u \equiv -\frac{17}{12} g_1^2 - \frac{9}{4} g_2^2 - 8g_3^2 + \Tr(X_e + 3(X_u+X_d))$ and $\gamma_d \equiv -\frac{5}{12} g_1^2 - \frac{9}{4} g_2^2 - 8g_3^2 + \Tr(X_e + 3(X_u+X_d))$. 

Apart from $I_{u\vphantom{d}}^{(4)},I_{d}^{(4)}$ which are defined and can be decomposed in the same way as $I_e^{(4)}$ we find two more invariants, $I_{u}^{'} = \Re\Tr\left( (X_u X_d X_u + \{X_d^{\vphantom{2}},X_{u\vphantom{d}}^2\}) \tilde{Y}_u^{\mathstrut} Y_u^{\dagger} + X_{u\vphantom{d}}^3 \tilde{Y}_d^{\vphantom{3}} Y_d^{\vphantom{3}\smash{\dagger}}\right)$ and $I_{d}^{'}$ ($= I_{u}^{'}(u \leftrightarrow d)$), which can be decomposed into invariants which are already in the set and therefore vanish iff the couplings come from a shift invariant axion. For details on the decomposition and the form of $I_{u,d}^{'}$ in terms of invariants in the set, see App.~\ref{appendix:RGinvariants}. 

We also want to highlight the form of the RGE of the CP even invariant $I_{ud}^{(4)}$ which is strongly constrained since $I_{ud}^{(4)}$ is the only CP conserving invariant in the set. The invariant can only flow into itself and a set of CP odd invariants multiplied by the Jarlskog invariant $J_4 = \Im\Tr([X_u,X_d]^3)$ where the set of CP odd invariants is further constrained by the mass dimension of $I_{ud}^{(4)}$. This is exactly what we find in \eq{eq:QuarkRG}.

The minimal set in \eq{eq:MinSet}, which gives a full rank transfer matrix even for degenerate fermion masses, contains the sum of $I_{u\vphantom{d}}^{(3)}$ and $I_{d}^{(3)}$ which evolve by themselves under RG flow as can be seen in \eq{eq:QuarkRG}. Therefore, the RG evolution will not only generate $I_{u\vphantom{d}}^{(3)}+I_{d}^{(3)}$, which is contained in the minimal set in \eq{eq:MinSet}, but also $I_{u\vphantom{d}}^{(3)}-I_{d}^{(3)}$ and the set only closes under RG flow if the difference can be decomposed in terms of invariants in the minimal set. Following techniques described in Refs.~\cite{Trautner:2018ipq,Wang:2021wdq,Yu:2021cco}, we indeed find a CP-odd relation including all 11 invariants in the redundant set at dimension 12\footnote{Where the dimension is defined such that $\text{dim}(X_{u,d})=1$ as well as $\text{dim}(\tilde{Y}^{\phantom{3}}_{u,d} Y_{u,d}^{\vphantom{3}\smash{\dagger}})=1$.} of a similar form as \eq{eq:Ie4} that allows us to decompose the difference of $I_{u\vphantom{d}}^{(3)}$ and $I_{d}^{(3)}$ in terms of the remaining invariants. However, since at dimension 12 many combinations of the parameters are possible the relation is very complicated and we will not present it here.
With this relation we can always find a minimal set of invariants that is closed under RG flow. This is still the case for some degenerate cases where the relation becomes trivial, since the number of necessary primary coefficients is sufficiently small and we can start with a smaller minimal set for which it is straightforward to compute that it is closed under RG flow.

\subsection{RG running below the electroweak scale and EDM bounds}\label{IRRG}

We now discuss the RG running of shift-breaking flavor-invariants in the quark sector below the electroweak scale. The conditions for shift-invariance in the IR were established in \eq{IRshiftconditions}, and, using the formulae of~\cite{Chala:2020wvs}, it is straightforward to work out the RGEs of the associated set of flavor-invariants (assuming a single axion-vertex insertion, i.e. working at $\cO(1/f)$),
\beq
\dot I_x^{(n,IR)}=-12(1+n)(q_x^2e^2+C_Fg_3^2)I_x^{(n,IR)} \ ,
\eeq
with $x=u,d$, $q_x$ the electric charge and $C_F=\frac{N_c^2-1}{2N_c}=\frac{4}{3}$, with $N_c$ the number of colors. $e$ and $g_3$ are the electromagnetic and the $SU(3)_C$ coupling constants, respectively. The running is therefore consistent with the fact that, assuming an exact axion shift symmetry, $I_x^{(n,IR)}$ must vanish above and below the matching scale.

We explained in Section~\ref{section:IREFT} that the IR conditions for shift symmetry do not correlate the up and down sectors. Nevertheless, in a top-down approach assuming a matching to a linear phase of the EW symmetry and an approximate axion shift symmetry, it is possible to tune at the matching scale all UV flavor-invariants in \eq{eq:QuarkInv}, including those which do not belong to the IR set, to very small values. As we will see, we then find that the RG flow to smaller energies will respect the power counting imposed at the matching scale by the shift symmetry, at leading order.

Since we integrate out the top, it is convenient to use a flavor basis which describes mass eigenstates in the up sector, such as the basis of \eq{upBasis}. The matching relations between the couplings in Eqs.~\eqref{eq:SMEFTaxion}-\eqref{eq:SMlag} and those in the IR basis of Eqs.~\eqref{IRmasses}-\eqref{eq:SMEFTaxionLowE} then read
\beq
\label{matchingUVIR}
m_d=\frac{v}{\sqrt 2}\text{diag}(y_d,y_s,y_b) \ , \quad m_u=\frac{v}{\sqrt 2}\text{diag}(y_u,y_c) \ , \quad \tilde m_d=\frac{v}{\sqrt 2}V_\CKM^\dagger\tilde Y_d \ , \quad \tilde m_u= \frac{v}{\sqrt 2}\tilde Y_u^{2\times 2} \ , 
\eeq
where $v$ is the Higgs vev and $M^{2\times 2}$ refers to the first two rows and columns of any matrix $M$. When the shift symmetry is exact in the UV, namely that the relations in \eq{eq:ShiftSymCoup} hold, the matching conditions in \eq{matchingUVIR} imply that, at the matching scale, all the UV invariants keep on vanishing when one replaces $Y_u,\tilde Y_u$ with $Y_u^\text{IR},\tilde Y_u^\text{IR}$, where
\beq
Y_u^\text{IR}\equiv\bmat Y_u^{2\times 2} & 0 \\ 0 & 0 \emat \ , \quad \tilde Y_u^\text{IR}\equiv\bmat \tilde Y_u^{2\times 2} & 0 \\ 0 & 0 \emat \ .
\eeq
This follows from the fact that, in a basis such as that of \eq{upBasis} where $Y_u=\bmat Y_u^{2\times 2} & 0 \\ 0 & y_t \emat$, 
\beq
i\(\tilde Y_u^\text{IR}Y_u^\text{IR}{}^\dagger+h.c.\)=\[c_Q,Y_u^\text{IR} Y_u^\text{IR}{}^\dagger\] \ ,
\eeq
which has the same form as \eq{eq:CommRel}. 
As an example, we apply this replacement to the simplest UV invariant which connects the up and down sectors, $I_{ud}^{(1)}$, which yields
,
\beq
I_{ud}^{(1,IR)}\equiv\Re \Tr(V_\CKM^{\mathstrut} m_d^{\mathstrut} m_d^\dagger V_\CKM^\dagger\tilde m_{u\vphantom{d}}^{\mathstrut}m_{u\vphantom{d}}^\dagger+V_\CKM^{\mathstrut} \tilde m_d^{\mathstrut}m_d^\dagger V_\CKM^\dagger m_{u\vphantom{d}}^{\mathstrut}m_{u\vphantom{d}}^\dagger)=0 \ ,
\eeq
where one only sums over the two first rows of the CKM matrix. 

In order to study the fate of $I_{ud}^{(1,IR)}$ under RG running, we need to reinterpret its content in terms of IR data. In particular, there remains $V_\CKM$ in its expression, which is not an IR coupling per se but can be mapped to explicit IR couplings. Indeed, integrating out the heavy particles at tree-level in the SMEFT (supplemented by the couplings of \eq{eq:SMEFTaxion}), one finds at $\cO(1/v^2)$ only one four-quark operator which depends on the CKM matrix\footnote{After performing spinor and colour Fierz identities, we can identify the operator $\cO^{V1,LL}_{uddu,prst}$ and its octet form with the operators $\cO^{V1,LL}_{ud,prst},\cO^{V8,LL}_{ud,prst}$ of the LEFT basis of~\cite{Jenkins:2017jig}, and their Wilson coefficients with $L^{V1,LL}_{ud,prst},L^{V8,LL}_{ud,prst}$. More precisely, $\cO^{V1,LL}_{uddu,prst} = \frac{1}{N_c} \cO^{V1,LL}_{ud,ptsr} + 2 \cO^{V8,LL}_{ud,ptsr}$ and $\cO^{V8,LL}_{uddu,prst} = \frac{c_{F}}{N_c} \cO^{V1,LL}_{ud,ptsr} - \frac{1}{N_c} \cO^{V8,LL}_{ud,ptsr}$, where the octet operator $\cO^{V8,LL}_{uddu,prst}$ is not generated by integrating out the $W$ at tree level but will appear in the RG running of $\cO^{V1,LL}_{uddu}$.},
\beq
\cL\supset -\frac{4}{v^2}V_{\CKM,pr}^{\vphantom{*}}V_{\CKM,ts}^*\bar u_{L,p} \gamma^\mu d_{L,r} \bar d_{L,s} \gamma^\mu u_{L,t} \equiv L^{V1,LL}_{uddu,prst}\cO^{V1,LL}_{uddu,prst} \ ,
\eeq
as well as a semi-leptonic operator containing one CKM matrix. We can therefore reexpress our UV invariant $I_{ud}^{(1,IR)}$ as
\beq
\label{defIud1IR}
I_{ud}^{(1,IR)}\propto \Re\(L^{V1,LL}_{uddu,prst}\[\(m_d^{\mathstrut}m_d^\dagger\)_{rs}\(\tilde m_u^{\mathstrut}m_u^\dagger\)_{tp}+\(\tilde m_d^{\mathstrut}m_d^\dagger\)_{rs}\(m_u^{\mathstrut}m_u^\dagger\)_{tp}\]\)
\eeq
where every quantity appearing is now a genuine IR coupling. Assuming an axion shift symmetry, tree-level matching imposes $I_{ud}^{(1,IR)}=0$ at the electroweak scale, and it turns out that it remains zero at lower energies (at least to the one-loop, leading-log and $\cO({1}/{(fv^2)})$ order that we checked). Indeed, RG running of the axion couplings at $\cO(1/f)$~\cite{Chala:2020wvs} implies that
\beq
\bead
\mu\frac{d}{d\mu}&\[\(m_d^{\mathstrut}m_d^\dagger\)_{rs}\(\tilde m_u^{\mathstrut}m_u^\dagger\)_{tp}+\(\tilde m_d^{\mathstrut}m_d^\dagger\)_{rs}\(m_u^{\mathstrut}m_u^\dagger\)_{tp}\]\\
&=-12\(2C_Fg_3^2+(q_u^2+q_d^2)e^2\)\[\(m_d^{\mathstrut}m_d^\dagger\)_{rs}\(\tilde m_u^{\mathstrut}m_u^\dagger\)_{tp}+\(\tilde m_d^{\mathstrut}m_d^\dagger\)_{rs}\(m_u^{\mathstrut}m_u^\dagger\)_{tp}\] \ ,
\eead
\eeq 
while the running of the four-fermion operators at $\cO(1/v^2)$ can be restricted to operators of the type vector current-vector current~\cite{Jenkins:2017jig,Jenkins:2017dyc}, provided the matching  at the electroweak scale only generates those, which is the case here at tree level. Indeed, the vector-vector Wilson coefficients, which we generically denote $L_{prst}$, do not contribute to the running of coefficients of another kind (such as scalar, tensor or dipole operators). They run into themselves, other $L_{prst}$ and structures such as $\delta_{pt}L_{wrsx}$. Therefore, flavor-invariants of the form
\beq
\label{usefulForRG1}
\Re\(L_{prst}\[\(m_d^{\mathstrut}m_d^\dagger\)_{rs}\(\tilde m_u^{\mathstrut}m_u^\dagger\)_{tp}+\(\tilde m_d^{\mathstrut}m_d^\dagger\)_{rs}\(m_u^{\mathstrut}m_u^\dagger\)_{tp}\]\)
\eeq
run into themselves, identical invariants with other $L_{prst}$ as well as invariants of the form
\beq
\label{usefulForRG2}
\Re\(\delta_{pt}L_{wrsw}\[\(m_d^{\mathstrut}m_d^\dagger\)_{rs}\(\tilde m_u^{\mathstrut}m_u^\dagger\)_{tp}+\(\tilde m_d^{\mathstrut}m_d^\dagger\)_{rs}\(m_u^{\mathstrut}m_u^\dagger\)_{tp}\]\) \ .
\eeq
This second kind of invariants runs into itself as well as structures where $\delta_{pt}L_{wrsw}\to \delta_{pt}\delta_{rs}L_{wxxw}$, where the last factor can be taken out of the real part due to the hermitian properties of the vector-vector operators. Finally, invariants of the form $L_{wxxw}$ only run into themselves. 
Therefore, taking into account the running and the boundary conditions of the 4-fermion operators at the matching scale, these IR flavor-invariant structures form a RG-closed space. In particular, if they all vanish at the matching scale, they all remain equal to zero at lower energies (at the order that we checked). Similarly, if they are suppressed by a small parameter at the matching scale, they remain suppressed by that parameter at lower energies. In addition, as anticipated above, the fact that they vanish (or are suppressed) follows from an exact (or approximate) axion shift symmetry. Indeed, at the matching scale, all $L_{prst}$ have a flavor structure given by combinations of $V_{\CKM,pr}^{\mathstrut}V_{\CKM,ts}^*$, $V_{\CKM,pr} \delta_{st}$ and $\delta_{pt}\delta_{rs}$. Hence, at the matching scale, the above IR invariants are proportional to combinations of $I_{ud}^{(1,IR)}$ and $\Tr(m_d^{\mathstrut}m_d^\dagger)I_{u}^{(1,IR)}+\Tr(m_{u\vphantom{d}}^{\mathstrut}m_{u\vphantom{d}}^\dagger)I_{d}^{(1,IR)}$ and all vanish for an exact shift symmetry. Therefore, assuming an exact (or approximate) axion shift symmetry in the UV makes all the above IR invariants, in particular $I_{ud}^{(1,IR)}$, vanishing (or small) at the matching scale as well as at any lower energy. 

The stability of the constraints under RG flow allows us to use them at low-energies, and to idenfity the impact of an approximate shift symmetry on bounds on the couplings of \eq{eq:SMEFTaxionLowE} derived from observables. The consequences are twofold: {\it (i)} the fundamental parameter space constrained by the bounds is reduced, and {\it (ii)} sum rules between different observables are predicted. We illustrate these two aspects by reanalysing the bounds derived in~\cite{DiLuzio:2020oah}, where the authors study electric dipole moments and allow for shift-breaking couplings in the generic basis of \eq{eq:SMEFTaxionLowE}. 

Bounds on the ALP couplings follow from bounds on the spin-precession frequency $\omega_\text{ThO}$ of the polar molecule ThO, the neutron EDM $d_n$ and the EDM $d_\text{Hg}$ of the diamagnetic atom ${}^{199}$Hg,
\beq
\label{boundsLuca}
\bead
\omega_\text{ThO} < 1.3\,\text{mrad/s ($90\%$ C.L.)} \ , \\ 
d_n< 1.8\times 10^{-26}\,e\,\text{cm ($90\%$ C.L.)} \ , \\
d_\text{Hg}<6\times 10^{-30}\,e\,\text{cm ($90\%$ C.L.)}\ .
\eead
\eeq
The expressions of these quantities are given in~\cite{DiLuzio:2020oah} in terms of the couplings\footnote{Matching to our notations, $y_S$ and $y_P$ of~\cite{DiLuzio:2020oah} are respectively the hermitian and anti-hermitian parts of $\tilde m$, for each kind of fermion.} $\tilde m$ of \eq{eq:SMEFTaxionLowE}, as well as CP-even and odd couplings of the axion to gluons and photons, under the assumption that the axion mass is of order a few GeV's so that QCD can be treated perturbatively. 

Let us study the fate of these bounds when the axion shift symmetry is approximate. More precisely, we assume that any shift-breaking coupling is generated by particles at the PQ scale $f$ and that a single spurion $\epsilon$ breaks the PQ symmetry. For instance, this suggests writing $m_a^2=\cO(\epsilon f^2)$ for the axion mass, or $\cI=\cO(\epsilon)$ for any of our shift-breaking invariants $\cI$. Working in the basis where masses are diagonal and real, the $\epsilon$-scaling of the IR invariants imply for instance that
\beq
\label{IRconstraintDiagonal}
(\tilde m_x^{\mathstrut}+\tilde m_x^\dagger)_{ii}=\cO(\epsilon) \ ,
\eeq
for any fermion species $x$. This reduces the number of fundamental coefficients which contribute and are bounded. Additionally, CP-odd axion-gluon or axion-photon couplings break the shift symmetry, and are therefore $\cO(\epsilon)$. Working at order $\cO(\epsilon^0/f^2)$ (we also assume $v^2/f^2\lesssim \epsilon$ and count $\log\epsilon=\cO(1)$), one obtain from~\cite{DiLuzio:2020oah} that
\beq
d_\text{Hg}\simeq 4\times 10^{-4}d_n \ ,
\eeq
which is an example of a sum-rule between observables following from the axion shift-symmetry. It follows from the fact that the only non-vanishing CP-odd interactions induced at one-loop by a shift-symmetric axion above the QCD scale are the axion-induced EDMs $d_i$ or chromo-EDMs $d_i^C$ of fundamental fermions $\psi_i$, which read~\cite{DiLuzio:2020oah},
\beq
\frac{d_i}{e}=Q_id_i^C \ , \quad d_i^C\simeq \frac{2}{16\pi^2f^2}\text{Im}\(\tilde mm^{-1}\tilde m\)_{ii} \ ,
\eeq
with $Q_i$ the electric charge. The same combination of quark (chromo-)EDMs enters $d_n$ and $d_\text{Hg}$, hence the related bounds yield constraints on the same combination of fundamental parameters $\tilde m$. These constraints turn out to be of very similar magnitude (see below).

More generally, the bounds of \eq{boundsLuca} are bound on these EDMs of fundamental fermions, for instance the bound on $\omega_\text{ThO}$ turns into a bound on the electron EDM,
\beq
d_e\lesssim 10^{-29}\,e\,\text{cm} \ .
\eeq
It is natural to expect that the contribution of the tau lepton dominates\footnote{This follows from the generic form of axion couplings at $\cO(\epsilon^0)$ given in \eq{eq:ShiftSymCoup}, and it can be checked in the Froggatt--Nielsen or 2HDM examples above that, generically, $\tilde m_{e,13}\sim \cO(1)m_\tau$.}, and one finds
\beq
\abs{\frac{\text{Im}\(\tilde m_{e\tau}\tilde m_{\tau e}\)}{m_\tau^2}}=\abs{\text{Im}\(c_{L,13}c_{e,31}\)} \lesssim 1.4\(\frac{f}{10^8 \text{GeV}}\)^2 \ ,
\eeq
where we expressed the bound in terms of the CP-violating couplings of \eq{eq:SMEFTa}, which can be used at $\cO(\epsilon^0)$ we are interested in, while at higher orders in $\epsilon$ also potential shift-breaking terms can contribute which are not captured by the $c_{Q,u,d,L,e}$. 
Similarly, the bounds on $d_n$ and $d_\text{Hg}$ reduce to bounds  of similar magnitude on the same combination of quark EDMs,
,
\beq
\(\frac{2}{3}0.784(28)-0.55(28)\)  d_u^C-\(1.10(55)-\frac{1}{3}0.294(11)\) d_d^C\lesssim 10^{-26} \text{cm} \ .
\eeq
Although it suffers from large uncertainties, the coefficient factoring the up-contribution is numerically suppressed ($\sim 0.03$), so that its top-mediated component does not dominate over the bottom contribution, due to the numerical value of $m_b/m_t$ which is $\sim 0.02$. Therefore, we expect the generic bound to combine the top and bottom contribution. For the sake of illustration, let us assume that the bottom contribution dominates, which at $\cO(\epsilon^0)$ leads to 
\beq
\abs{\frac{\text{Im}\(\tilde m_{db}\tilde m_{bd}\)}{m_b^2}}=\abs{\text{Im}\(c_{Q,13}c_{d,31}\)} \lesssim 1.1\times 10\(\frac{f}{10^7 \text{GeV}}\)^2 \ .
\eeq
In addition, the $\epsilon$-scaling of $I_{ud}^{1,IR}$ of \eq{defIud1IR} further correlates the entries of $\tilde m$ which contribute to the different EDMs. However, this does not generate more sum-rules between observables at $\cO(\epsilon^0/f^2)$.

\subsection{ALP-SMEFT interference and sum rules}\label{section:SMEFTRGrunning}

If the presence of an axion is detected, it will be crucial to learn more about its couplings, in particular to SM fermions. However, those may be difficult to probe, and indirect probes will play an important role in constraining them. The axion-induced RG running of couplings between SM particles is a good example, as it will generically deviate from that of a situation without any axion.

Assuming no further light degree of freedom, the couplings of SM particles can be captured by the Standard Model Effective Field Theory (SMEFT). The presence of an axion induces an RG evolution driven by the dimension-five axion couplings~\cite{Galda:2021hbr}, which deviates from that in the pure SMEFT~\cite{Grojean:2013kd,Jenkins:2013zja,Jenkins:2013wua,Alonso:2013hga}. In this section, we ask: how can one extract from these deviations clear information regarding the axion couplings? Our answer illustrates the use of having algebraic and explicit conditions for an axion shift-symmetry. Indeed, our invariants allow us to immediately identify implications
of the axion shift-symmetry, in the form of flavor-invariant sum-rules on the RG evolution of the SMEFT Wilson coefficients.

We define the terms sourcing the deviations from the SMEFT RGEs induced by the ALP EFT as follows~\cite{Galda:2021hbr},
\beq
\mu\frac{d C_i^{\text{SMEFT}}}{d\mu} - \gamma_{ji}^{\text{SMEFT}} C_j^{\text{SMEFT}} \equiv \frac{S_i}{(4\pi f)^2} \ .
\eeq
We can use the following source terms~\cite{Galda:2021hbr}
\begin{equation}
\label{eq:DipoleSourceTerms}
\begin{split}
S_{uG} = -4i g_s \tilde{Y}_u C_{GG}, \qquad & \qquad S_{dG} = -4i g_s \tilde{Y}_d C_{GG}, \\
S_{uW} = -i g_2 \tilde{Y}_u C_{WW}, \qquad & \qquad S_{dW} = -i g_2 \tilde{Y}_d C_{WW}, \\
S_{uB} = -2i g_1 (y_Q+y_u) \tilde{Y}_u C_{BB}, \qquad & \qquad S_{dB} = -2i g_1 (y_Q+y_d) \tilde{Y}_d C_{BB}, 
\end{split}
\end{equation}
to write
\begin{equation}
\begin{split}
& \Im\Tr\left( X_{x}^{0,1,2} S_{xG}^{\mathstrut} Y_{x}^{\dagger} \right) = -4g_s C_{GG} I_{x}^{(1,2,3)}  \\
& \Im\Tr\left( X_{x}^{0,1,2} S_{xW}^{\mathstrut} Y_{x}^{\dagger} \right) = -g_2 C_{WW} I_{x}^{(1,2,3)} \\
& \Im\Tr\left( X_{x}^{0,1,2} S_{xB}^{\mathstrut} Y_{x}^{\dagger} \right) = -g_1 (y_Q+y_x) C_{BB} I_{x}^{(1,2,3)} \\
\end{split}
\end{equation}
with $x=u,d$ and all Wilson coefficients which have not been defined previously can be found in Tab.~\ref{tab:addOperators}. Furthermore, we can use the source terms in \eq{eq:DipoleSourceTerms} which only depend on the type of the fermion through the dimension-5 Yukawa to write down relations for the mixed invariants. For the gluon source terms we find e.g.
\begin{equation}
\begin{split}
& \Im\Tr\left( X_d^{\mathstrut}X_{u\vphantom{d}}^{\mathstrut}X_d^{\mathstrut} S_{uG}^{\mathstrut} Y_{u\vphantom{d}}^{\dagger} + X_{u\vphantom{d}}^{\mathstrut}X_d^{\mathstrut}X_{u\vphantom{d}}^{\mathstrut} S_{dG}^{\mathstrut} Y_d^{\dagger} \right) = -4 g_s C_{GG} I_{ud}^{(3)}
\end{split}
\end{equation}
and similar expressions for the $W$-boson source terms. Furthermore, we can find relations in terms of our invariants where the ALP-fermion couplings appear in tensor products. This is the case for some 4-fermion operators. E.g. combining the source terms~\cite{Galda:2021hbr}
\begin{equation}
\begin{split}
\(S_{qx}^{(1)}\)_{prst} = \frac{1}{N_c} \(\tilde{Y}_x^{\mathstrut}\)_{pt} \(\tilde{Y}_x^{\dagger\mathstrut}\)_{sr} + \frac{16}{3} g_1^2 y_Q y_x C_{BB}^2 \delta_{pr} \delta_{st}
\end{split}
\end{equation}
with $x=u,d$ we can find the following relation
\begin{equation}
\begin{split}
    \Re\Bigg( \(S_{qu}^{(1)}\)_{prst} \(Y_u^{\dagger\mathstrut}\)_{tp} \(Y_u^{\mathstrut}\)_{rs} - \frac{y_u \Tr(X_u)}{y_d \Tr(X_d)} & \(S_{qd}^{(1)}\)_{prst} \(Y_d^{\dagger\mathstrut}\)_{tp} \(Y_d^{\mathstrut}\)_{rs} \Bigg) \\
    & = \frac{1}{N_c} \( \(I_u^{(1)}\)^2 - \frac{y_u \Tr(X_u)}{y_d \Tr(X_d)} \(I_d^{(1)}\)^2 \).
\end{split}
\end{equation}
Finally, with the source term~\cite{Galda:2021hbr}
\begin{equation}
\(S_{ledq}^{\mathstrut}\)_{prst} = -2 \(\tilde{Y}_{e\vphantom{d}}^{\mathstrut}\)_{pr} \(\tilde{Y}_d^{\dagger}\)_{st}
\end{equation}
we can write
\begin{equation} 
\Re \( \( S_{ledq}^{\mathstrut} \)_{prst} \( Y_{e\vphantom{d}}^{\dagger\mathstrut} \)_{rp} \( Y_d^{\mathstrut} \)_{ts} \right) = - 2 I_{e}^{(1)} I_d^{(1)}.
\end{equation}
\renewcommand{\arraystretch}{1.35}
\begin{table}[t!]
  \begin{center}{}
  \begin{tabular}{c}
  \textbf{SMEFT} \\
$\cO_{q_RG} = \bar{Q} \sigma^{\mu\nu} T^a q_R H_{(q_R)} G_{\mu\nu}^a$\\
$\cO_{q_RW} = \bar{Q} \sigma^{\mu\nu} \sigma^I q_R H_{(q_R)} W_{\mu\nu}^I$ \\
$\cO_{q_RB} = \bar{Q} \sigma^{\mu\nu} u H_{(q_R)} B_{\mu\nu}$\\
$\cO_{qq_R}^{(1)} = \(\bar{Q} \gamma_{\mu} Q\) \(\bar{q_R}\gamma^{\mu}q_R \)$ \\
$\cO_{ledq} = \(\bar{L}e\) \(\bar{d}Q\)$\\
  \end{tabular}
  \hspace{5mm}
  \begin{tabular}{c}
    \textbf{ALP EFT} \\
	$\cO_{GG} = \frac{a}{f} G_{\mu\nu}^a \tilde{G}^{\mu\nu,a}$\\
	$\cO_{WW} = \frac{a}{f} W_{\mu\nu}^I \tilde{W}^{\mu\nu,I}$\\
	$\cO_{BB} = \frac{a}{f} B_{\mu\nu} \tilde{B}^{\mu\nu}$\\
	\\
	\\
  \end{tabular}
  \hspace{5mm}
  \end{center}
  \caption{Additional EFT operators of the SMEFT and ALP EFT as defined in Ref.~\cite{Galda:2021hbr} that enter in the sum rules and have not been defined previously in this paper. $q_R$ stands for $u,d$ and $H_{(u)}\equiv \tilde H,H_{(d)}\equiv H$.} 
  \label{tab:addOperators}
\end{table}
\renewcommand{\arraystretch}{1}
The sum rules of this type give zeroes in the RG evolution of the SMEFT Wilson coefficients if the ALP is shift symmetric, i.e. the RG evolution of the precise combination of SMEFT Wilson coefficients appearing in the sum-rule is completely determined by SMEFT Wilson coefficients. Said differently, observing RGEs compatible with the SMEFT for the combinations of Wilson coefficients entering the above sum-rules suggests that the axion shift symmetry is weakly broken. The uncertainty in the measurements of the SMEFT coefficients quantifies which room there remains for non-vanishing invariants, i.e. for shift-symmetry breaking.

\section{Couplings to gluons and non-perturbative shift-invariance}
\label{section:thetaQCD}

Previously, we have focussed on the breaking of shift-invariance which arises at the perturbative level. This is for instance relevant for interactions which induce axion potentials at the tree or loop levels, as is often discussed in the axion quality problem or relaxion literature. We have, however, neglected axion couplings to gauge bosons of the SM gauge group $SU(3)_C\times SU(2)_W\times U(1)_Y$, and in particular to gluons. The latter do not break the shift symmetry at the perturbative level, but they do so \textit{non-perturbatively}. Furthermore, in PQ-symmetric models, they are related via mixed PQ-$SU(3)_C$ anomalies to axion couplings to fermions, which is the main focus of this paper. Therefore, we work out in this appendix conditions for the axion couplings to remain shift-symmetric when one also considers gluons.

Let us thus add to the Lagrangian of \eq{eq:SMEFTaxion} a term
\beq
\label{eq:gluonCouplings}
-\frac{C_gg_3^2}{16\pi^2}\frac{a}{f}\Tr( G_{\mu\nu}\tilde G^{\mu\nu}) \ ,
\eeq
and to that of \eqref{eq:SMEFTa} the same term with $C_g\to C_g^{(s)}$. We use explicitly different notations for clarity, since both couplings will appear in the same relations when we match between the two operator bases. $G$ is the gluon field strength, $\tilde G^{\mu\nu}\equiv \frac{\epsilon^{\mu\nu\rho\sigma}}{2}G_{\rho\sigma}$ is its dual, $g_3$ is the $SU(3)_C$ coupling constant and we chose the overall normalization consistently with naive dimensional analysis, and with the origin of $C_g,C_g^{(s)}$ in UV theories with heavy anomalous fermions, which are such that $C_g,C_g^{(s)}=\cO(1)$. 

The gluon coupling breaks the shift symmetry \textit{non-perturbatively}, unless at least one quark is massless. In that case, a shift of the axion field $a\to 2\pi\alpha_\text{PQ}f$ is equivalent to a shift of $\theta_\text{QCD}\to 2\pi\alpha_\text{PQ}$, which can be absorbed with an appropriate chiral transformation of the massless quark. Therefore, we assume here that all quarks are massive, so that there are no chiral symmetries of the spectrum, and $\theta_\text{QCD}$ is physical. I.e. the theory differentiates between different values of the axion vev $\langle a \rangle$ and the shift symmetry is broken, generating an axion potential. 

We now follow the same logic as in the perturbative case: we look for quantities which must vanish for the shift invariance to hold non-perturbatively, which are therefore order parameters for the non-perturbative breaking.

\subsection{Non-perturbative order parameter}
\label{section:thetaQCDOrder}

We thus assume that the axion shift symmetry is exact. It is therefore unbroken at the perturbative level and one can work in the basis of \eq{eq:SMEFTa} for axion-fermion couplings, where all fermion couplings are unchanged by a shift of $a$. At the non-perturbative level, one needs to require $C_g^{(s)}=0$ to cancel the gluon-induced shift-breaking contributions, for instance to the axion potential. 

However, as in previous sections, we want to identify order parameters in the most general operator basis, where they could be non-zero in realistic models. This means that we want to derive the equivalent of the condition $C_g^{(s)}=0$ in terms of the couplings of \eq{eq:SMEFTaxion}. For that, we need to account for anomalies when matching from \eq{eq:SMEFTa} to \eq{eq:SMEFTaxion}, which is achieved by the following field redefinition,
\beq
\psi'\equiv e^{-i\frac{a}{f}c_\psi}\psi \ ,
\eeq
for each chiral fermion field $\psi$. This transformation is anomalous and generates the following matching relation between the coupling to gluons, 
\beq
\label{anomalousShift}
C_g=C_g^{(s)}+\Tr(2c_Q-c_u-c_d) \ .
\eeq
When the gluon couplings do not break the PQ symmetry, $C_g^{(s)}=0$ and one finds\footnote{Notice that the slight redundancy of the basis \eq{eq:SMEFTa} is irrelevant here: the freedom to add at will $\alpha_B(\partial_\mu a/f) J^\mu_B$, where $J_B$ is the baryon number current, discussed in section~\ref{section:counting}, shifts $c_{Q,u,d}\to c_{Q,u,d}+(\alpha_B/3)\mathbb{1}$ and leaves $\Tr(2c_Q-c_u-c_d)$ unchanged. This is nothing else but the statement that the baryon number has no mixed anomalies with $SU(3)_C$.}
\beq
\label{relationNoAnomaly}
C_g=\Tr(2c_Q-c_u-c_d)
\ .
\eeq
Using the matching conditions of \eq{eq:ShiftSymCoup}, we can substitute the coefficients $c$ for the $\tilde Y$ and obtain
\beq
\label{constraintNonPerturbativeNonShiftBasis}
C_g-i\Tr\(Y_{u\vphantom{d}}^{-1}\tilde Y_{u\vphantom{d}}^{\vphantom{-1}}+Y_d^{-1}\tilde Y_d^{\vphantom{-1}}\)=0 \ .
\eeq 
Note that our assumption of massive quarks make the Yukawa matrices invertible and the expression meaningful. This expression yields an extra condition for a perturbative shift symmetry to remain valid even non-perturbatively in $g_3$, in the basis of \eq{eq:SMEFTaxion}. It can be rewritten in terms of positive powers of the Yukawa matrices, upon using a relation valid for $3\times 3$ matrices, used for instance in~\cite{Yu:2022ttm},
\beq
X^{-1}=\frac{X^2-\(\Tr X\) X+\frac{1}{2}\[\(\Tr X\)^2-\Tr X^2\]\mathbb{1}}{\det X}\ .
\eeq
Taking $X=YY^\dagger$, we can use it in order to keep the flavor-invariant nature of the constraint explicit,
\beq
\Tr\(Y^{-1}\tilde Y\)=\Tr\(X^{-1}\tilde YY^\dagger\)=\frac{\Tr\(X^2\tilde YY^\dagger\)-\Tr X\Tr\(X\tilde YY^\dagger\)+\frac{1}{2}\[\(\Tr X\)^2-\Tr X^2\]\Tr\(\tilde YY^\dagger\)}{\det X} \ .
\eeq
The right-handside of the above expression is constrained to be imaginary due to our conditions for perturbative shift-invariance of \eq{eq:QuarkInv}, so we find that the new condition is CP-even and reads $I_g=0$ for
\beq
I_g \equiv C_g+\Im\Tr\(Y_{u\vphantom{d}}^{-1}\tilde Y_{u\vphantom{d}}^{\vphantom{-1}}+Y_d^{-1}\tilde Y_d^{\vphantom{-1}}\) \ .
\eeq
When all the perturbative invariants of Eqs.~\eqref{eq:QuarkInv}-\eqref{eq:LeptonInv} vanish, i.e. when there exists a PQ symmetry at the perturbative level, $I_g$ captures the mixed anomaly polynomial of that symmetry with $SU(3)_C$. This can easily be seen in the axiflavon/flaxion model of \eq{eq:axiflavon}, where
\beq
I_g=\sum_i\(2q_{Q_i}-q_{u_i}-q_{d_i}\) \ .
\eeq
The invariant $I_g$, which features couplings from the up and down sectors, highlights a new kind of collective breaking at the non-perturbative level, which is consistent with the fact that mixed PQ anomalies can be cancelled by balancing non-vanishing contributions in different quark sectors.

In addition, the derivation never referred to the invariants which correlate the up and down sectors in Eq.~\eqref{eq:QuarkInv} and are absent in \eq{IRshiftconditions}, therefore it is valid below the electroweak scale, up to the replacements $Y,\tilde Y\to m,\tilde m$ to match the notations of section~\ref{section:IREFT}.

\subsection{RG running}
\label{section:thetaQCDRunning}

By, once more, using the RGEs of the Standard Model and axion Yukawa couplings above the electroweak scale~\cite{Bauer:2020jbp,Chala:2020wvs}, we can show that all contributions to the running of this invariant cancel at the one-loop level
\beq
\label{eq:Igrunning}
\mu \frac{dI_g}{d\mu} = 0 \ .
\eeq
Let us stress that we chose a scaling in \eq{eq:gluonCouplings} similar to that of~\cite{Bauer:2020jbp}, where $C_g$ already comes with a one-loop factor $g_3^2/(16\pi^2)$. This allowed us to account for the anomalous shift without loop-factor hierarchies in \eq{relationNoAnomaly}. However, when working out the RGEs as in~\cite{Chala:2020wvs}, that implies that we also need to account for anomalies and their contribution to the running of $C_g$. They yield an extra running of $C_g$,
\beq
\label{eq:CgRunning}
\mu\frac{dC_g}{d\mu}=\frac{1}{16\pi^2}\[-4\,\Im\Tr\(\tilde Y_{u\vphantom{d}}^{\mathstrut}Y_{u\vphantom{d}}^\dagger+\tilde Y_d^{\mathstrut}Y_d^\dagger\)+\frac{22g_1^4}{16\pi^2}C_B+\frac{27g_2^4}{16\pi^2}C_W+\frac{128g_3^4}{16\pi^2}C_g\] \ ,
\eeq
where $C_{W,B}$ are the equivalent of $C_g$ for gauge fields of $SU(2)_W$ and $U(1)_Y$, respectively\footnote{This result reproduces that of appendix~A of~\cite{Bauer:2020jbp}.}. The running of $C_g$ is then cancelled by the additional Yukawa contribution to $I_g$. We give more details on this running as well as other possible contributions of similar magnitude, i.e. at $\cO\(\tilde YY^{\dagger},g^4C\)$, in appendix~\ref{appendix:higherOrderRunningCg}. If one instead absorbs a factor $g_3^2/(16\pi^2)$ in $C_g\equiv \frac{16\pi^2}{g_3^2}\tilde C_g$, and works like in~\cite{Chala:2020wvs} with RGEs at $\cO(g_3^2/(16\pi^2))$ under the assumption that $\tilde C_g=\cO(1)$, one is led to define the invariant
\beq
\tilde I_g\equiv \frac{g_3^2}{16\pi^2}I_g=\tilde C_g+\frac{g_3^2}{16\pi^2}\Im\Tr\(Y_{u\vphantom{d}}^{-1}\tilde Y_{u\vphantom{d}}^{\vphantom{-1}}+Y_d^{-1}\tilde Y_d^{\vphantom{-1}}\) \ ,
\eeq
which runs into itself at the one-loop order~\cite{Chala:2020wvs},
\beq
\mu \frac{d\tilde I_g}{d\mu}=\mu \frac{d\tilde C_g}{d\mu}=-\frac{14g_3^2}{16\pi^2} \tilde I_g
\ ,
\eeq
where the running is fully induced by the running of the strong coupling.

The situation below the electroweak scale is very similar, with the exception that there are no more Higgs loops to consider in the running, which results in the absence of Yukawa contributions above.

\section{Conclusions}\label{conclusions}

In this paper, we have investigated the implications of an axion shift-symmetry on the dimension-5 axion couplings to the Standard Model fermions. In particular, we have found explicit and algebraic conditions implied by the shift symmetry on these couplings, instead of the implicit relations that are well-known in the literature. The set of constraints is formulated in a flavor-invariant way and gives necessary and sufficient conditions for shift-symmetry to hold, hence yielding a set of 13 order parameters for shift symmetry in the dimension-5 axion EFT.  Our results make it explicit that the axion shift symmetry is a collective effect: our set of invariants connects the up- and down-quark sectors, which can be traced to the presence of electroweak gauge interactions. Upon calculating the RGEs of all invariants, we showed that the set is closed under RG flow, consistently with the fact that it captures a complete set of order parameters for shift symmetry breaking.

We have then illustrated how our set of flavor-invariants capture the sources of shift-symmetry breaking by studying explicit UV scenarios matched onto the ALP EFT. In particular, we checked that our set of invariants vanishes for a UV completion that respects a PQ symmetry -- i.e. a shift symmetry -- for the axion, and we highlighted that the presence of collective effects is made particularly transparent in our framework: UV scenarios which give rise to collective effects only make non-zero the invariants that include both up- and down-type couplings. Furthermore, we scrutinized the connection to CP violation, and we found that all but one order parameters for the axion shift symmetry are CP-odd. We have also studied the impact of taking limits of degenerate masses or texture zeros, and we showed that in all cases our set of invariants captures the sources of shift-symmetry breaking which can interfere with the SM parameters. Finally, we emphasized that the collective aspects of the axion shift symmetry are lost when one assumes a non-linear realization of electroweak symmetry breaking, which explains why it is also absent in the low-energy EFT where all heavy particles -- in particular the $W$ boson -- are integrated out. Nevertheless, assuming a UV completion in terms of a linear realization of the electroweak symmetry imposes extra constraints, which are shown to be stable under RG flow in the IR. This allows one to use the constraints at low-energy, where the coefficients are constrained by low-energy observables.

Finally, we have extended the discussion to the non-perturbative breaking of the PQ symmetry induced by the axion couplings to gluons, and have identified the order parameter for this breaking in various operator bases. In situations with a PQ symmetry at the classical level, this order parameter naturally captures the mixed anomaly between the PQ and the $SU(3)_C$ symmetries. We have also argued that it does not run at the one-loop order.

There are several ways in which our work can be extended. One could first make further connections to phenomenology, in particular related to CP-odd observables. Indeed, we argued that all but one parameters which break the axion shift-symmetry are CP-odd, hence they can contribute to CP-odd observables. Relatedly, it would be interesting to identify the impact of the collective nature of shift-symmetry breaking at the level of observables, and to further study the interplay between the flavorful axion couplings studied in this paper, shift-breaking bosonic couplings of the axion and the axion mass. One could also compute the RG running below the electroweak scale at next-to-leading orders, in order to see to which precision the matching conditions to a linear realization of the electroweak symmetry are preserved by the RG flow. Finite matching contributions at loop-level could also be included.

\section*{Acknowledgments}

We thank Emanuele Gendy and Josh Ruderman for numerous discussions on flavor invariants. We also thank Emanuele Gendy for providing us with a code which automatizes the search for relations between flavor invariants. This work is supported by the Deutsche Forschungsgemeinschaft under Germany’s Excellence Strategy EXC 2121 “Quantum Universe” - 390833306. This work has been also partially funded by the Deutsche Forschungsgemeinschaft under the grant 491245950.

\appendix
\section{Useful matrix relations}\label{appendix:matrixRelations}

\subsection{Commutator relations used in section~\ref{section:quarkInvariants}}\label{appendix:commutatorRelations}

The simplest commutator identity one can write down using three matrices $A,B,C$ is
\beq
\label{eq:Comm3}
\[A, BC\] = \[A,B\]C + B\[A,C\] \ .
\eeq
Using \eq{eq:CommRel} and the fact that the trace of a commutator vanishes, we obtain
\beq
-i \Tr\(X_d \[c_Q,X_u^{\mathstrut}\] + X_u \[c_Q,X_d^{\mathstrut}\] \) = \Tr\(X_d^{\mathstrut} \( \tilde{Y}_{u\vphantom{d}}^{\mathstrut} Y_{u\vphantom{d}}^{\dagger} + Y_{u\vphantom{d}}^{\mathstrut} \tilde{Y}_{u\vphantom{d}}^{\dagger} \) + \( u \leftrightarrow d\) \) = 0 \ .
\eeq
It is straightforward to generalise the identity in \eq{eq:Comm3} to four and five matrices and obtain identities at higher order in $X_{u,d}$. For any four matrices $A,B,C,D$ we have
\beq
\[A,BCD \] = \[A,B\]CD + B\[A,C\]D + BC\[A,D\].
\eeq
Identifying $A=c_Q,B=X_u,C=D=X_d$ and tracing over both sides gives 
\beq
\Tr\(X_d^{2\mathstrut} \( \tilde{Y}_{u\vphantom{d}}^{\mathstrut} Y_{u\vphantom{d}}^{\dagger} + Y_{u\vphantom{d}}^{\mathstrut} \tilde{Y}_{u\vphantom{d}}^{\dagger} \) + \{ X_{d}^{\mathstrut},X_{u\vphantom{d}}^{\mathstrut} \} \( \tilde{Y}_d^{\mathstrut} Y_d^{\dagger} + Y_d^{\mathstrut} \tilde{Y}_d^{\dagger} \) \) = 0 \ .
\eeq
This expression is not symmetric under $u \leftrightarrow d$, allowing us to find another independent condition by exchanging $u \leftrightarrow d$,
\beq
\Tr\(X_{u\vphantom{d}}^{2\mathstrut} \( \tilde{Y}_d^{\mathstrut} Y_d^{\dagger} + Y_d^{\mathstrut} \tilde{Y}_d^{\dagger} \) + \{X_d^{\mathstrut},X_{u\vphantom{d}}^{\mathstrut} \} \( \tilde{Y}_{u\vphantom{d}}^{\mathstrut} Y_{u\vphantom{d}}^{\dagger} + Y_{u\vphantom{d}}^{\mathstrut} \tilde{Y}_{u\vphantom{d}}^{\dagger} \) \) = 0 \ .
\eeq
The following identity involving five generic matrices $A,B,C,D,E$,
\beq
\label{eq:Comm5}
\[A,BCDE \] = \[A,B\]CDE + B\[A,C\]DE + BC\[A,D\]E + BCD\[A,E\] \ ,
\eeq
allows us to find a fourth condition
\beq
\Tr\(X_d^{\mathstrut} X_{u\vphantom{d}}^{\mathstrut} X_d^{\mathstrut} \( \tilde{Y}_{u\vphantom{d}}^{\mathstrut} Y_{u\vphantom{d}}^{\dagger} + Y_{u\vphantom{d}}^{\mathstrut} \tilde{Y}_{u\vphantom{d}}^{\dagger} \) + X_{u\vphantom{d}}^{\mathstrut} X_d^{\mathstrut} X_{u\vphantom{d}}^{\mathstrut} \(\tilde{Y}_d^{\mathstrut} Y_d^{\dagger} + Y_d^{\mathstrut} \tilde{Y}_d^{\dagger}\)\) \ ,
\eeq
and the final condition which we consider derives from applying the Jacobi identity on \eq{eq:CommRel},
\beq
\begin{split}
\[X_{u\vphantom{d}}^{\mathstrut},\tilde Y_d^{\mathstrut}Y_d^\dagger+Y_d^{\mathstrut}\tilde Y_d^\dagger\]-\[X_d^{\mathstrut},\tilde Y_{u\vphantom{d}}^{\mathstrut}Y_{u\vphantom{d}}^\dagger+Y_{u\vphantom{d}}^{\mathstrut}\tilde Y_{u\vphantom{d}}^\dagger\]=-i \, \bigg(\[X_{u\vphantom{d}}^{\mathstrut},\[c_Q^{\mathstrut},X_d^{\mathstrut}\]\]+&\[X_d^{\mathstrut},\[ X_{u\vphantom{d}}^{\mathstrut},c_Q^{\mathstrut}\]\]\bigg) \\
& =i\[c_Q^{\mathstrut},\[X_d^{\mathstrut},X_u^{\mathstrut}\]\] \ ,
\end{split}
\eeq
so that \eq{twoMatIdentity} yields
\beq
\Tr\(\[X_{u\vphantom{d}}^{\mathstrut},X_d^{\mathstrut}\]^n\(\[X_{u\vphantom{d}}^{\mathstrut},\tilde Y_d^{\mathstrut}Y_d^\dagger+Y_d^{\mathstrut}\tilde Y_d^\dagger\]-\[X_d^{\mathstrut},\tilde Y_{u\vphantom{d}}^{\mathstrut}Y_{u\vphantom{d}}^\dagger+Y_{u\vphantom{d}}^{\mathstrut}\tilde Y_{u\vphantom{d}}^\dagger\]\)\)=0 \ .
\eeq
We only make use of the condition where $n=2$.

The above expressions can be made more compact by noticing that, for any two hermitian matrices $H_u,H_d$, we can write
\beq
\begin{split}
 \Tr \left(H_{u\vphantom{d}}^{\mathstrut} (\tilde{Y}_{u\vphantom{d}}^{\mathstrut} Y_{u\vphantom{d}}^{\dagger} + Y_{u\vphantom{d}}^{\mathstrut} \tilde{Y}_{u\vphantom{d}}^{\dagger}) + H_d^{\mathstrut} (\tilde{Y}_d^{\mathstrut} Y_d^{\dagger} + Y_d^{\mathstrut} \tilde{Y}_d^{\dagger}) \right) & = \Tr\left( H_{u\vphantom{d}}^{\mathstrut} \tilde{Y}_{u\vphantom{d}}^{\mathstrut} Y_{u\vphantom{d}}^{\dagger} + Y_{u\vphantom{d}}^{\mathstrut} \tilde{Y}_{u\vphantom{d}}^{\dagger} H_{u\vphantom{d}}^{\dagger} + (u \leftrightarrow d) \right) \\
= \Tr\left( H_{u\vphantom{d}}^{\mathstrut} \tilde{Y}_{u\vphantom{d}}^{\mathstrut} Y_{u\vphantom{d}}^{\dagger} \right) + \Tr\left( H_{u\vphantom{d}}^{\mathstrut} \tilde{Y}_{u\vphantom{d}}^{\mathstrut} Y_{u\vphantom{d}}^{\dagger} \right)^* + (u \leftrightarrow d) & = 2 \Re\Tr\left( H_{u\vphantom{d}}^{\mathstrut} \tilde{Y}_{u\vphantom{d}}^{\mathstrut} Y_{u\vphantom{d}}^{\dagger} \right) + (u \leftrightarrow d) \ .
\end{split} 
\eeq
For two anti-hermitian matrices $A_u,A_d$, one similarly finds
\beq
 \Tr \left(A_{u\vphantom{d}}^{\mathstrut} (\tilde{Y}_{u\vphantom{d}}^{\mathstrut} Y_{u\vphantom{d}}^{\dagger} + Y_{u\vphantom{d}}^{\mathstrut} \tilde{Y}_{u\vphantom{d}}^{\dagger}) + A_d^{\mathstrut} (\tilde{Y}_d^{\mathstrut} Y_d^{\dagger} + Y_d^{\mathstrut} \tilde{Y}_d^{\dagger}) \right) = 2 i \Im\Tr\left( A_{u\vphantom{d}}^{\mathstrut} \tilde{Y}_{u\vphantom{d}}^{\mathstrut} Y_{u\vphantom{d}}^{\dagger} \right) + (u \leftrightarrow d) \ .
\eeq

\subsection{Details on decomposition of invariants generated by RG flow}
\label{appendix:RGinvariants}

In the RGEs of the invariants we find invariants which naively are not in the minimal set. However they can be decomposed into invariants in the set which we will show here in detail. 
Apart from $I_{u\vphantom{d}}^{(4)},I_{d}^{(4)}$ which can be decomposed in an analogous way as $I_e^{(4)}$, the RG evolution also generates, $I_{u}^{'} = \Re\Tr\( \(X_{u\vphantom{d}}^{\mathstrut} X_d^{\mathstrut} X_{u\vphantom{d}}^{\mathstrut} + \{X_d^{\mathstrut},X_{u\vphantom{d}}^{2\mathstrut}\}\) \tilde{Y}_{u\vphantom{d}}^{\mathstrut} Y_{u\vphantom{d}}^{\dagger} + X_{u\vphantom{d}}^3 \tilde{Y}_d^{\vphantom{3}} Y_d^{\vphantom{3}\smash{\dagger}}\)$ and $I_{d}^{'}$ ($= I_{u\vphantom{d}}^{'}(u \leftrightarrow d)$), which are redundant as we will see now. As before we can construct the invariants generated by the RG flow of the original set by using again the commutator relation in \eq{eq:Comm5} with $A=c_Q, B=C=E= X_u,D=X_d$ for $I_u^{'}$ and $A=c_Q, B=C=E= X_d,D=X_u$ for $I_d^{'}$. To see that the invariants are not independent of the invariants in \eq{eq:QuarkInv} we have to employ the Cayley-Hamilton theorem. Multiplying \eq{eq:CHthm} by $A$, taking the trace, replacing $A \rightarrow A+B+C$ and only keeping terms of order $A^2BC$ we find the following relation~\cite{Jenkins:2009dy}
\begin{equation}
\label{eq:CHthm2}
\begin{split}
0 = & \Tr(A)^2\Tr(B)\Tr(C)-\Tr(BC)\Tr(A)^2-2\Tr(AB)\Tr(A)\Tr(C)-2\Tr(AC)\Tr(A)\Tr(B) \\
& +2\Tr(ABC)\Tr(A) + 2\Tr(ACB)\Tr(A)-\Tr(A^2)\Tr(B)\Tr(C)+2\Tr(AB)\Tr(AC) \\
& +\Tr(A^2)\Tr(BC) +2\Tr(C)\Tr(A^2B)+2\Tr(B)\Tr(A^2C)-2\Tr(A^2BC)-2\Tr(A^2CB) \\
& -2\Tr(ABAC).
\end{split}
\end{equation}
By identifying $A=X_u,B=X_d,C=\tilde{Y}_u Y_u^{\dagger}$ the last three single trace terms in \eq{eq:CHthm2} are the same as the terms containing $\tilde{Y}_u Y_u^{\dagger}$ in $I_u^{'}$. Using this decomposition and \eq{eq:CHthm} for the $X_u^3$ term in $I_u^{'}$ we find
\begin{equation}
\label{eq:Iuprime}
\begin{split}
I_u^{'} = & \frac{1}{2} I_u^{(1)} \left( \Tr(X_u)^2 \Tr(X_d) - \Tr(X_u^2) \Tr(X_d) + 2 \Tr(X_u^2X_d) - 2 \Tr(X_u)\Tr(X_uX_d) \right) \\
& + 2 I_u^{(2)} \left( \Tr(X_uX_d) - \Tr(X_u) \Tr(X_d) \right) + 2 \Tr(X_d) I_u^{(3)} + \frac{1}{2} \left( \Tr(X_u^2) - \Tr(X_u)^2 \right) I_{ud}^{(1)} \\
& + \Tr(X_u) I_{ud,u}^{(2)} + \frac{1}{6} \left( \Tr(X_u)^3 - 3 \Tr(X_u^2)\Tr(X_u) + 2 \Tr(X_u^3) \right) I_d^{(1)} \ ,
\end{split}
\end{equation}
and a similar decomposition for $I_d^{'}$. 

\section{Parameter counting in degenerate cases}
\label{appendix:degenerateCases}

In section~\ref{section:shiftSymmetry}, we focused on the case where the fermion masses were non-degenerate. We also evaluated the number of independent conditions for shift symmetry when the CKM matrix is generic. We now relax those assumptions, which can enlarge the conserved flavor symmetry at the level of the dimension-four Lagrangian (we still assume non-vanishing masses). The main outcome of this analysis is that our set of invariants \eq{eq:MinSet} captures all conditions for shift-invariance at leading order, even in those degenerate cases.

Let us clarify what we mean by "conditions for shift-invariance at leading order" in the previous sentence. The EFT power counting allows us to perform a splitting between the different sources of shift-symmetry breaking, according to whether they can be observed in observables computed at a given order in the EFT expansion. In this section, we will be interested in observables computed at leading $\cO(1/f)$  order, namely in SM amplitudes squared and in interference between the SM (dimension-four) and $\cO(1/f)$ contribution to the amplitudes. It turns out that not all coefficients of the EFT can be probed at leading order, even if they are associated to dimension-five operators. The reason is that observables are invariant under flavor symmetries of the scattering states, hence they can only depend on combinations of coefficients which are invariant under these symmetries. A good example is provided by lepton numbers. EFT coefficients which are charged under those can only contribute to observables when they are combined with other charged coefficients. However, since there is not any such charged coefficient in the SM, charged EFT coefficients cannot interfere with the SM and cannot contribute to observables at $\cO(1/f)$. We use the vocabulary of~\cite{Bonnefoy:2021tbt} and call the kind of coefficients which can interfere at $\cO(1/f)$ {\it primary coefficients}, while the others are called {\it secondary coefficients}.

We can apply this decomposition to physical sources of shift-symmetry breaking, in which case the primary sources are in one-to-one correspondence with flavor-invariants linear in the EFT Wilson coefficients at $\cO(1/f)$. All our invariants verify this criterion, therefore the discussion of the previous section shows that all sources of shift-symmetry breaking are primary when fermion masses and mixings are generic (which corresponds to a SM flavor symmetry $U(1)_B\times U(1)_{L_i}$). It is a straightforward extension of the discussion of the previous section to check that they capture all primary sources of shift-symmetry breaking, for any flavor symmetry of the dimension-four Lagrangian\footnote{When masses are non-degenerate, flavor symmetries are abelian and depend on the texture zeros in the CKM matrix. When masses are non-degenerate, flavor symmetries can be non-abelian. See~\cite{Bonnefoy:2021tbt} for more details.}. We provide in Table~\ref{tableRanksDegenerate} a counting of the physical quantities in the EFTs of \eq{eq:SMEFTaxion} and \eq{eq:SMEFTa}, as well as the primary ones, and upon comparing the numbers we derive the expected rank of an appropriate set of linear flavor-invariant order parameters for shift-symmetry breaking. As said above, our set fulfills the criterion of capturing all primary sources of shift-symmetry breaking in all degenerate cases. For completeness, we sketch in appendix~\ref{appendix:nonLinearDegenerate} how one can capture all sources of shift-symmetry breaking, primary and secondary, using non-linear flavor invariants.

\begin{table}[h!]
\small
\centering
\renewcommand{\arraystretch}{1.3}
\resizebox{\columnwidth}{!}{%
\begin{tabular}{c|c|c|c|c|c|c|c|c|c|c|c|c}
&\multicolumn{4}{c|}{Shift-symmetric Wilson coefficients $c_{Q,u,d}$}&\multicolumn{4}{c|}{Generic Wilson coefficients $\tilde Y_{u,d}$}&\multicolumn{4}{c}{Number of constraints}\\
\cline{2-13}
&\multicolumn{2}{c|}{All}&\multicolumn{2}{c|}{Primary}&\multicolumn{2}{c|}{All}&\multicolumn{2}{c|}{Primary}&\multicolumn{2}{c|}{All}&\multicolumn{2}{c}{\begin{tabular}{c}Primary\\(\# of indep. invariants)\end{tabular}}\\
\hline
$\begin{matrix} \text{Flavor symmetries of}\\\text{the quark sector of the SM}\end{matrix}$&CP-even&CP-odd&CP-even&CP-odd&CP-even&CP-odd&CP-even&CP-odd&CP-even&CP-odd&CP-even&CP-odd\\
\hline
$U(1)_B$&$17$&$9$&$17$&$9$&$18$&$18$&$18$&$18$&1&9&1&9\\
$U(1)^2$&$16$&$8$&$10$&$3$&$18$&$17$&$10$&$10$&$2$&$9$&0&7\\
$U(1)^3$&$15$&$7$&$6$&0&$18$&$16$&$6$&$6$&$3$&$9$&0&6\\
$U(2)\times U(1)$&$13$&$5$&$4$&0&$17$&$15$&$4$&$4$&$4$&$10$&0&4\\
$U(3)$&$9$&$1$&$2$&0&$15$&$13$&$2$&$2$&$6$&$12$&0&2\\
\end{tabular}
}\\
\vspace{10pt}
\resizebox{\columnwidth}{!}{%
\begin{tabular}{c|c|c|c|c|c|c|c|c|c|c|c|c}
&\multicolumn{4}{c|}{Shift-symmetric Wilson coefficients $c_{L,e}$}&\multicolumn{4}{c|}{Generic Wilson coefficients $\tilde Y_e$}&\multicolumn{4}{c}{Number of constraints}\\
\cline{2-13}
&\multicolumn{2}{c|}{All}&\multicolumn{2}{c|}{Primary}&\multicolumn{2}{c|}{All}&\multicolumn{2}{c|}{Primary}&\multicolumn{2}{c|}{All}&\multicolumn{2}{c}{\begin{tabular}{c}Primary\\(\# of indep. invariants)\end{tabular}}\\
\hline
$\begin{matrix} \text{Flavor symmetries of}\\\text{the lepton sector of the SM}\end{matrix}$&CP-even&CP-odd&CP-even&CP-odd&CP-even&CP-odd&CP-even&CP-odd&CP-even&CP-odd&CP-even&CP-odd\\
\hline
$U(1)^3$&$9$&$4$&$3$&0&$9$&$7$&$3$&$3$&$0$&$3$&0&3\\
$U(2)\times U(1)$&$7$&$2$&$2$&0&$8$&$6$&$2$&$2$&$1$&$4$&0&2\\
$U(3)$&$3$&$0$&$1$&0&$6$&$4$&$1$&$1$&$3$&$4$&0&1\\
\end{tabular}
}
\caption{Counting of the physical coefficients at dimension-five in the EFTs of \eq{eq:SMEFTaxion} and \eq{eq:SMEFTa}, as a function of the flavor symmetry of the dimension-four Lagrangian. See the text for details.}
\label{tableRanksDegenerate}
\end{table}

Let us give an explicit example of how the numbers in Table~\ref{tableRanksDegenerate} are obtained, focusing on the flavor symmetry $U(2)\times U(1)$ in the lepton sector of the SM. This flavor symmetry arises when two lepton masses are degenerate, which we take to be $m_e=m_\mu$ without loss of generality. We work in the mass basis, where the symmetry acts as
\beq
\label{U2U1action}
L/e \to \bmat U^{(2)} & 0 \\ 0 & e^{i\xi_3} \emat L/e \ , \quad U^{(2)}U^{(2)}{}^\dagger=\mathbb{1} \ , \ \xi_3 \in \mathbb{R} \ .
\eeq
Starting with the shift-symmetric EFT of \eq{eq:SMEFTa}, we first use the freedom to add to the theory a conserved $U(2)\times U(1)$ current,
\beq
A_{ij}\partial_\mu a\(\bar L_i \gamma^\mu L_j +\bar e_i \gamma^\mu e_j\) \ , \quad A=\bmat \alpha^{k=0,...,3}\sigma^k & 0 \\ 0 & \alpha \emat \ .
\eeq
With this, one shifts $c_{L/e}\to c_{L/e}+A$, allowing us to choose $c_{L,ij=1,2}=-c_{e,ij=1,2},c_{L,33}=-c_{e,33}$. Then, one can act on the fields with \eq{U2U1action} and choose $c_{L,12}=c_{e,12}=0$. After this flavor basis choice, one can still act with the $U(1)^3$ part of \eq{U2U1action}, bringing us back to the discussion of the beginning of section~\ref{section:counting}. We then consider the $U(1)^3$-rephasing-invariant quantities $c_{L/e,ii},\abs{c_{L/e,ij}},\arg(c_{e,ij}c_{L,ji})$ ($i<j$) and $\arg(c_{L,12}c_{L,23}c_{L,31})$, but according to our basis choice, only $c_{L,ii},\abs{c_{L/e,13/23}},\arg(c_{e,13/23}c_{L,31/32})$ remain, corresponding to 7 CP-even and 2 CP-odd quantities. Among those, only two can interfere with the $U(2)\times U(1)$-invariant SM coefficients, i.e. only two are primary, $\delta^{ij=1,2}c_{L,ij},c_{L,33}$. Turning to the EFT of \eq{eq:SMEFTaxion}, the freedom to add a conserved current coupled to the axion current is not present anymore, but \eq{U2U1action} can still be used to ensure that $\tilde Y_{e,12}=\pm \tilde Y_{e,21}^*$, which leaves behind the $U(1)^3$ part of $U(2)\times U(1)$. The rephasing-invariant given in section~\ref{section:counting} are $\tilde Y_{e,ii}$, $\abs{\tilde Y_{e,ij}},\arg(\tilde Y_{e,ij}\tilde Y_{e,ji})$ ($i< j$) and $\arg(\tilde Y_{e,12}\tilde Y_{e,23}\tilde Y_{e,31})$, among which $\abs{\tilde Y_{e,21}}$ and $\arg(\tilde Y_{e,12}\tilde Y_{e,21})$ become redundant due to our choice of flavor basis. Therefore, there are $8$ CP-even and $6$ CP-odd physical quantities, among which $2$ and $2$ are primary, corresponding to the real and imaginary parts of $\delta^{ij=1,2}\tilde Y_{e,ij},\tilde Y_{e,33}$. All those numbers are consistent with Table~\ref{tableRanksDegenerate}.

\section{Non-linear flavor-invariants in degenerate cases}
\label{appendix:nonLinearDegenerate}

Focusing on the lepton case for simplicity, we discuss how to identify flavor-invariant order parameters which account for any kind of physical source of shift-symmetry breaking, be it primary or secondary. 

We already said that, for non-degenerate lepton masses, the invariants of \eq{eq:diagonalConditionsLeptons} are sufficient. Let us now consider the case of two degenerate lepton masses. This condition on the masses can be algebraically encoded in the two following flavor-invariant conditions,
\beq
\Delta(\chi)=0 \ , \quad \Delta_0\(\chi\)\neq 0 \ ,
\eeq
where $\Delta$ is the discriminant and $\Delta_0$ a resultant\footnote{For a cubic polynomial $P=ax^3+bx^2+cx+d$, one has
\beq
\Delta=18abcd-4b^3d+b^2c^2-4ac^3-27a^2d^2 \ , \quad \Delta_0=b^2-3ac \ .
\eeq} of $\chi_e$, the characteristic polynomial of $X_e$ (whose coefficients are flavor-invariants).

Using the freedom to appropriately define $c_L$, the constraints from \eq{eq:conditionsLeptonsMassBasis} which remains on $\tilde Y_e$ read, in the basis where $Y_e$ is diagonal and the two first flavors are degenerate in mass,
\beq
\bmat \tilde Y_{e,11}&\tilde Y_{e,12}\\ \tilde Y_{e,21}&\tilde Y_{e,22}\emat \text{ is anti-hermitian and } \tilde Y_{e,33} \in i\mathbb{R} \ ,
\eeq 
i.e. the two first rows and columns of the hermitian part of $\tilde Y_e$ vanishes. However, there are only three genuinely physical conditions on $\tilde Y_e$. Indeed, in this degenerate limit, the flavor symmetry of the SM Yukawas increases to $U(2)\times U(1)$, acting as
\beq
L\to \bmat U_e^{(2)}&0\\0&e^{i\xi_e}\emat \cdot L \ , \quad e\to \bmat U_e^{(2)}&0\\0&e^{i\xi_e}\emat \cdot e \ ,
\eeq
Therefore, $U_e^{(2)}$ can be used to make the two first rows and columns of the hermitian part of $\tilde Y_e$ diagonal. Said differently, the only physical components of this upper-left $2\times 2$ submatrix, which should vanish for an exact shift-symmetry, are its eigenvalues. 

The conditions in \eq{eq:diagonalConditionsLeptons} are no longer sufficient to obtain an exact shift-symmetry: they collapse to two independent conditions in the limit of two degenerate masses. This is understood as follows: to get a quantity which violates the shift-symmetry, we need to build a non-zero $U(2)$-invariant quantity out of the two first rows and columns of the hermitian part of $\tilde Y_e$. Such an expression, if it is linear with respect to $\tilde Y_e$, can only be the trace, and indeed that is what is captured by the invariants in \eq{eq:diagonalConditionsLeptons}. To obtain the second constraint on that part of $\tilde Y_e$, one can form
\beq
\label{eq:firstLeptonNonLinearConstraint}
\Tr[M_e^2]=0 \ , \quad M_e\equiv X_e\(\tilde Y_e^{\mathstrut}Y_e^\dagger+h.c.\)X_e-\frac{m_{1,2}^2m_3^2}{(m_{1,2}^2-m_3^2)^2}\[\[\tilde Y_e^{\mathstrut}Y_e^\dagger+h.c.,X_e\],X_e\] \ ,
\eeq
where $m_{1,2}$ and $m_3$ are the degenerate and non-degenerate masses respectively, which can be expressed in terms of the (flavor-invariant) coefficients of $\chi_e$\footnote{Precisely, for $\chi=ax^3+bx^2+cx+d$ with a double root $m_{12}^2$ and a single root $m_3^2$, we have
\beq
m_{1,2}^2=\frac{9ad-bc}{2\Delta_0} \ , \quad m_3^2=\frac{4abc-9a^2d-b^3}{2\Delta_0} \ .
\eeq}.
As we discussed earlier, there is a qualitative difference between constraints which are linear with respect to $\tilde Y_e$ and the other ones: the linear invariants are the only sources of shift-symmetry breaking which can be probed in observables computed at $\cO(1/f)$. Therefore, the constraint of \eq{eq:firstLeptonNonLinearConstraint} can only be accessible at $\cO(1/f^2)$. 

When all electron-type masses are degenerate (which is encoded in the conditions $\Delta(\chi)=\Delta_0\(\chi\)= 0$), one finds the constraint
\beq
\exists U_e \text{ s.t. }U_e^\dagger \(\tilde Y_e^{\mathstrut} Y_e^{\dagger}+h.c.\) U_e^{\mathstrut}=0 \ ,
\eeq
which simply means that $\tilde Y_eY_e^\dagger$ is anti-hermitian. The conditions in \eq{eq:diagonalConditionsLeptons} reduce to one condition. To capture all conditions, it suffices to impose
\beq
\Tr[\(\tilde Y_e^{\mathstrut}Y_e^\dagger+h.c.\)^2]=0 \ ,
\eeq
but as previously, this cannot be accessed at $\cO(1/f)$.

\section{Anomalous and higher-order contributions to the RGEs}
\label{appendix:higherOrderRunningCg}

In this appendix, we give details on the RGE of $C_g$ in the basis of \eq{eq:SMEFTaxion} used in section~\ref{section:thetaQCDRunning}. We also argue that there is no additional contribution of the same magnitude arising from two-loop diagrams.

We first justify the anomalous RGE of \eq{eq:CgRunning}. Although it does not appear explicitly in~\cite{Chala:2020wvs}, as it corresponds to a higher-order effect in the power counting of this reference, it can be straightforwardly derived given the formulae contained there. In particular, the anomalous RGE arises when one accounts for anomalies in the equations of motion (eom) used to reduce the Green basis to the non-redundant basis of \eq{eq:SMEFTaxion}. More precisely, \cite{Chala:2020wvs} makes use of the operators
\beq
\label{redundantGreenOps}
\cL\supset r_{sq}^{\alpha\beta}\cR_{sq}^{\alpha\beta}+r_{\tilde{sq}}^{\alpha\beta}\cR_{\tilde{sq}}^{\alpha\beta} \ ,
\eeq
where $r_{sq}^{\alpha\beta},r_{\tilde{sq}}^{\alpha\beta}$ are arbitrary real coefficients, $\alpha,\beta$ are summed over and
\beq
\cR_{sq}^{\alpha\beta} \equiv a\(\bar Q^\alpha \slashed D Q^\beta+\bar Q^\beta \overleftarrow{\slashed D} Q^\alpha\) \ , \quad \cR_{\tilde{sq}}^{\alpha\beta}\equiv ia\(\bar Q^\alpha \slashed D Q^\beta-\bar Q^\beta \overleftarrow{\slashed D} Q^\alpha\) \ .
\eeq
Other chiral fields are treated identically. Via the eom of $Q$, one can trade these operators for those of \eq{eq:SMEFTaxion}. However, there are anomalous contributions to these eom, which we derive using the equivalent language of field redefinitions. The operators in \eq{redundantGreenOps} are removed by redefining 
\beq
Q\to e^{iX_Qa}Q \ , \quad X_Q\equiv -(r_{sq}+ir_{\tilde{sq}})^\dagger \ ,
\eeq
which, when inserted in the Dirac Lagrangian, reproduces the implications of the $Q$ eom presented in~\cite{Chala:2020wvs}. However, this field redefinition is also anomalous and generates axion-gauge boson couplings. Focussing on gluons and using the expression of the axial anomaly (see e.g.~\cite{Bilal:2008qx}), one obtains a contribution to $C_g$ equal to
\beq
\delta C_g=\frac{1}{2}\Tr\(2X_Q-X_u-X_d+h.c.\) \ .
\eeq
Replacing $X_Q$ by the couplings $r_{sq},r_{\tilde{sq}}$ and similarly for $u,d$, and plugging the UV divergences of these Wilson coefficients derived in~\cite{Chala:2020wvs}, one finds exactly \eq{eq:CgRunning}.

\begin{figure}
\centering
\includegraphics[width=0.9\textwidth]{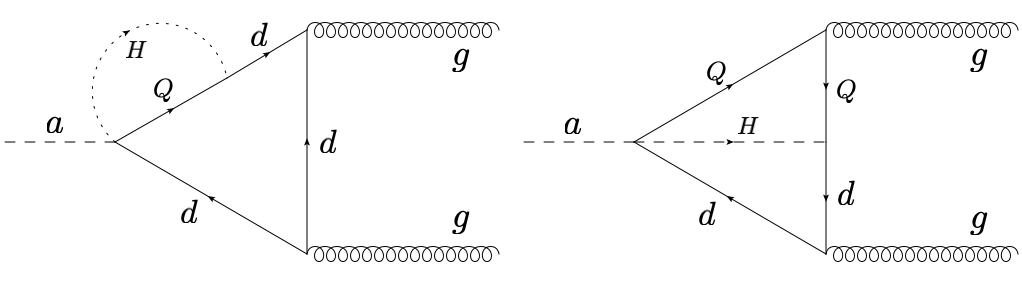}
\caption{Example of two-loop contributions to axion-gluon couplings of order $\cO\(\frac{g_3^2\tilde Y_d Y_d^\dagger}{(16\pi^2)^2}\)$}
\label{twoLoopDiags}
\end{figure}

This result reproduces that of appendix A of~\cite{Bauer:2020jbp}, which required to include contributions of similar magnitude up to two-loop order in the RGEs of the fermion couplings. In order to conclude that we identified the full running of $I_g$, given by \eq{eq:Igrunning}, at this order, it remains therefore to argue that such two-loop contributions are not relevant in the basis of \eq{eq:SMEFTaxion}. For that, we first notice that all diagrams which are proportional to bosonic axion couplings exist identically in \eq{eq:SMEFTaxion} and \eq{eq:SMEFTa}. Therefore, they have been accounted for in~\cite{Bauer:2020jbp} up to one-loop order in Yukawa couplings and two-loop order in gauge couplings, whose result we reproduce fully, hence we do not expect any new contribution of this kind. There remains the question of divergent diagrams proportional to $\tilde Y$. Those could contribute in three ways: {\it i)} as direct contributions to $\tilde Y_{u,d}$, which enter $I_g$, but any two-loop graph would be higher-order than the terms appearing in \eq{eq:CgRunning}, {\it ii)} as indirect contributions via $R_{sq,\tilde sq}$ and similar operators, but again all induced contributions to $\tilde Y_{u,d}$ are of higher order, and so would be the anomalous contributions, {\it iii)} as direct contributions to $C_g$. There exist diagrams at two-loop with the right magnitude (see examples in Fig.~\ref{twoLoopDiags}), but the part proportional to the Levi--Civita tensor is found to always multiply the invariants $\Im\Tr \tilde Y Y^\dagger$, while the invariants $\Re\Tr \tilde Y Y^\dagger$ multiply structures built with the Minkowski metric. Since only $\Re\Tr \tilde Y Y^\dagger$ breaks the shift-invariance perturbatively, we conclude that the contributions proportional to the Levi--Civita tensor can be computed in the basis of \eq{eq:SMEFTa} and have therefore been taken into account in~\cite{Bauer:2020jbp}. We leave complete computations for future work.

\clearpage
\bibliographystyle{apsrev4-1_title}
\bibliography{biblio.bib}

\begin{thebibliography}{85}%
\makeatletter
\providecommand \@ifxundefined [1]{%
 \@ifx{#1\undefined}
}%
\providecommand \@ifnum [1]{%
 \ifnum #1\expandafter \@firstoftwo
 \else \expandafter \@secondoftwo
 \fi
}%
\providecommand \@ifx [1]{%
 \ifx #1\expandafter \@firstoftwo
 \else \expandafter \@secondoftwo
 \fi
}%
\providecommand \natexlab [1]{#1}%
\providecommand \enquote  [1]{``#1''}%
\providecommand \bibnamefont  [1]{#1}%
\providecommand \bibfnamefont [1]{#1}%
\providecommand \citenamefont [1]{#1}%
\providecommand \href@noop [0]{\@secondoftwo}%
\providecommand \href [0]{\begingroup \@sanitize@url \@href}%
\providecommand \@href[1]{\@@startlink{#1}\@@href}%
\providecommand \@@href[1]{\endgroup#1\@@endlink}%
\providecommand \@sanitize@url [0]{\catcode `\\12\catcode `\$12\catcode
  `\&12\catcode `\#12\catcode `\^12\catcode `\_12\catcode `\%12\relax}%
\providecommand \@@startlink[1]{}%
\providecommand \@@endlink[0]{}%
\providecommand \url  [0]{\begingroup\@sanitize@url \@url }%
\providecommand \@url [1]{\endgroup\@href {#1}{\urlprefix }}%
\providecommand \urlprefix  [0]{URL }%
\providecommand \Eprint [0]{\href }%
\providecommand \doibase [0]{http://dx.doi.org/}%
\providecommand \selectlanguage [0]{\@gobble}%
\providecommand \bibinfo [0]{\@secondoftwo}%
\providecommand \bibfield [0]{\@secondoftwo}%
\providecommand \translation [1]{[#1]}%
\providecommand \BibitemOpen [0]{}%
\providecommand \bibitemStop [0]{}%
\providecommand \bibitemNoStop [0]{.\EOS\space}%
\providecommand \EOS [0]{\spacefactor3000\relax}%
\providecommand \BibitemShut  [1]{\csname bibitem#1\endcsname}%
\let\auto@bib@innerbib\@empty
\bibitem [{\citenamefont{Peccei} and
  \citenamefont{Quinn}(1977{\natexlab{a}})}]{Peccei:1977hh}%
  \BibitemOpen
  \bibfield{author}{\bibinfo{author}{\bibfnamefont{R.~D.} \bibnamefont{Peccei}}
  and \bibinfo{author}{\bibfnamefont{H.~R.} \bibnamefont{Quinn}},
  }\bibfield{title}{\emph {\bibinfo{title}{{CP Conservation in the Presence of
  Instantons}}}, }\href {\doibase 10.1103/PhysRevLett.38.1440}
  {\bibfield{journal}{\bibinfo{journal}{Phys. Rev.
  Lett.}\,}\textbf{\bibinfo{volume}{38}}\,(\bibinfo{year}{1977}{\natexlab{a}})\,\bibinfo{pages}{1440}}\BibitemShut
  {NoStop}%
\bibitem [{\citenamefont{Peccei} and
  \citenamefont{Quinn}(1977{\natexlab{b}})}]{Peccei:1977ur}%
  \BibitemOpen
  \bibfield{author}{\bibinfo{author}{\bibfnamefont{R.~D.} \bibnamefont{Peccei}}
  and \bibinfo{author}{\bibfnamefont{H.~R.} \bibnamefont{Quinn}},
  }\bibfield{title}{\emph {\bibinfo{title}{{Constraints Imposed by CP
  Conservation in the Presence of Instantons}}}, }\href {\doibase
  10.1103/PhysRevD.16.1791} {\bibfield{journal}{\bibinfo{journal}{Phys. Rev.
  D}\,}\textbf{\bibinfo{volume}{16}}\,(\bibinfo{year}{1977}{\natexlab{b}})\,\bibinfo{pages}{1791}}\BibitemShut
  {NoStop}%
\bibitem [{\citenamefont{Weinberg}(1978)}]{Weinberg:1977ma}%
  \BibitemOpen
  \bibfield{author}{\bibinfo{author}{\bibfnamefont{S.}\,\bibnamefont{Weinberg}},
  }\bibfield{title}{\emph {\bibinfo{title}{{A New Light Boson?}}}, }\href
  {\doibase 10.1103/PhysRevLett.40.223}
  {\bibfield{journal}{\bibinfo{journal}{Phys. Rev.
  Lett.}\,}\textbf{\bibinfo{volume}{40}}\,(\bibinfo{year}{1978})\,\bibinfo{pages}{223}}\BibitemShut
  {NoStop}%
\bibitem [{\citenamefont{Wilczek}(1978)}]{Wilczek:1977pj}%
  \BibitemOpen
  \bibfield{author}{\bibinfo{author}{\bibfnamefont{F.}\,\bibnamefont{Wilczek}},
  }\bibfield{title}{\emph {\bibinfo{title}{{Problem of Strong $P$ and $T$
  Invariance in the Presence of Instantons}}}, }\href {\doibase
  10.1103/PhysRevLett.40.279} {\bibfield{journal}{\bibinfo{journal}{Phys. Rev.
  Lett.}\,}\textbf{\bibinfo{volume}{40}}\,(\bibinfo{year}{1978})\,\bibinfo{pages}{279}}\BibitemShut
  {NoStop}%
\bibitem [{\citenamefont{Kim}(1979)}]{Kim:1979if}%
  \BibitemOpen
  \bibfield{author}{\bibinfo{author}{\bibfnamefont{J.~E.} \bibnamefont{Kim}},
  }\bibfield{title}{\emph {\bibinfo{title}{{Weak Interaction Singlet and Strong
  CP Invariance}}}, }\href {\doibase 10.1103/PhysRevLett.43.103}
  {\bibfield{journal}{\bibinfo{journal}{Phys. Rev.
  Lett.}\,}\textbf{\bibinfo{volume}{43}}\,(\bibinfo{year}{1979})\,\bibinfo{pages}{103}}\BibitemShut
  {NoStop}%
\bibitem [{\citenamefont{Shifman} \emph {et\,al.}(1980)\citenamefont{Shifman},
  \citenamefont{Vainshtein}, and \citenamefont{Zakharov}}]{Shifman:1979if}%
  \BibitemOpen
  \bibfield{author}{\bibinfo{author}{\bibfnamefont{M.~A.}
  \bibnamefont{Shifman}},
  \bibinfo{author}{\bibfnamefont{A.}\,\bibnamefont{Vainshtein}},  and
  \bibinfo{author}{\bibfnamefont{V.~I.} \bibnamefont{Zakharov}},
  }\bibfield{title}{\emph {\bibinfo{title}{{Can Confinement Ensure Natural CP
  Invariance of Strong Interactions?}}}, }\href {\doibase
  10.1016/0550-3213(80)90209-6} {\bibfield{journal}{\bibinfo{journal}{Nucl.
  Phys.
  B}\,}\textbf{\bibinfo{volume}{166}}\,(\bibinfo{year}{1980})\,\bibinfo{pages}{493}}\BibitemShut
  {NoStop}%
\bibitem [{\citenamefont{Dine} \emph {et\,al.}(1981)\citenamefont{Dine},
  \citenamefont{Fischler}, and \citenamefont{Srednicki}}]{Dine:1981rt}%
  \BibitemOpen
  \bibfield{author}{\bibinfo{author}{\bibfnamefont{M.}\,\bibnamefont{Dine}},
  \bibinfo{author}{\bibfnamefont{W.}\,\bibnamefont{Fischler}},  and
  \bibinfo{author}{\bibfnamefont{M.}\,\bibnamefont{Srednicki}},
  }\bibfield{title}{\emph {\bibinfo{title}{{A Simple Solution to the Strong CP
  Problem with a Harmless Axion}}}, }\href {\doibase
  10.1016/0370-2693(81)90590-6} {\bibfield{journal}{\bibinfo{journal}{Phys.
  Lett.
  B}\,}\textbf{\bibinfo{volume}{104}}\,(\bibinfo{year}{1981})\,\bibinfo{pages}{199}}\BibitemShut
  {NoStop}%
\bibitem [{\citenamefont{Zhitnitsky}(1980)}]{Zhitnitsky:1980tq}%
  \BibitemOpen
  \bibfield{author}{\bibinfo{author}{\bibfnamefont{A.}\,\bibnamefont{Zhitnitsky}},
  }\bibfield{title}{\emph {\bibinfo{title}{{On Possible Suppression of the
  Axion Hadron Interactions. (In Russian)}}}, }\href@noop {}
  {\bibfield{journal}{\bibinfo{journal}{Sov. J. Nucl.
  Phys.}\,}\textbf{\bibinfo{volume}{31}}\,(\bibinfo{year}{1980})\,\bibinfo{pages}{260}}\BibitemShut
  {NoStop}%
\bibitem [{\citenamefont{Preskill} \emph
  {et\,al.}(1983)\citenamefont{Preskill}, \citenamefont{Wise}, and
  \citenamefont{Wilczek}}]{Preskill:1982cy}%
  \BibitemOpen
  \bibfield{author}{\bibinfo{author}{\bibfnamefont{J.}\,\bibnamefont{Preskill}},
  \bibinfo{author}{\bibfnamefont{M.~B.} \bibnamefont{Wise}},  and
  \bibinfo{author}{\bibfnamefont{F.}\,\bibnamefont{Wilczek}},
  }\bibfield{title}{\emph {\bibinfo{title}{{Cosmology of the Invisible
  Axion}}}, }\href {\doibase 10.1016/0370-2693(83)90637-8}
  {\bibfield{journal}{\bibinfo{journal}{Phys. Lett.
  B}\,}\textbf{\bibinfo{volume}{120}}\,(\bibinfo{year}{1983})\,\bibinfo{pages}{127}}\BibitemShut
  {NoStop}%
\bibitem [{\citenamefont{Abbott} and
  \citenamefont{Sikivie}(1983)}]{Abbott:1982af}%
  \BibitemOpen
  \bibfield{author}{\bibinfo{author}{\bibfnamefont{L.~F.} \bibnamefont{Abbott}}
  and \bibinfo{author}{\bibfnamefont{P.}\,\bibnamefont{Sikivie}},
  }\bibfield{title}{\emph {\bibinfo{title}{{A Cosmological Bound on the
  Invisible Axion}}}, }\href {\doibase 10.1016/0370-2693(83)90638-X}
  {\bibfield{journal}{\bibinfo{journal}{Phys. Lett.
  B}\,}\textbf{\bibinfo{volume}{120}}\,(\bibinfo{year}{1983})\,\bibinfo{pages}{133}}\BibitemShut
  {NoStop}%
\bibitem [{\citenamefont{Dine} and
  \citenamefont{Fischler}(1983)}]{Dine:1982ah}%
  \BibitemOpen
  \bibfield{author}{\bibinfo{author}{\bibfnamefont{M.}\,\bibnamefont{Dine}} and
  \bibinfo{author}{\bibfnamefont{W.}\,\bibnamefont{Fischler}},
  }\bibfield{title}{\emph {\bibinfo{title}{{The Not So Harmless Axion}}},
  }\href {\doibase 10.1016/0370-2693(83)90639-1}
  {\bibfield{journal}{\bibinfo{journal}{Phys. Lett.
  B}\,}\textbf{\bibinfo{volume}{120}}\,(\bibinfo{year}{1983})\,\bibinfo{pages}{137}}\BibitemShut
  {NoStop}%
\bibitem [{\citenamefont{Di~Luzio} \emph
  {et\,al.}(2020)\citenamefont{Di~Luzio}, \citenamefont{Giannotti},
  \citenamefont{Nardi}, and \citenamefont{Visinelli}}]{DiLuzio:2020wdo}%
  \BibitemOpen
  \bibfield{author}{\bibinfo{author}{\bibfnamefont{L.}\,\bibnamefont{Di~Luzio}},
  \bibinfo{author}{\bibfnamefont{M.}\,\bibnamefont{Giannotti}},
  \bibinfo{author}{\bibfnamefont{E.}\,\bibnamefont{Nardi}},  and
  \bibinfo{author}{\bibfnamefont{L.}\,\bibnamefont{Visinelli}},
  }\bibfield{title}{\emph {\bibinfo{title}{{The landscape of QCD axion
  models}}}, }\href {\doibase 10.1016/j.physrep.2020.06.002}
  {\bibfield{journal}{\bibinfo{journal}{Phys.
  Rept.}\,}\textbf{\bibinfo{volume}{870}}\,(\bibinfo{year}{2020})\,\bibinfo{pages}{1}},
  \Eprint {http://arxiv.org/abs/2003.01100}{arXiv:2003.01100
  [hep-ph]}\BibitemShut {NoStop}%
\bibitem [{\citenamefont{Hawking}(1987)}]{Hawking:1987mz}%
  \BibitemOpen
  \bibfield{author}{\bibinfo{author}{\bibfnamefont{S.~W.}
  \bibnamefont{Hawking}}, }\bibfield{title}{\emph {\bibinfo{title}{{Quantum
  Coherence Down the Wormhole}}}, }\href {\doibase
  10.1016/0370-2693(87)90028-1} {\bibfield{journal}{\bibinfo{journal}{Phys.
  Lett.
  B}\,}\textbf{\bibinfo{volume}{195}}\,(\bibinfo{year}{1987})\,\bibinfo{pages}{337}}\BibitemShut
  {NoStop}%
\bibitem [{\citenamefont{Giddings} and
  \citenamefont{Strominger}(1988)}]{Giddings:1988cx}%
  \BibitemOpen
  \bibfield{author}{\bibinfo{author}{\bibfnamefont{S.~B.}
  \bibnamefont{Giddings}} and
  \bibinfo{author}{\bibfnamefont{A.}\,\bibnamefont{Strominger}},
  }\bibfield{title}{\emph {\bibinfo{title}{{Loss of Incoherence and
  Determination of Coupling Constants in Quantum Gravity}}}, }\href {\doibase
  10.1016/0550-3213(88)90109-5} {\bibfield{journal}{\bibinfo{journal}{Nucl.
  Phys.
  B}\,}\textbf{\bibinfo{volume}{307}}\,(\bibinfo{year}{1988})\,\bibinfo{pages}{854}}\BibitemShut
  {NoStop}%
\bibitem [{\citenamefont{Banks} and
  \citenamefont{Seiberg}(2011)}]{Banks:2010zn}%
  \BibitemOpen
  \bibfield{author}{\bibinfo{author}{\bibfnamefont{T.}\,\bibnamefont{Banks}}
  and \bibinfo{author}{\bibfnamefont{N.}\,\bibnamefont{Seiberg}},
  }\bibfield{title}{\emph {\bibinfo{title}{{Symmetries and Strings in Field
  Theory and Gravity}}}, }\href {\doibase 10.1103/PhysRevD.83.084019}
  {\bibfield{journal}{\bibinfo{journal}{Phys. Rev.
  D}\,}\textbf{\bibinfo{volume}{83}}\,(\bibinfo{year}{2011})\,\bibinfo{pages}{084019}},
  \Eprint {http://arxiv.org/abs/1011.5120}{arXiv:1011.5120
  [hep-th]}\BibitemShut {NoStop}%
\bibitem [{\citenamefont{Graham} \emph {et\,al.}(2015)\citenamefont{Graham},
  \citenamefont{Kaplan}, and \citenamefont{Rajendran}}]{Graham:2015cka}%
  \BibitemOpen
  \bibfield{author}{\bibinfo{author}{\bibfnamefont{P.~W.}
  \bibnamefont{Graham}}, \bibinfo{author}{\bibfnamefont{D.~E.}
  \bibnamefont{Kaplan}},  and
  \bibinfo{author}{\bibfnamefont{S.}\,\bibnamefont{Rajendran}},
  }\bibfield{title}{\emph {\bibinfo{title}{{Cosmological Relaxation of the
  Electroweak Scale}}}, }\href {\doibase 10.1103/PhysRevLett.115.221801}
  {\bibfield{journal}{\bibinfo{journal}{Phys. Rev.
  Lett.}\,}\textbf{\bibinfo{volume}{115}}\,(\bibinfo{year}{2015})\,\bibinfo{pages}{221801}},
  \Eprint {http://arxiv.org/abs/1504.07551}{arXiv:1504.07551
  [hep-ph]}\BibitemShut {NoStop}%
\bibitem [{\citenamefont{Georgi} \emph {et\,al.}(1986)\citenamefont{Georgi},
  \citenamefont{Kaplan}, and \citenamefont{Randall}}]{Georgi:1986df}%
  \BibitemOpen
  \bibfield{author}{\bibinfo{author}{\bibfnamefont{H.}\,\bibnamefont{Georgi}},
  \bibinfo{author}{\bibfnamefont{D.~B.} \bibnamefont{Kaplan}},  and
  \bibinfo{author}{\bibfnamefont{L.}\,\bibnamefont{Randall}},
  }\bibfield{title}{\emph {\bibinfo{title}{{Manifesting the Invisible Axion at
  Low-energies}}}, }\href {\doibase 10.1016/0370-2693(86)90688-X}
  {\bibfield{journal}{\bibinfo{journal}{Phys. Lett.
  B}\,}\textbf{\bibinfo{volume}{169}}\,(\bibinfo{year}{1986})\,\bibinfo{pages}{73}}\BibitemShut
  {NoStop}%
\bibitem [{\citenamefont{Srednicki}(1985)}]{Srednicki:1985xd}%
  \BibitemOpen
  \bibfield{author}{\bibinfo{author}{\bibfnamefont{M.}\,\bibnamefont{Srednicki}},
  }\bibfield{title}{\emph {\bibinfo{title}{{Axion Couplings to Matter. 1. CP
  Conserving Parts}}}, }\href {\doibase 10.1016/0550-3213(85)90054-9}
  {\bibfield{journal}{\bibinfo{journal}{Nucl. Phys.
  B}\,}\textbf{\bibinfo{volume}{260}}\,(\bibinfo{year}{1985})\,\bibinfo{pages}{689}}\BibitemShut
  {NoStop}%
\bibitem [{\citenamefont{Izaguirre} \emph
  {et\,al.}(2017)\citenamefont{Izaguirre}, \citenamefont{Lin}, and
  \citenamefont{Shuve}}]{Izaguirre:2016dfi}%
  \BibitemOpen
  \bibfield{author}{\bibinfo{author}{\bibfnamefont{E.}\,\bibnamefont{Izaguirre}},
  \bibinfo{author}{\bibfnamefont{T.}\,\bibnamefont{Lin}},  and
  \bibinfo{author}{\bibfnamefont{B.}\,\bibnamefont{Shuve}},
  }\bibfield{title}{\emph {\bibinfo{title}{{Searching for Axionlike Particles
  in Flavor-Changing Neutral Current Processes}}}, }\href {\doibase
  10.1103/PhysRevLett.118.111802} {\bibfield{journal}{\bibinfo{journal}{Phys.
  Rev.
  Lett.}\,}\textbf{\bibinfo{volume}{118}}\,(\bibinfo{year}{2017})\,\bibinfo{pages}{111802}},
  \Eprint {http://arxiv.org/abs/1611.09355}{arXiv:1611.09355
  [hep-ph]}\BibitemShut {NoStop}%
\bibitem [{\citenamefont{Bj\"orkeroth} \emph
  {et\,al.}(2018)\citenamefont{Bj\"orkeroth}, \citenamefont{Chun}, and
  \citenamefont{King}}]{Bjorkeroth:2018dzu}%
  \BibitemOpen
  \bibfield{author}{\bibinfo{author}{\bibfnamefont{F.}\,\bibnamefont{Bj\"orkeroth}},
  \bibinfo{author}{\bibfnamefont{E.~J.} \bibnamefont{Chun}},  and
  \bibinfo{author}{\bibfnamefont{S.~F.} \bibnamefont{King}},
  }\bibfield{title}{\emph {\bibinfo{title}{{Flavourful Axion Phenomenology}}},
  }\href {\doibase 10.1007/JHEP08(2018)117}
  {\bibfield{journal}{\bibinfo{journal}{JHEP}\,}\textbf{\bibinfo{volume}{08}}\,(\bibinfo{year}{2018})\,\bibinfo{pages}{117}},
  \Eprint {http://arxiv.org/abs/1806.00660}{arXiv:1806.00660
  [hep-ph]}\BibitemShut {NoStop}%
\bibitem [{\citenamefont{Gavela} \emph {et\,al.}(2019)\citenamefont{Gavela},
  \citenamefont{Houtz}, \citenamefont{Quilez}, \citenamefont{Del~Rey}, and
  \citenamefont{Sumensari}}]{Gavela:2019wzg}%
  \BibitemOpen
  \bibfield{author}{\bibinfo{author}{\bibfnamefont{M.~B.}
  \bibnamefont{Gavela}},
  \bibinfo{author}{\bibfnamefont{R.}\,\bibnamefont{Houtz}},
  \bibinfo{author}{\bibfnamefont{P.}\,\bibnamefont{Quilez}},
  \bibinfo{author}{\bibfnamefont{R.}\,\bibnamefont{Del~Rey}},  and
  \bibinfo{author}{\bibfnamefont{O.}\,\bibnamefont{Sumensari}},
  }\bibfield{title}{\emph {\bibinfo{title}{{Flavor constraints on electroweak
  ALP couplings}}}, }\href {\doibase 10.1140/epjc/s10052-019-6889-y}
  {\bibfield{journal}{\bibinfo{journal}{Eur. Phys. J.
  C}\,}\textbf{\bibinfo{volume}{79}}\,(\bibinfo{year}{2019})\,\bibinfo{pages}{369}},
  \Eprint {http://arxiv.org/abs/1901.02031}{arXiv:1901.02031
  [hep-ph]}\BibitemShut {NoStop}%
\bibitem [{\citenamefont{Bauer} \emph {et\,al.}(2020)\citenamefont{Bauer},
  \citenamefont{Neubert}, \citenamefont{Renner}, \citenamefont{Schnubel}, and
  \citenamefont{Thamm}}]{Bauer:2019gfk}%
  \BibitemOpen
  \bibfield{author}{\bibinfo{author}{\bibfnamefont{M.}\,\bibnamefont{Bauer}},
  \bibinfo{author}{\bibfnamefont{M.}\,\bibnamefont{Neubert}},
  \bibinfo{author}{\bibfnamefont{S.}\,\bibnamefont{Renner}},
  \bibinfo{author}{\bibfnamefont{M.}\,\bibnamefont{Schnubel}},  and
  \bibinfo{author}{\bibfnamefont{A.}\,\bibnamefont{Thamm}},
  }\bibfield{title}{\emph {\bibinfo{title}{{Axionlike Particles, Lepton-Flavor
  Violation, and a New Explanation of $a_\mu$ and $a_e$}}}, }\href {\doibase
  10.1103/PhysRevLett.124.211803} {\bibfield{journal}{\bibinfo{journal}{Phys.
  Rev.
  Lett.}\,}\textbf{\bibinfo{volume}{124}}\,(\bibinfo{year}{2020})\,\bibinfo{pages}{211803}},
  \Eprint {http://arxiv.org/abs/1908.00008}{arXiv:1908.00008
  [hep-ph]}\BibitemShut {NoStop}%
\bibitem [{\citenamefont{Albrecht} \emph
  {et\,al.}(2020)\citenamefont{Albrecht}, \citenamefont{Stamou},
  \citenamefont{Ziegler}, and \citenamefont{Zwicky}}]{Albrecht:2019zul}%
  \BibitemOpen
  \bibfield{author}{\bibinfo{author}{\bibfnamefont{J.}\,\bibnamefont{Albrecht}},
  \bibinfo{author}{\bibfnamefont{E.}\,\bibnamefont{Stamou}},
  \bibinfo{author}{\bibfnamefont{R.}\,\bibnamefont{Ziegler}},  and
  \bibinfo{author}{\bibfnamefont{R.}\,\bibnamefont{Zwicky}},
  }\bibfield{title}{\emph {\bibinfo{title}{{Flavoured axions in the tail of
  B$_{q}$ \textrightarrow{} \ensuremath{\mu}$^{+}$\ensuremath{\mu}$^{-}$ and B
  \textrightarrow{} \ensuremath{\gamma}$^{*}$ form factors}}}, }\href {\doibase
  10.1007/JHEP09(2021)139}
  {\bibfield{journal}{\bibinfo{journal}{JHEP}\,}\textbf{\bibinfo{volume}{21}}\,(\bibinfo{year}{2020})\,\bibinfo{pages}{139}},
  \Eprint {http://arxiv.org/abs/1911.05018}{arXiv:1911.05018
  [hep-ph]}\BibitemShut {NoStop}%
\bibitem [{\citenamefont{Martin~Camalich} \emph
  {et\,al.}(2020)\citenamefont{Martin~Camalich}, \citenamefont{Pospelov},
  \citenamefont{Vuong}, \citenamefont{Ziegler}, and
  \citenamefont{Zupan}}]{MartinCamalich:2020dfe}%
  \BibitemOpen
  \bibfield{author}{\bibinfo{author}{\bibfnamefont{J.}\,\bibnamefont{Martin~Camalich}},
  \bibinfo{author}{\bibfnamefont{M.}\,\bibnamefont{Pospelov}},
  \bibinfo{author}{\bibfnamefont{P.~N.~H.} \bibnamefont{Vuong}},
  \bibinfo{author}{\bibfnamefont{R.}\,\bibnamefont{Ziegler}},  and
  \bibinfo{author}{\bibfnamefont{J.}\,\bibnamefont{Zupan}},
  }\bibfield{title}{\emph {\bibinfo{title}{{Quark Flavor Phenomenology of the
  QCD Axion}}}, }\href {\doibase 10.1103/PhysRevD.102.015023}
  {\bibfield{journal}{\bibinfo{journal}{Phys. Rev.
  D}\,}\textbf{\bibinfo{volume}{102}}\,(\bibinfo{year}{2020})\,\bibinfo{pages}{015023}},
  \Eprint {http://arxiv.org/abs/2002.04623}{arXiv:2002.04623
  [hep-ph]}\BibitemShut {NoStop}%
\bibitem [{\citenamefont{Endo} \emph {et\,al.}(2020)\citenamefont{Endo},
  \citenamefont{Iguro}, and \citenamefont{Kitahara}}]{Endo:2020mev}%
  \BibitemOpen
  \bibfield{author}{\bibinfo{author}{\bibfnamefont{M.}\,\bibnamefont{Endo}},
  \bibinfo{author}{\bibfnamefont{S.}\,\bibnamefont{Iguro}},  and
  \bibinfo{author}{\bibfnamefont{T.}\,\bibnamefont{Kitahara}},
  }\bibfield{title}{\emph {\bibinfo{title}{{Probing $e\mu$ flavor-violating ALP
  at Belle II}}}, }\href {\doibase 10.1007/JHEP06(2020)040}
  {\bibfield{journal}{\bibinfo{journal}{JHEP}\,}\textbf{\bibinfo{volume}{06}}\,(\bibinfo{year}{2020})\,\bibinfo{pages}{040}},
  \Eprint {http://arxiv.org/abs/2002.05948}{arXiv:2002.05948
  [hep-ph]}\BibitemShut {NoStop}%
\bibitem [{\citenamefont{Ishida} \emph {et\,al.}(2021)\citenamefont{Ishida},
  \citenamefont{Matsuzaki}, and \citenamefont{Shigekami}}]{Ishida:2020oxl}%
  \BibitemOpen
  \bibfield{author}{\bibinfo{author}{\bibfnamefont{H.}\,\bibnamefont{Ishida}},
  \bibinfo{author}{\bibfnamefont{S.}\,\bibnamefont{Matsuzaki}},  and
  \bibinfo{author}{\bibfnamefont{Y.}\,\bibnamefont{Shigekami}},
  }\bibfield{title}{\emph {\bibinfo{title}{{New perspective in searching for
  axionlike particles from flavor physics}}}, }\href {\doibase
  10.1103/PhysRevD.103.095022} {\bibfield{journal}{\bibinfo{journal}{Phys. Rev.
  D}\,}\textbf{\bibinfo{volume}{103}}\,(\bibinfo{year}{2021})\,\bibinfo{pages}{095022}},
  \Eprint {http://arxiv.org/abs/2006.02725}{arXiv:2006.02725
  [hep-ph]}\BibitemShut {NoStop}%
\bibitem [{\citenamefont{Calibbi} \emph {et\,al.}(2021)\citenamefont{Calibbi},
  \citenamefont{Redigolo}, \citenamefont{Ziegler}, and
  \citenamefont{Zupan}}]{Calibbi:2020jvd}%
  \BibitemOpen
  \bibfield{author}{\bibinfo{author}{\bibfnamefont{L.}\,\bibnamefont{Calibbi}},
  \bibinfo{author}{\bibfnamefont{D.}\,\bibnamefont{Redigolo}},
  \bibinfo{author}{\bibfnamefont{R.}\,\bibnamefont{Ziegler}},  and
  \bibinfo{author}{\bibfnamefont{J.}\,\bibnamefont{Zupan}},
  }\bibfield{title}{\emph {\bibinfo{title}{{Looking forward to
  lepton-flavor-violating ALPs}}}, }\href {\doibase 10.1007/JHEP09(2021)173}
  {\bibfield{journal}{\bibinfo{journal}{JHEP}\,}\textbf{\bibinfo{volume}{09}}\,(\bibinfo{year}{2021})\,\bibinfo{pages}{173}},
  \Eprint {http://arxiv.org/abs/2006.04795}{arXiv:2006.04795
  [hep-ph]}\BibitemShut {NoStop}%
\bibitem [{\citenamefont{Carmona} \emph {et\,al.}(2021)\citenamefont{Carmona},
  \citenamefont{Scherb}, and \citenamefont{Schwaller}}]{Carmona:2021seb}%
  \BibitemOpen
  \bibfield{author}{\bibinfo{author}{\bibfnamefont{A.}\,\bibnamefont{Carmona}},
  \bibinfo{author}{\bibfnamefont{C.}\,\bibnamefont{Scherb}},  and
  \bibinfo{author}{\bibfnamefont{P.}\,\bibnamefont{Schwaller}},
  }\bibfield{title}{\emph {\bibinfo{title}{{Charming ALPs}}}, }\href {\doibase
  10.1007/JHEP08(2021)121}
  {\bibfield{journal}{\bibinfo{journal}{JHEP}\,}\textbf{\bibinfo{volume}{08}}\,(\bibinfo{year}{2021})\,\bibinfo{pages}{121}},
  \Eprint {http://arxiv.org/abs/2101.07803}{arXiv:2101.07803
  [hep-ph]}\BibitemShut {NoStop}%
\bibitem [{\citenamefont{Chakraborty} \emph
  {et\,al.}(2021)\citenamefont{Chakraborty}, \citenamefont{Kraus},
  \citenamefont{Loladze}, \citenamefont{Okui}, and
  \citenamefont{Tobioka}}]{Chakraborty:2021wda}%
  \BibitemOpen
  \bibfield{author}{\bibinfo{author}{\bibfnamefont{S.}\,\bibnamefont{Chakraborty}},
  \bibinfo{author}{\bibfnamefont{M.}\,\bibnamefont{Kraus}},
  \bibinfo{author}{\bibfnamefont{V.}\,\bibnamefont{Loladze}},
  \bibinfo{author}{\bibfnamefont{T.}\,\bibnamefont{Okui}},  and
  \bibinfo{author}{\bibfnamefont{K.}\,\bibnamefont{Tobioka}},
  }\bibfield{title}{\emph {\bibinfo{title}{{Heavy QCD axion in
  b\textrightarrow{}s transition: Enhanced limits and projections}}}, }\href
  {\doibase 10.1103/PhysRevD.104.055036}
  {\bibfield{journal}{\bibinfo{journal}{Phys. Rev.
  D}\,}\textbf{\bibinfo{volume}{104}}\,(\bibinfo{year}{2021})\,\bibinfo{pages}{055036}},
  \Eprint {http://arxiv.org/abs/2102.04474}{arXiv:2102.04474
  [hep-ph]}\BibitemShut {NoStop}%
\bibitem [{\citenamefont{Davoudiasl} \emph
  {et\,al.}(2021)\citenamefont{Davoudiasl}, \citenamefont{Marcarelli},
  \citenamefont{Miesch}, and \citenamefont{Neil}}]{Davoudiasl:2021haa}%
  \BibitemOpen
  \bibfield{author}{\bibinfo{author}{\bibfnamefont{H.}\,\bibnamefont{Davoudiasl}},
  \bibinfo{author}{\bibfnamefont{R.}\,\bibnamefont{Marcarelli}},
  \bibinfo{author}{\bibfnamefont{N.}\,\bibnamefont{Miesch}},  and
  \bibinfo{author}{\bibfnamefont{E.~T.} \bibnamefont{Neil}},
  }\bibfield{title}{\emph {\bibinfo{title}{{Searching for flavor-violating ALPs
  in Higgs boson decays}}}, }\href {\doibase 10.1103/PhysRevD.104.055022}
  {\bibfield{journal}{\bibinfo{journal}{Phys. Rev.
  D}\,}\textbf{\bibinfo{volume}{104}}\,(\bibinfo{year}{2021})\,\bibinfo{pages}{055022}},
  \Eprint {http://arxiv.org/abs/2105.05866}{arXiv:2105.05866
  [hep-ph]}\BibitemShut {NoStop}%
\bibitem [{\citenamefont{Bauer} \emph
  {et\,al.}(2021{\natexlab{a}})\citenamefont{Bauer}, \citenamefont{Neubert},
  \citenamefont{Renner}, \citenamefont{Schnubel}, and
  \citenamefont{Thamm}}]{Bauer:2021mvw}%
  \BibitemOpen
  \bibfield{author}{\bibinfo{author}{\bibfnamefont{M.}\,\bibnamefont{Bauer}},
  \bibinfo{author}{\bibfnamefont{M.}\,\bibnamefont{Neubert}},
  \bibinfo{author}{\bibfnamefont{S.}\,\bibnamefont{Renner}},
  \bibinfo{author}{\bibfnamefont{M.}\,\bibnamefont{Schnubel}},  and
  \bibinfo{author}{\bibfnamefont{A.}\,\bibnamefont{Thamm}},
  }\bibfield{title}{\emph {\bibinfo{title}{{Flavor probes of axion-like
  particles}}}, }\href@noop {} {\,(\bibinfo{year}{2021}{\natexlab{a}})},
  \Eprint {http://arxiv.org/abs/2110.10698}{arXiv:2110.10698
  [hep-ph]}\BibitemShut {NoStop}%
\bibitem [{\citenamefont{Coloma} \emph {et\,al.}(2022)\citenamefont{Coloma},
  \citenamefont{Hern\'andez}, and \citenamefont{Urrea}}]{Coloma:2022hlv}%
  \BibitemOpen
  \bibfield{author}{\bibinfo{author}{\bibfnamefont{P.}\,\bibnamefont{Coloma}},
  \bibinfo{author}{\bibfnamefont{P.}\,\bibnamefont{Hern\'andez}},  and
  \bibinfo{author}{\bibfnamefont{S.}\,\bibnamefont{Urrea}},
  }\bibfield{title}{\emph {\bibinfo{title}{{New bounds on axion-like particles
  from MicroBooNE}}}, }\href@noop {} {\,(\bibinfo{year}{2022})}, \Eprint
  {http://arxiv.org/abs/2202.03447}{arXiv:2202.03447 [hep-ph]}\BibitemShut
  {NoStop}%
\bibitem [{\citenamefont{Gori} \emph {et\,al.}(2020)\citenamefont{Gori},
  \citenamefont{Perez}, and \citenamefont{Tobioka}}]{Gori:2020xvq}%
  \BibitemOpen
  \bibfield{author}{\bibinfo{author}{\bibfnamefont{S.}\,\bibnamefont{Gori}},
  \bibinfo{author}{\bibfnamefont{G.}\,\bibnamefont{Perez}},  and
  \bibinfo{author}{\bibfnamefont{K.}\,\bibnamefont{Tobioka}},
  }\bibfield{title}{\emph {\bibinfo{title}{{KOTO vs. NA62 Dark Scalar
  Searches}}}, }\href {\doibase 10.1007/JHEP08(2020)110}
  {\bibfield{journal}{\bibinfo{journal}{JHEP}\,}\textbf{\bibinfo{volume}{08}}\,(\bibinfo{year}{2020})\,\bibinfo{pages}{110}},
  \Eprint {http://arxiv.org/abs/2005.05170}{arXiv:2005.05170
  [hep-ph]}\BibitemShut {NoStop}%
\bibitem [{\citenamefont{Mimasu} and
  \citenamefont{Sanz}(2015)}]{Mimasu:2014nea}%
  \BibitemOpen
  \bibfield{author}{\bibinfo{author}{\bibfnamefont{K.}\,\bibnamefont{Mimasu}}
  and \bibinfo{author}{\bibfnamefont{V.}\,\bibnamefont{Sanz}},
  }\bibfield{title}{\emph {\bibinfo{title}{{ALPs at Colliders}}}, }\href
  {\doibase 10.1007/JHEP06(2015)173}
  {\bibfield{journal}{\bibinfo{journal}{JHEP}\,}\textbf{\bibinfo{volume}{06}}\,(\bibinfo{year}{2015})\,\bibinfo{pages}{173}},
  \Eprint {http://arxiv.org/abs/1409.4792}{arXiv:1409.4792
  [hep-ph]}\BibitemShut {NoStop}%
\bibitem [{\citenamefont{Jaeckel} and
  \citenamefont{Spannowsky}(2016)}]{Jaeckel:2015jla}%
  \BibitemOpen
  \bibfield{author}{\bibinfo{author}{\bibfnamefont{J.}\,\bibnamefont{Jaeckel}}
  and \bibinfo{author}{\bibfnamefont{M.}\,\bibnamefont{Spannowsky}},
  }\bibfield{title}{\emph {\bibinfo{title}{{Probing MeV to 90 GeV axion-like
  particles with LEP and LHC}}}, }\href {\doibase
  10.1016/j.physletb.2015.12.037} {\bibfield{journal}{\bibinfo{journal}{Phys.
  Lett.
  B}\,}\textbf{\bibinfo{volume}{753}}\,(\bibinfo{year}{2016})\,\bibinfo{pages}{482}},
  \Eprint {http://arxiv.org/abs/1509.00476}{arXiv:1509.00476
  [hep-ph]}\BibitemShut {NoStop}%
\bibitem [{\citenamefont{Brivio} \emph {et\,al.}(2017)\citenamefont{Brivio},
  \citenamefont{Gavela}, \citenamefont{Merlo}, \citenamefont{Mimasu},
  \citenamefont{No}, \citenamefont{del\,Rey}, and
  \citenamefont{Sanz}}]{Brivio:2017ije}%
  \BibitemOpen
  \bibfield{author}{\bibinfo{author}{\bibfnamefont{I.}\,\bibnamefont{Brivio}},
  \bibinfo{author}{\bibfnamefont{M.~B.} \bibnamefont{Gavela}},
  \bibinfo{author}{\bibfnamefont{L.}\,\bibnamefont{Merlo}},
  \bibinfo{author}{\bibfnamefont{K.}\,\bibnamefont{Mimasu}},
  \bibinfo{author}{\bibfnamefont{J.~M.} \bibnamefont{No}},
  \bibinfo{author}{\bibfnamefont{R.}\,\bibnamefont{del\,Rey}},  and
  \bibinfo{author}{\bibfnamefont{V.}\,\bibnamefont{Sanz}},
  }\bibfield{title}{\emph {\bibinfo{title}{{ALPs Effective Field Theory and
  Collider Signatures}}}, }\href {\doibase 10.1140/epjc/s10052-017-5111-3}
  {\bibfield{journal}{\bibinfo{journal}{Eur. Phys. J.
  C}\,}\textbf{\bibinfo{volume}{77}}\,(\bibinfo{year}{2017})\,\bibinfo{pages}{572}},
  \Eprint {http://arxiv.org/abs/1701.05379}{arXiv:1701.05379
  [hep-ph]}\BibitemShut {NoStop}%
\bibitem [{\citenamefont{Bauer} \emph
  {et\,al.}(2017{\natexlab{a}})\citenamefont{Bauer}, \citenamefont{Neubert},
  and \citenamefont{Thamm}}]{Bauer:2017nlg}%
  \BibitemOpen
  \bibfield{author}{\bibinfo{author}{\bibfnamefont{M.}\,\bibnamefont{Bauer}},
  \bibinfo{author}{\bibfnamefont{M.}\,\bibnamefont{Neubert}},  and
  \bibinfo{author}{\bibfnamefont{A.}\,\bibnamefont{Thamm}},
  }\bibfield{title}{\emph {\bibinfo{title}{{LHC as an Axion Factory: Probing an
  Axion Explanation for $(g-2)_\mu$ with Exotic Higgs Decays}}}, }\href
  {\doibase 10.1103/PhysRevLett.119.031802}
  {\bibfield{journal}{\bibinfo{journal}{Phys. Rev.
  Lett.}\,}\textbf{\bibinfo{volume}{119}}\,(\bibinfo{year}{2017}{\natexlab{a}})\,\bibinfo{pages}{031802}},
  \Eprint {http://arxiv.org/abs/1704.08207}{arXiv:1704.08207
  [hep-ph]}\BibitemShut {NoStop}%
\bibitem [{\citenamefont{Bauer} \emph
  {et\,al.}(2017{\natexlab{b}})\citenamefont{Bauer}, \citenamefont{Neubert},
  and \citenamefont{Thamm}}]{Bauer:2017ris}%
  \BibitemOpen
  \bibfield{author}{\bibinfo{author}{\bibfnamefont{M.}\,\bibnamefont{Bauer}},
  \bibinfo{author}{\bibfnamefont{M.}\,\bibnamefont{Neubert}},  and
  \bibinfo{author}{\bibfnamefont{A.}\,\bibnamefont{Thamm}},
  }\bibfield{title}{\emph {\bibinfo{title}{{Collider Probes of Axion-Like
  Particles}}}, }\href {\doibase 10.1007/JHEP12(2017)044}
  {\bibfield{journal}{\bibinfo{journal}{JHEP}\,}\textbf{\bibinfo{volume}{12}}\,(\bibinfo{year}{2017}{\natexlab{b}})\,\bibinfo{pages}{044}},
  \Eprint {http://arxiv.org/abs/1708.00443}{arXiv:1708.00443
  [hep-ph]}\BibitemShut {NoStop}%
\bibitem [{\citenamefont{Alonso-\'Alvarez} \emph
  {et\,al.}(2019)\citenamefont{Alonso-\'Alvarez}, \citenamefont{Gavela}, and
  \citenamefont{Quilez}}]{Alonso-Alvarez:2018irt}%
  \BibitemOpen
  \bibfield{author}{\bibinfo{author}{\bibfnamefont{G.}\,\bibnamefont{Alonso-\'Alvarez}},
  \bibinfo{author}{\bibfnamefont{M.~B.} \bibnamefont{Gavela}},  and
  \bibinfo{author}{\bibfnamefont{P.}\,\bibnamefont{Quilez}},
  }\bibfield{title}{\emph {\bibinfo{title}{{Axion couplings to electroweak
  gauge bosons}}}, }\href {\doibase 10.1140/epjc/s10052-019-6732-5}
  {\bibfield{journal}{\bibinfo{journal}{Eur. Phys. J.
  C}\,}\textbf{\bibinfo{volume}{79}}\,(\bibinfo{year}{2019})\,\bibinfo{pages}{223}},
  \Eprint {http://arxiv.org/abs/1811.05466}{arXiv:1811.05466
  [hep-ph]}\BibitemShut {NoStop}%
\bibitem [{\citenamefont{Ebadi} \emph {et\,al.}(2019)\citenamefont{Ebadi},
  \citenamefont{Khatibi}, and
  \citenamefont{Mohammadi~Najafabadi}}]{Ebadi:2019gij}%
  \BibitemOpen
  \bibfield{author}{\bibinfo{author}{\bibfnamefont{J.}\,\bibnamefont{Ebadi}},
  \bibinfo{author}{\bibfnamefont{S.}\,\bibnamefont{Khatibi}},  and
  \bibinfo{author}{\bibfnamefont{M.}\,\bibnamefont{Mohammadi~Najafabadi}},
  }\bibfield{title}{\emph {\bibinfo{title}{{New probes for axionlike particles
  at hadron colliders}}}, }\href {\doibase 10.1103/PhysRevD.100.015016}
  {\bibfield{journal}{\bibinfo{journal}{Phys. Rev.
  D}\,}\textbf{\bibinfo{volume}{100}}\,(\bibinfo{year}{2019})\,\bibinfo{pages}{015016}},
  \Eprint {http://arxiv.org/abs/1901.03061}{arXiv:1901.03061
  [hep-ph]}\BibitemShut {NoStop}%
\bibitem [{\citenamefont{D\"obrich} \emph
  {et\,al.}(2019)\citenamefont{D\"obrich}, \citenamefont{Jaeckel}, and
  \citenamefont{Spadaro}}]{Dobrich:2019dxc}%
  \BibitemOpen
  \bibfield{author}{\bibinfo{author}{\bibfnamefont{B.}\,\bibnamefont{D\"obrich}},
  \bibinfo{author}{\bibfnamefont{J.}\,\bibnamefont{Jaeckel}},  and
  \bibinfo{author}{\bibfnamefont{T.}\,\bibnamefont{Spadaro}},
  }\bibfield{title}{\emph {\bibinfo{title}{{Light in the beam dump - ALP
  production from decay photons in proton beam-dumps}}}, }\href {\doibase
  10.1007/JHEP05(2019)213}
  {\bibfield{journal}{\bibinfo{journal}{JHEP}\,}\textbf{\bibinfo{volume}{05}}\,(\bibinfo{year}{2019})\,\bibinfo{pages}{213}},
  \bibinfo{note}{[Erratum: JHEP 10, 046 (2020)]}, \Eprint
  {http://arxiv.org/abs/1904.02091}{arXiv:1904.02091 [hep-ph]}\BibitemShut
  {NoStop}%
\bibitem [{\citenamefont{Coelho} \emph {et\,al.}(2020)\citenamefont{Coelho},
  \citenamefont{Goncalves}, \citenamefont{Martins}, and
  \citenamefont{Rangel}}]{Coelho:2020saz}%
  \BibitemOpen
  \bibfield{author}{\bibinfo{author}{\bibfnamefont{R.~O.}
  \bibnamefont{Coelho}}, \bibinfo{author}{\bibfnamefont{V.~P.}
  \bibnamefont{Goncalves}}, \bibinfo{author}{\bibfnamefont{D.~E.}
  \bibnamefont{Martins}},  and \bibinfo{author}{\bibfnamefont{M.~S.}
  \bibnamefont{Rangel}}, }\bibfield{title}{\emph {\bibinfo{title}{{Production
  of axionlike particles in $PbPb$ collisions at the LHC, HE\textendash{}LHC
  and FCC: A phenomenological analysis}}}, }\href {\doibase
  10.1016/j.physletb.2020.135512} {\bibfield{journal}{\bibinfo{journal}{Phys.
  Lett.
  B}\,}\textbf{\bibinfo{volume}{806}}\,(\bibinfo{year}{2020})\,\bibinfo{pages}{135512}},
  \Eprint {http://arxiv.org/abs/2002.06027}{arXiv:2002.06027
  [hep-ph]}\BibitemShut {NoStop}%
\bibitem [{\citenamefont{Haghighat} \emph
  {et\,al.}(2020)\citenamefont{Haghighat}, \citenamefont{Haji~Raissi}, and
  \citenamefont{Mohammadi~Najafabadi}}]{Haghighat:2020nuh}%
  \BibitemOpen
  \bibfield{author}{\bibinfo{author}{\bibfnamefont{G.}\,\bibnamefont{Haghighat}},
  \bibinfo{author}{\bibfnamefont{D.}\,\bibnamefont{Haji~Raissi}},  and
  \bibinfo{author}{\bibfnamefont{M.}\,\bibnamefont{Mohammadi~Najafabadi}},
  }\bibfield{title}{\emph {\bibinfo{title}{{New collider searches for axionlike
  particles coupling to gluons}}}, }\href {\doibase
  10.1103/PhysRevD.102.115010} {\bibfield{journal}{\bibinfo{journal}{Phys. Rev.
  D}\,}\textbf{\bibinfo{volume}{102}}\,(\bibinfo{year}{2020})\,\bibinfo{pages}{115010}},
  \Eprint {http://arxiv.org/abs/2006.05302}{arXiv:2006.05302
  [hep-ph]}\BibitemShut {NoStop}%
\bibitem [{\citenamefont{Bonnefoy} \emph
  {et\,al.}(2021{\natexlab{a}})\citenamefont{Bonnefoy},
  \citenamefont{Di~Luzio}, \citenamefont{Grojean}, \citenamefont{Paul}, and
  \citenamefont{Rossia}}]{Bonnefoy:2020gyh}%
  \BibitemOpen
  \bibfield{author}{\bibinfo{author}{\bibfnamefont{Q.}\,\bibnamefont{Bonnefoy}},
  \bibinfo{author}{\bibfnamefont{L.}\,\bibnamefont{Di~Luzio}},
  \bibinfo{author}{\bibfnamefont{C.}\,\bibnamefont{Grojean}},
  \bibinfo{author}{\bibfnamefont{A.}\,\bibnamefont{Paul}},  and
  \bibinfo{author}{\bibfnamefont{A.~N.} \bibnamefont{Rossia}},
  }\bibfield{title}{\emph {\bibinfo{title}{{The anomalous case of axion EFTs
  and massive chiral gauge fields}}}, }\href {\doibase 10.1007/JHEP07(2021)189}
  {\bibfield{journal}{\bibinfo{journal}{JHEP}\,}\textbf{\bibinfo{volume}{07}}\,(\bibinfo{year}{2021}{\natexlab{a}})\,\bibinfo{pages}{189}},
  \Eprint {http://arxiv.org/abs/2011.10025}{arXiv:2011.10025
  [hep-ph]}\BibitemShut {NoStop}%
\bibitem [{\citenamefont{Fl\'orez} \emph
  {et\,al.}(2021)\citenamefont{Fl\'orez}, \citenamefont{Gurrola},
  \citenamefont{Johns}, \citenamefont{Sheldon}, \citenamefont{Sheridan},
  \citenamefont{Sinha}, and \citenamefont{Soubasis}}]{Florez:2021zoo}%
  \BibitemOpen
  \bibfield{author}{\bibinfo{author}{\bibfnamefont{A.}\,\bibnamefont{Fl\'orez}},
  \bibinfo{author}{\bibfnamefont{A.}\,\bibnamefont{Gurrola}},
  \bibinfo{author}{\bibfnamefont{W.}\,\bibnamefont{Johns}},
  \bibinfo{author}{\bibfnamefont{P.}\,\bibnamefont{Sheldon}},
  \bibinfo{author}{\bibfnamefont{E.}\,\bibnamefont{Sheridan}},
  \bibinfo{author}{\bibfnamefont{K.}\,\bibnamefont{Sinha}},  and
  \bibinfo{author}{\bibfnamefont{B.}\,\bibnamefont{Soubasis}},
  }\bibfield{title}{\emph {\bibinfo{title}{{Probing axionlike particles with
  $\gamma\gamma$ final states from vector boson fusion processes at the LHC}}},
  }\href {\doibase 10.1103/PhysRevD.103.095001}
  {\bibfield{journal}{\bibinfo{journal}{Phys. Rev.
  D}\,}\textbf{\bibinfo{volume}{103}}\,(\bibinfo{year}{2021})\,\bibinfo{pages}{095001}},
  \Eprint {http://arxiv.org/abs/2101.11119}{arXiv:2101.11119
  [hep-ph]}\BibitemShut {NoStop}%
\bibitem [{\citenamefont{Wang} \emph
  {et\,al.}(2021{\natexlab{a}})\citenamefont{Wang}, \citenamefont{Wu},
  \citenamefont{Yang}, and \citenamefont{Zhang}}]{Wang:2021uyb}%
  \BibitemOpen
  \bibfield{author}{\bibinfo{author}{\bibfnamefont{D.}\,\bibnamefont{Wang}},
  \bibinfo{author}{\bibfnamefont{L.}\,\bibnamefont{Wu}},
  \bibinfo{author}{\bibfnamefont{J.~M.} \bibnamefont{Yang}},  and
  \bibinfo{author}{\bibfnamefont{M.}\,\bibnamefont{Zhang}},
  }\bibfield{title}{\emph {\bibinfo{title}{{Photon-jet events as a probe of
  axionlike particles at the LHC}}}, }\href {\doibase
  10.1103/PhysRevD.104.095016} {\bibfield{journal}{\bibinfo{journal}{Phys. Rev.
  D}\,}\textbf{\bibinfo{volume}{104}}\,(\bibinfo{year}{2021}{\natexlab{a}})\,\bibinfo{pages}{095016}},
  \Eprint {http://arxiv.org/abs/2102.01532}{arXiv:2102.01532
  [hep-ph]}\BibitemShut {NoStop}%
\bibitem [{\citenamefont{d'Enterria}(2021)}]{dEnterria:2021ljz}%
  \BibitemOpen
  \bibfield{author}{\bibinfo{author}{\bibfnamefont{D.}\,\bibnamefont{d'Enterria}},
  }\bibfield{title}{\emph {\bibinfo{title}{{Collider constraints on axion-like
  particles}}}, }in \href@noop {} {\emph {\bibinfo{booktitle}{{Workshop on
  Feebly Interacting Particles}}}}\,(\bibinfo{year}{2021})\,\Eprint
  {http://arxiv.org/abs/2102.08971}{arXiv:2102.08971 [hep-ex]}\BibitemShut
  {NoStop}%
\bibitem [{\citenamefont{Li} \emph {et\,al.}(2021)\citenamefont{Li},
  \citenamefont{Li}, \citenamefont{Lu}, and \citenamefont{Si}}]{Li:2021ygc}%
  \BibitemOpen
  \bibfield{author}{\bibinfo{author}{\bibfnamefont{S.-Y.} \bibnamefont{Li}},
  \bibinfo{author}{\bibfnamefont{Z.-Y.} \bibnamefont{Li}},
  \bibinfo{author}{\bibfnamefont{P.-C.} \bibnamefont{Lu}},  and
  \bibinfo{author}{\bibfnamefont{Z.-G.} \bibnamefont{Si}},
  }\bibfield{title}{\emph {\bibinfo{title}{{Precise evaluation of $h\to
  c\bar{c}$ and axion-like particle production}}}, }\href {\doibase
  10.1088/1674-1137/ac0c0d} {\bibfield{journal}{\bibinfo{journal}{Chin. Phys.
  C}\,}\textbf{\bibinfo{volume}{45}}\,(\bibinfo{year}{2021})\,\bibinfo{pages}{093105}},
  \Eprint {http://arxiv.org/abs/2103.00409}{arXiv:2103.00409
  [hep-ph]}\BibitemShut {NoStop}%
\bibitem [{\citenamefont{Goncalves} \emph
  {et\,al.}(2021)\citenamefont{Goncalves}, \citenamefont{Martins}, and
  \citenamefont{Rangel}}]{Goncalves:2021pdc}%
  \BibitemOpen
  \bibfield{author}{\bibinfo{author}{\bibfnamefont{V.~P.}
  \bibnamefont{Goncalves}}, \bibinfo{author}{\bibfnamefont{D.~E.}
  \bibnamefont{Martins}},  and \bibinfo{author}{\bibfnamefont{M.~S.}
  \bibnamefont{Rangel}}, }\bibfield{title}{\emph {\bibinfo{title}{{Searching
  for axionlike particles with low masses in pPb and PbPb collisions}}}, }\href
  {\doibase 10.1140/epjc/s10052-021-09314-2}
  {\bibfield{journal}{\bibinfo{journal}{Eur. Phys. J.
  C}\,}\textbf{\bibinfo{volume}{81}}\,(\bibinfo{year}{2021})\,\bibinfo{pages}{522}},
  \Eprint {http://arxiv.org/abs/2103.01862}{arXiv:2103.01862
  [hep-ph]}\BibitemShut {NoStop}%
\bibitem [{\citenamefont{Alves} \emph {et\,al.}(2021)\citenamefont{Alves},
  \citenamefont{Dias}, and \citenamefont{Lopes}}]{Alves:2021puo}%
  \BibitemOpen
  \bibfield{author}{\bibinfo{author}{\bibfnamefont{A.}\,\bibnamefont{Alves}},
  \bibinfo{author}{\bibfnamefont{A.~G.} \bibnamefont{Dias}},  and
  \bibinfo{author}{\bibfnamefont{D.~D.} \bibnamefont{Lopes}},
  }\bibfield{title}{\emph {\bibinfo{title}{{Jets and photons spectroscopy of
  Higgs-ALP interactions}}}, }\href {\doibase 10.1007/JHEP10(2021)012}
  {\bibfield{journal}{\bibinfo{journal}{JHEP}\,}\textbf{\bibinfo{volume}{10}}\,(\bibinfo{year}{2021})\,\bibinfo{pages}{012}},
  \Eprint {http://arxiv.org/abs/2105.01095}{arXiv:2105.01095
  [hep-ph]}\BibitemShut {NoStop}%
\bibitem [{\citenamefont{Bonilla} \emph {et\,al.}(2022)\citenamefont{Bonilla},
  \citenamefont{Brivio}, \citenamefont{Machado-Rodr\'\i{}guez}, and
  \citenamefont{de\,Troc\'oniz}}]{Bonilla:2022pxu}%
  \BibitemOpen
  \bibfield{author}{\bibinfo{author}{\bibfnamefont{J.}\,\bibnamefont{Bonilla}},
  \bibinfo{author}{\bibfnamefont{I.}\,\bibnamefont{Brivio}},
  \bibinfo{author}{\bibfnamefont{J.}\,\bibnamefont{Machado-Rodr\'\i{}guez}},
  and \bibinfo{author}{\bibfnamefont{J.~F.} \bibnamefont{de\,Troc\'oniz}},
  }\bibfield{title}{\emph {\bibinfo{title}{{Nonresonant Searches for Axion-Like
  Particles in Vector Boson Scattering Processes at the LHC}}}, }\href@noop {}
  {\,(\bibinfo{year}{2022})}, \Eprint
  {http://arxiv.org/abs/2202.03450}{arXiv:2202.03450 [hep-ph]}\BibitemShut
  {NoStop}%
\bibitem [{\citenamefont{Carmona} \emph {et\,al.}(2022)\citenamefont{Carmona},
  \citenamefont{Elahi}, \citenamefont{Scherb}, and
  \citenamefont{Schwaller}}]{Carmona:2022jid}%
  \BibitemOpen
  \bibfield{author}{\bibinfo{author}{\bibfnamefont{A.}\,\bibnamefont{Carmona}},
  \bibinfo{author}{\bibfnamefont{F.}\,\bibnamefont{Elahi}},
  \bibinfo{author}{\bibfnamefont{C.}\,\bibnamefont{Scherb}},  and
  \bibinfo{author}{\bibfnamefont{P.}\,\bibnamefont{Schwaller}},
  }\bibfield{title}{\emph {\bibinfo{title}{{The ALPs from the Top: Searching
  for long lived axion-like particles from exotic top decays}}}, }\href@noop {}
  {\,(\bibinfo{year}{2022})}, \Eprint
  {http://arxiv.org/abs/2202.09371}{arXiv:2202.09371 [hep-ph]}\BibitemShut
  {NoStop}%
\bibitem [{\citenamefont{Kling} and
  \citenamefont{Qu\'\i{}lez}(2022)}]{Kling:2022ehv}%
  \BibitemOpen
  \bibfield{author}{\bibinfo{author}{\bibfnamefont{F.}\,\bibnamefont{Kling}}
  and \bibinfo{author}{\bibfnamefont{P.}\,\bibnamefont{Qu\'\i{}lez}},
  }\bibfield{title}{\emph {\bibinfo{title}{{Axion Searches at the LHC: FASER as
  a Light Shining through Walls Experiment}}}, }\href@noop {}
  {\,(\bibinfo{year}{2022})}, \Eprint
  {http://arxiv.org/abs/2204.03599}{arXiv:2204.03599 [hep-ph]}\BibitemShut
  {NoStop}%
\bibitem [{\citenamefont{Chala} \emph {et\,al.}(2021)\citenamefont{Chala},
  \citenamefont{Guedes}, \citenamefont{Ramos}, and
  \citenamefont{Santiago}}]{Chala:2020wvs}%
  \BibitemOpen
  \bibfield{author}{\bibinfo{author}{\bibfnamefont{M.}\,\bibnamefont{Chala}},
  \bibinfo{author}{\bibfnamefont{G.}\,\bibnamefont{Guedes}},
  \bibinfo{author}{\bibfnamefont{M.}\,\bibnamefont{Ramos}},  and
  \bibinfo{author}{\bibfnamefont{J.}\,\bibnamefont{Santiago}},
  }\bibfield{title}{\emph {\bibinfo{title}{{Running in the ALPs}}}, }\href
  {\doibase 10.1140/epjc/s10052-021-08968-2}
  {\bibfield{journal}{\bibinfo{journal}{Eur. Phys. J.
  C}\,}\textbf{\bibinfo{volume}{81}}\,(\bibinfo{year}{2021})\,\bibinfo{pages}{181}},
  \Eprint {http://arxiv.org/abs/2012.09017}{arXiv:2012.09017
  [hep-ph]}\BibitemShut {NoStop}%
\bibitem [{\citenamefont{Bonilla} \emph {et\,al.}(2021)\citenamefont{Bonilla},
  \citenamefont{Brivio}, \citenamefont{Gavela}, and
  \citenamefont{Sanz}}]{Bonilla:2021ufe}%
  \BibitemOpen
  \bibfield{author}{\bibinfo{author}{\bibfnamefont{J.}\,\bibnamefont{Bonilla}},
  \bibinfo{author}{\bibfnamefont{I.}\,\bibnamefont{Brivio}},
  \bibinfo{author}{\bibfnamefont{M.~B.} \bibnamefont{Gavela}},  and
  \bibinfo{author}{\bibfnamefont{V.}\,\bibnamefont{Sanz}},
  }\bibfield{title}{\emph {\bibinfo{title}{{One-loop corrections to ALP
  couplings}}}, }\href {\doibase 10.1007/JHEP11(2021)168}
  {\bibfield{journal}{\bibinfo{journal}{JHEP}\,}\textbf{\bibinfo{volume}{11}}\,(\bibinfo{year}{2021})\,\bibinfo{pages}{168}},
  \Eprint {http://arxiv.org/abs/2107.11392}{arXiv:2107.11392
  [hep-ph]}\BibitemShut {NoStop}%
\bibitem [{\citenamefont{Bauer} \emph
  {et\,al.}(2021{\natexlab{b}})\citenamefont{Bauer}, \citenamefont{Neubert},
  \citenamefont{Renner}, \citenamefont{Schnubel}, and
  \citenamefont{Thamm}}]{Bauer:2020jbp}%
  \BibitemOpen
  \bibfield{author}{\bibinfo{author}{\bibfnamefont{M.}\,\bibnamefont{Bauer}},
  \bibinfo{author}{\bibfnamefont{M.}\,\bibnamefont{Neubert}},
  \bibinfo{author}{\bibfnamefont{S.}\,\bibnamefont{Renner}},
  \bibinfo{author}{\bibfnamefont{M.}\,\bibnamefont{Schnubel}},  and
  \bibinfo{author}{\bibfnamefont{A.}\,\bibnamefont{Thamm}},
  }\bibfield{title}{\emph {\bibinfo{title}{{The Low-Energy Effective Theory of
  Axions and ALPs}}}, }\href {\doibase 10.1007/JHEP04(2021)063}
  {\bibfield{journal}{\bibinfo{journal}{JHEP}\,}\textbf{\bibinfo{volume}{04}}\,(\bibinfo{year}{2021}{\natexlab{b}})\,\bibinfo{pages}{063}},
  \Eprint {http://arxiv.org/abs/2012.12272}{arXiv:2012.12272
  [hep-ph]}\BibitemShut {NoStop}%
\bibitem [{\citenamefont{Jarlskog}(1985{\natexlab{a}})}]{Jarlskog:1985ht}%
  \BibitemOpen
  \bibfield{author}{\bibinfo{author}{\bibfnamefont{C.}\,\bibnamefont{Jarlskog}},
  }\bibfield{title}{\emph {\bibinfo{title}{{Commutator of the Quark Mass
  Matrices in the Standard Electroweak Model and a Measure of Maximal CP
  Violation}}}, }\href {\doibase 10.1103/PhysRevLett.55.1039}
  {\bibfield{journal}{\bibinfo{journal}{Phys. Rev.
  Lett.}\,}\textbf{\bibinfo{volume}{55}}\,(\bibinfo{year}{1985}{\natexlab{a}})\,\bibinfo{pages}{1039}}\BibitemShut
  {NoStop}%
\bibitem [{\citenamefont{Jarlskog}(1985{\natexlab{b}})}]{Jarlskog:1985cw}%
  \BibitemOpen
  \bibfield{author}{\bibinfo{author}{\bibfnamefont{C.}\,\bibnamefont{Jarlskog}},
  }\bibfield{title}{\emph {\bibinfo{title}{{A Basis Independent Formulation of
  the Connection Between Quark Mass Matrices, CP Violation and Experiment}}},
  }\href {\doibase 10.1007/BF01565198} {\bibfield{journal}{\bibinfo{journal}{Z.
  Phys.
  C}\,}\textbf{\bibinfo{volume}{29}}\,(\bibinfo{year}{1985}{\natexlab{b}})\,\bibinfo{pages}{491}}\BibitemShut
  {NoStop}%
\bibitem [{\citenamefont{Bonnefoy} \emph
  {et\,al.}(2021{\natexlab{b}})\citenamefont{Bonnefoy}, \citenamefont{Gendy},
  \citenamefont{Grojean}, and \citenamefont{Ruderman}}]{Bonnefoy:2021tbt}%
  \BibitemOpen
  \bibfield{author}{\bibinfo{author}{\bibfnamefont{Q.}\,\bibnamefont{Bonnefoy}},
  \bibinfo{author}{\bibfnamefont{E.}\,\bibnamefont{Gendy}},
  \bibinfo{author}{\bibfnamefont{C.}\,\bibnamefont{Grojean}},  and
  \bibinfo{author}{\bibfnamefont{J.~T.} \bibnamefont{Ruderman}},
  }\bibfield{title}{\emph {\bibinfo{title}{{Beyond Jarlskog: 699 invariants for
  CP violation in SMEFT}}}, }\href@noop {}
  {\,(\bibinfo{year}{2021}{\natexlab{b}})}, \Eprint
  {http://arxiv.org/abs/2112.03889}{arXiv:2112.03889 [hep-ph]}\BibitemShut
  {NoStop}%
\bibitem [{\citenamefont{Davidson} and
  \citenamefont{Wali}(1982)}]{Davidson:1981zd}%
  \BibitemOpen
  \bibfield{author}{\bibinfo{author}{\bibfnamefont{A.}\,\bibnamefont{Davidson}}
  and \bibinfo{author}{\bibfnamefont{K.~C.} \bibnamefont{Wali}},
  }\bibfield{title}{\emph {\bibinfo{title}{{MINIMAL FLAVOR UNIFICATION VIA
  MULTIGENERATIONAL PECCEI-QUINN SYMMETRY}}}, }\href {\doibase
  10.1103/PhysRevLett.48.11} {\bibfield{journal}{\bibinfo{journal}{Phys. Rev.
  Lett.}\,}\textbf{\bibinfo{volume}{48}}\,(\bibinfo{year}{1982})\,\bibinfo{pages}{11}}\BibitemShut
  {NoStop}%
\bibitem [{\citenamefont{Calibbi} \emph {et\,al.}(2017)\citenamefont{Calibbi},
  \citenamefont{Goertz}, \citenamefont{Redigolo}, \citenamefont{Ziegler}, and
  \citenamefont{Zupan}}]{Calibbi:2016hwq}%
  \BibitemOpen
  \bibfield{author}{\bibinfo{author}{\bibfnamefont{L.}\,\bibnamefont{Calibbi}},
  \bibinfo{author}{\bibfnamefont{F.}\,\bibnamefont{Goertz}},
  \bibinfo{author}{\bibfnamefont{D.}\,\bibnamefont{Redigolo}},
  \bibinfo{author}{\bibfnamefont{R.}\,\bibnamefont{Ziegler}},  and
  \bibinfo{author}{\bibfnamefont{J.}\,\bibnamefont{Zupan}},
  }\bibfield{title}{\emph {\bibinfo{title}{{Minimal axion model from flavor}}},
  }\href {\doibase 10.1103/PhysRevD.95.095009}
  {\bibfield{journal}{\bibinfo{journal}{Phys. Rev.
  D}\,}\textbf{\bibinfo{volume}{95}}\,(\bibinfo{year}{2017})\,\bibinfo{pages}{095009}},
  \Eprint {http://arxiv.org/abs/1612.08040}{arXiv:1612.08040
  [hep-ph]}\BibitemShut {NoStop}%
\bibitem [{\citenamefont{Ema} \emph {et\,al.}(2017)\citenamefont{Ema},
  \citenamefont{Hamaguchi}, \citenamefont{Moroi}, and
  \citenamefont{Nakayama}}]{Ema:2016ops}%
  \BibitemOpen
  \bibfield{author}{\bibinfo{author}{\bibfnamefont{Y.}\,\bibnamefont{Ema}},
  \bibinfo{author}{\bibfnamefont{K.}\,\bibnamefont{Hamaguchi}},
  \bibinfo{author}{\bibfnamefont{T.}\,\bibnamefont{Moroi}},  and
  \bibinfo{author}{\bibfnamefont{K.}\,\bibnamefont{Nakayama}},
  }\bibfield{title}{\emph {\bibinfo{title}{{Flaxion: a minimal extension to
  solve puzzles in the standard model}}}, }\href {\doibase
  10.1007/JHEP01(2017)096}
  {\bibfield{journal}{\bibinfo{journal}{JHEP}\,}\textbf{\bibinfo{volume}{01}}\,(\bibinfo{year}{2017})\,\bibinfo{pages}{096}},
  \Eprint {http://arxiv.org/abs/1612.05492}{arXiv:1612.05492
  [hep-ph]}\BibitemShut {NoStop}%
\bibitem [{\citenamefont{Froggatt} and
  \citenamefont{Nielsen}(1979)}]{Froggatt:1978nt}%
  \BibitemOpen
  \bibfield{author}{\bibinfo{author}{\bibfnamefont{C.~D.}
  \bibnamefont{Froggatt}} and \bibinfo{author}{\bibfnamefont{H.~B.}
  \bibnamefont{Nielsen}}, }\bibfield{title}{\emph {\bibinfo{title}{{Hierarchy
  of Quark Masses, Cabibbo Angles and CP Violation}}}, }\href {\doibase
  10.1016/0550-3213(79)90316-X} {\bibfield{journal}{\bibinfo{journal}{Nucl.
  Phys.
  B}\,}\textbf{\bibinfo{volume}{147}}\,(\bibinfo{year}{1979})\,\bibinfo{pages}{277}}\BibitemShut
  {NoStop}%
\bibitem [{\citenamefont{Bonnefoy} \emph
  {et\,al.}(2020)\citenamefont{Bonnefoy}, \citenamefont{Dudas}, and
  \citenamefont{Pokorski}}]{Bonnefoy:2019lsn}%
  \BibitemOpen
  \bibfield{author}{\bibinfo{author}{\bibfnamefont{Q.}\,\bibnamefont{Bonnefoy}},
  \bibinfo{author}{\bibfnamefont{E.}\,\bibnamefont{Dudas}},  and
  \bibinfo{author}{\bibfnamefont{S.}\,\bibnamefont{Pokorski}},
  }\bibfield{title}{\emph {\bibinfo{title}{{Chiral Froggatt-Nielsen models,
  gauge anomalies and flavourful axions}}}, }\href {\doibase
  10.1007/JHEP01(2020)191}
  {\bibfield{journal}{\bibinfo{journal}{JHEP}\,}\textbf{\bibinfo{volume}{01}}\,(\bibinfo{year}{2020})\,\bibinfo{pages}{191}},
  \Eprint {http://arxiv.org/abs/1909.05336}{arXiv:1909.05336
  [hep-ph]}\BibitemShut {NoStop}%
\bibitem [{\citenamefont{Branco} \emph {et\,al.}(2012)\citenamefont{Branco},
  \citenamefont{Ferreira}, \citenamefont{Lavoura}, \citenamefont{Rebelo},
  \citenamefont{Sher}, and \citenamefont{Silva}}]{Branco:2011iw}%
  \BibitemOpen
  \bibfield{author}{\bibinfo{author}{\bibfnamefont{G.~C.}
  \bibnamefont{Branco}}, \bibinfo{author}{\bibfnamefont{P.~M.}
  \bibnamefont{Ferreira}},
  \bibinfo{author}{\bibfnamefont{L.}\,\bibnamefont{Lavoura}},
  \bibinfo{author}{\bibfnamefont{M.~N.} \bibnamefont{Rebelo}},
  \bibinfo{author}{\bibfnamefont{M.}\,\bibnamefont{Sher}},  and
  \bibinfo{author}{\bibfnamefont{J.~P.} \bibnamefont{Silva}},
  }\bibfield{title}{\emph {\bibinfo{title}{{Theory and phenomenology of
  two-Higgs-doublet models}}}, }\href {\doibase 10.1016/j.physrep.2012.02.002}
  {\bibfield{journal}{\bibinfo{journal}{Phys.
  Rept.}\,}\textbf{\bibinfo{volume}{516}}\,(\bibinfo{year}{2012})\,\bibinfo{pages}{1}},
  \Eprint {http://arxiv.org/abs/1106.0034}{arXiv:1106.0034
  [hep-ph]}\BibitemShut {NoStop}%
\bibitem [{\citenamefont{Di~Luzio} \emph
  {et\,al.}(2021)\citenamefont{Di~Luzio}, \citenamefont{Gr\"ober}, and
  \citenamefont{Paradisi}}]{DiLuzio:2020oah}%
  \BibitemOpen
  \bibfield{author}{\bibinfo{author}{\bibfnamefont{L.}\,\bibnamefont{Di~Luzio}},
  \bibinfo{author}{\bibfnamefont{R.}\,\bibnamefont{Gr\"ober}},  and
  \bibinfo{author}{\bibfnamefont{P.}\,\bibnamefont{Paradisi}},
  }\bibfield{title}{\emph {\bibinfo{title}{{Hunting for $CP$-violating
  axionlike particle interactions}}}, }\href {\doibase
  10.1103/PhysRevD.104.095027} {\bibfield{journal}{\bibinfo{journal}{Phys. Rev.
  D}\,}\textbf{\bibinfo{volume}{104}}\,(\bibinfo{year}{2021})\,\bibinfo{pages}{095027}},
  \Eprint {http://arxiv.org/abs/2010.13760}{arXiv:2010.13760
  [hep-ph]}\BibitemShut {NoStop}%
\bibitem [{\citenamefont{Dekens} \emph {et\,al.}(2022)\citenamefont{Dekens},
  \citenamefont{de\,Vries}, and \citenamefont{Shain}}]{Dekens:2022gha}%
  \BibitemOpen
  \bibfield{author}{\bibinfo{author}{\bibfnamefont{W.}\,\bibnamefont{Dekens}},
  \bibinfo{author}{\bibfnamefont{J.}\,\bibnamefont{de\,Vries}},  and
  \bibinfo{author}{\bibfnamefont{S.}\,\bibnamefont{Shain}},
  }\bibfield{title}{\emph {\bibinfo{title}{{CP-violating axion interactions in
  effective field theory}}}, }\href@noop {} {\,(\bibinfo{year}{2022})}, \Eprint
  {http://arxiv.org/abs/2203.11230}{arXiv:2203.11230 [hep-ph]}\BibitemShut
  {NoStop}%
\bibitem [{\citenamefont{Coleman} \emph {et\,al.}(1969)\citenamefont{Coleman},
  \citenamefont{Wess}, and \citenamefont{Zumino}}]{Coleman:1969sm}%
  \BibitemOpen
  \bibfield{author}{\bibinfo{author}{\bibfnamefont{S.~R.}
  \bibnamefont{Coleman}},
  \bibinfo{author}{\bibfnamefont{J.}\,\bibnamefont{Wess}},  and
  \bibinfo{author}{\bibfnamefont{B.}\,\bibnamefont{Zumino}},
  }\bibfield{title}{\emph {\bibinfo{title}{{Structure of phenomenological
  Lagrangians. 1.}}}, }\href {\doibase 10.1103/PhysRev.177.2239}
  {\bibfield{journal}{\bibinfo{journal}{Phys.
  Rev.}\,}\textbf{\bibinfo{volume}{177}}\,(\bibinfo{year}{1969})\,\bibinfo{pages}{2239}}\BibitemShut
  {NoStop}%
\bibitem [{\citenamefont{Callan} \emph {et\,al.}(1969)\citenamefont{Callan},
  \citenamefont{Coleman}, \citenamefont{Wess}, and
  \citenamefont{Zumino}}]{Callan:1969sn}%
  \BibitemOpen
  \bibfield{author}{\bibinfo{author}{\bibfnamefont{C.~G.} \bibnamefont{Callan},
  \bibfnamefont{Jr.}}, \bibinfo{author}{\bibfnamefont{S.~R.}
  \bibnamefont{Coleman}},
  \bibinfo{author}{\bibfnamefont{J.}\,\bibnamefont{Wess}},  and
  \bibinfo{author}{\bibfnamefont{B.}\,\bibnamefont{Zumino}},
  }\bibfield{title}{\emph {\bibinfo{title}{{Structure of phenomenological
  Lagrangians. 2.}}}, }\href {\doibase 10.1103/PhysRev.177.2247}
  {\bibfield{journal}{\bibinfo{journal}{Phys.
  Rev.}\,}\textbf{\bibinfo{volume}{177}}\,(\bibinfo{year}{1969})\,\bibinfo{pages}{2247}}\BibitemShut
  {NoStop}%
\bibitem [{\citenamefont{Feruglio}(1993)}]{Feruglio:1992wf}%
  \BibitemOpen
  \bibfield{author}{\bibinfo{author}{\bibfnamefont{F.}\,\bibnamefont{Feruglio}},
  }\bibfield{title}{\emph {\bibinfo{title}{{The Chiral approach to the
  electroweak interactions}}}, }\href {\doibase 10.1142/S0217751X93001946}
  {\bibfield{journal}{\bibinfo{journal}{Int. J. Mod. Phys.
  A}\,}\textbf{\bibinfo{volume}{8}}\,(\bibinfo{year}{1993})\,\bibinfo{pages}{4937}},
  \Eprint
  {http://arxiv.org/abs/hep-ph/9301281}{arXiv:hep-ph/9301281}\BibitemShut
  {NoStop}%
\bibitem [{\citenamefont{Feldmann} \emph
  {et\,al.}(2015)\citenamefont{Feldmann}, \citenamefont{Mannel}, and
  \citenamefont{Schwertfeger}}]{Feldmann:2015nia}%
  \BibitemOpen
  \bibfield{author}{\bibinfo{author}{\bibfnamefont{T.}\,\bibnamefont{Feldmann}},
  \bibinfo{author}{\bibfnamefont{T.}\,\bibnamefont{Mannel}},  and
  \bibinfo{author}{\bibfnamefont{S.}\,\bibnamefont{Schwertfeger}},
  }\bibfield{title}{\emph {\bibinfo{title}{{Renormalization Group Evolution of
  Flavour Invariants}}}, }\href {\doibase 10.1007/JHEP10(2015)007}
  {\bibfield{journal}{\bibinfo{journal}{JHEP}\,}\textbf{\bibinfo{volume}{10}}\,(\bibinfo{year}{2015})\,\bibinfo{pages}{007}},
  \Eprint {http://arxiv.org/abs/1507.00328}{arXiv:1507.00328
  [hep-ph]}\BibitemShut {NoStop}%
\bibitem [{\citenamefont{Yu} and
  \citenamefont{Zhou}(2021{\natexlab{a}})}]{Yu:2020gre}%
  \BibitemOpen
  \bibfield{author}{\bibinfo{author}{\bibfnamefont{B.}\,\bibnamefont{Yu}} and
  \bibinfo{author}{\bibfnamefont{S.}\,\bibnamefont{Zhou}},
  }\bibfield{title}{\emph {\bibinfo{title}{{Sufficient and Necessary Conditions
  for CP Conservation in the Case of Degenerate Majorana Neutrino Masses}}},
  }\href {\doibase 10.1103/PhysRevD.103.035017}
  {\bibfield{journal}{\bibinfo{journal}{Phys. Rev.
  D}\,}\textbf{\bibinfo{volume}{103}}\,(\bibinfo{year}{2021}{\natexlab{a}})\,\bibinfo{pages}{035017}},
  \Eprint {http://arxiv.org/abs/2009.12347}{arXiv:2009.12347
  [hep-ph]}\BibitemShut {NoStop}%
\bibitem [{\citenamefont{Wang} \emph
  {et\,al.}(2021{\natexlab{b}})\citenamefont{Wang}, \citenamefont{Yu}, and
  \citenamefont{Zhou}}]{Wang:2021wdq}%
  \BibitemOpen
  \bibfield{author}{\bibinfo{author}{\bibfnamefont{Y.}\,\bibnamefont{Wang}},
  \bibinfo{author}{\bibfnamefont{B.}\,\bibnamefont{Yu}},  and
  \bibinfo{author}{\bibfnamefont{S.}\,\bibnamefont{Zhou}},
  }\bibfield{title}{\emph {\bibinfo{title}{{Flavor invariants and
  renormalization-group equations in the leptonic sector with massive Majorana
  neutrinos}}}, }\href {\doibase 10.1007/JHEP09(2021)053}
  {\bibfield{journal}{\bibinfo{journal}{JHEP}\,}\textbf{\bibinfo{volume}{09}}\,(\bibinfo{year}{2021}{\natexlab{b}})\,\bibinfo{pages}{053}},
  \Eprint {http://arxiv.org/abs/2107.06274}{arXiv:2107.06274
  [hep-ph]}\BibitemShut {NoStop}%
\bibitem [{\citenamefont{Jenkins} and
  \citenamefont{Manohar}(2009)}]{Jenkins:2009dy}%
  \BibitemOpen
  \bibfield{author}{\bibinfo{author}{\bibfnamefont{E.~E.}
  \bibnamefont{Jenkins}} and \bibinfo{author}{\bibfnamefont{A.~V.}
  \bibnamefont{Manohar}}, }\bibfield{title}{\emph {\bibinfo{title}{{Algebraic
  Structure of Lepton and Quark Flavor Invariants and CP Violation}}}, }\href
  {\doibase 10.1088/1126-6708/2009/10/094}
  {\bibfield{journal}{\bibinfo{journal}{JHEP}\,}\textbf{\bibinfo{volume}{10}}\,(\bibinfo{year}{2009})\,\bibinfo{pages}{094}},
  \Eprint {http://arxiv.org/abs/0907.4763}{arXiv:0907.4763
  [hep-ph]}\BibitemShut {NoStop}%
\bibitem [{\citenamefont{Trautner}(2019)}]{Trautner:2018ipq}%
  \BibitemOpen
  \bibfield{author}{\bibinfo{author}{\bibfnamefont{A.}\,\bibnamefont{Trautner}},
  }\bibfield{title}{\emph {\bibinfo{title}{{Systematic construction of basis
  invariants in the 2HDM}}}, }\href {\doibase 10.1007/JHEP05(2019)208}
  {\bibfield{journal}{\bibinfo{journal}{JHEP}\,}\textbf{\bibinfo{volume}{05}}\,(\bibinfo{year}{2019})\,\bibinfo{pages}{208}},
  \Eprint {http://arxiv.org/abs/1812.02614}{arXiv:1812.02614
  [hep-ph]}\BibitemShut {NoStop}%
\bibitem [{\citenamefont{Yu} and
  \citenamefont{Zhou}(2021{\natexlab{b}})}]{Yu:2021cco}%
  \BibitemOpen
  \bibfield{author}{\bibinfo{author}{\bibfnamefont{B.}\,\bibnamefont{Yu}} and
  \bibinfo{author}{\bibfnamefont{S.}\,\bibnamefont{Zhou}},
  }\bibfield{title}{\emph {\bibinfo{title}{{Hilbert Series for Leptonic Flavor
  Invariants in the Minimal Seesaw Model}}}, }\href@noop {}
  {\,(\bibinfo{year}{2021}{\natexlab{b}})}, \Eprint
  {http://arxiv.org/abs/2107.11928}{arXiv:2107.11928 [hep-ph]}\BibitemShut
  {NoStop}%
\bibitem [{\citenamefont{Jenkins} \emph
  {et\,al.}(2018{\natexlab{a}})\citenamefont{Jenkins}, \citenamefont{Manohar},
  and \citenamefont{Stoffer}}]{Jenkins:2017jig}%
  \BibitemOpen
  \bibfield{author}{\bibinfo{author}{\bibfnamefont{E.~E.}
  \bibnamefont{Jenkins}}, \bibinfo{author}{\bibfnamefont{A.~V.}
  \bibnamefont{Manohar}},  and
  \bibinfo{author}{\bibfnamefont{P.}\,\bibnamefont{Stoffer}},
  }\bibfield{title}{\emph {\bibinfo{title}{{Low-Energy Effective Field Theory
  below the Electroweak Scale: Operators and Matching}}}, }\href {\doibase
  10.1007/JHEP03(2018)016}
  {\bibfield{journal}{\bibinfo{journal}{JHEP}\,}\textbf{\bibinfo{volume}{03}}\,(\bibinfo{year}{2018}{\natexlab{a}})\,\bibinfo{pages}{016}},
  \Eprint {http://arxiv.org/abs/1709.04486}{arXiv:1709.04486
  [hep-ph]}\BibitemShut {NoStop}%
\bibitem [{\citenamefont{Jenkins} \emph
  {et\,al.}(2018{\natexlab{b}})\citenamefont{Jenkins}, \citenamefont{Manohar},
  and \citenamefont{Stoffer}}]{Jenkins:2017dyc}%
  \BibitemOpen
  \bibfield{author}{\bibinfo{author}{\bibfnamefont{E.~E.}
  \bibnamefont{Jenkins}}, \bibinfo{author}{\bibfnamefont{A.~V.}
  \bibnamefont{Manohar}},  and
  \bibinfo{author}{\bibfnamefont{P.}\,\bibnamefont{Stoffer}},
  }\bibfield{title}{\emph {\bibinfo{title}{{Low-Energy Effective Field Theory
  below the Electroweak Scale: Anomalous Dimensions}}}, }\href {\doibase
  10.1007/JHEP01(2018)084}
  {\bibfield{journal}{\bibinfo{journal}{JHEP}\,}\textbf{\bibinfo{volume}{01}}\,(\bibinfo{year}{2018}{\natexlab{b}})\,\bibinfo{pages}{084}},
  \Eprint {http://arxiv.org/abs/1711.05270}{arXiv:1711.05270
  [hep-ph]}\BibitemShut {NoStop}%
\bibitem [{\citenamefont{Galda} \emph {et\,al.}(2021)\citenamefont{Galda},
  \citenamefont{Neubert}, and \citenamefont{Renner}}]{Galda:2021hbr}%
  \BibitemOpen
  \bibfield{author}{\bibinfo{author}{\bibfnamefont{A.~M.} \bibnamefont{Galda}},
  \bibinfo{author}{\bibfnamefont{M.}\,\bibnamefont{Neubert}},  and
  \bibinfo{author}{\bibfnamefont{S.}\,\bibnamefont{Renner}},
  }\bibfield{title}{\emph {\bibinfo{title}{{ALP \textemdash{} SMEFT
  interference}}}, }\href {\doibase 10.1007/JHEP06(2021)135}
  {\bibfield{journal}{\bibinfo{journal}{JHEP}\,}\textbf{\bibinfo{volume}{06}}\,(\bibinfo{year}{2021})\,\bibinfo{pages}{135}},
  \Eprint {http://arxiv.org/abs/2105.01078}{arXiv:2105.01078
  [hep-ph]}\BibitemShut {NoStop}%
\bibitem [{\citenamefont{Grojean} \emph {et\,al.}(2013)\citenamefont{Grojean},
  \citenamefont{Jenkins}, \citenamefont{Manohar}, and
  \citenamefont{Trott}}]{Grojean:2013kd}%
  \BibitemOpen
  \bibfield{author}{\bibinfo{author}{\bibfnamefont{C.}\,\bibnamefont{Grojean}},
  \bibinfo{author}{\bibfnamefont{E.~E.} \bibnamefont{Jenkins}},
  \bibinfo{author}{\bibfnamefont{A.~V.} \bibnamefont{Manohar}},  and
  \bibinfo{author}{\bibfnamefont{M.}\,\bibnamefont{Trott}},
  }\bibfield{title}{\emph {\bibinfo{title}{{Renormalization Group Scaling of
  Higgs Operators and \textbackslash{}Gamma(h -\ensuremath{>}
  \textbackslash{}gamma \textbackslash{}gamma)}}}, }\href {\doibase
  10.1007/JHEP04(2013)016}
  {\bibfield{journal}{\bibinfo{journal}{JHEP}\,}\textbf{\bibinfo{volume}{04}}\,(\bibinfo{year}{2013})\,\bibinfo{pages}{016}},
  \Eprint {http://arxiv.org/abs/1301.2588}{arXiv:1301.2588
  [hep-ph]}\BibitemShut {NoStop}%
\bibitem [{\citenamefont{Jenkins} \emph {et\,al.}(2013)\citenamefont{Jenkins},
  \citenamefont{Manohar}, and \citenamefont{Trott}}]{Jenkins:2013zja}%
  \BibitemOpen
  \bibfield{author}{\bibinfo{author}{\bibfnamefont{E.~E.}
  \bibnamefont{Jenkins}}, \bibinfo{author}{\bibfnamefont{A.~V.}
  \bibnamefont{Manohar}},  and
  \bibinfo{author}{\bibfnamefont{M.}\,\bibnamefont{Trott}},
  }\bibfield{title}{\emph {\bibinfo{title}{{Renormalization Group Evolution of
  the Standard Model Dimension Six Operators I: Formalism and lambda
  Dependence}}}, }\href {\doibase 10.1007/JHEP10(2013)087}
  {\bibfield{journal}{\bibinfo{journal}{JHEP}\,}\textbf{\bibinfo{volume}{10}}\,(\bibinfo{year}{2013})\,\bibinfo{pages}{087}},
  \Eprint {http://arxiv.org/abs/1308.2627}{arXiv:1308.2627
  [hep-ph]}\BibitemShut {NoStop}%
\bibitem [{\citenamefont{Jenkins} \emph {et\,al.}(2014)\citenamefont{Jenkins},
  \citenamefont{Manohar}, and \citenamefont{Trott}}]{Jenkins:2013wua}%
  \BibitemOpen
  \bibfield{author}{\bibinfo{author}{\bibfnamefont{E.~E.}
  \bibnamefont{Jenkins}}, \bibinfo{author}{\bibfnamefont{A.~V.}
  \bibnamefont{Manohar}},  and
  \bibinfo{author}{\bibfnamefont{M.}\,\bibnamefont{Trott}},
  }\bibfield{title}{\emph {\bibinfo{title}{{Renormalization Group Evolution of
  the Standard Model Dimension Six Operators II: Yukawa Dependence}}}, }\href
  {\doibase 10.1007/JHEP01(2014)035}
  {\bibfield{journal}{\bibinfo{journal}{JHEP}\,}\textbf{\bibinfo{volume}{01}}\,(\bibinfo{year}{2014})\,\bibinfo{pages}{035}},
  \Eprint {http://arxiv.org/abs/1310.4838}{arXiv:1310.4838
  [hep-ph]}\BibitemShut {NoStop}%
\bibitem [{\citenamefont{Alonso} \emph {et\,al.}(2014)\citenamefont{Alonso},
  \citenamefont{Jenkins}, \citenamefont{Manohar}, and
  \citenamefont{Trott}}]{Alonso:2013hga}%
  \BibitemOpen
  \bibfield{author}{\bibinfo{author}{\bibfnamefont{R.}\,\bibnamefont{Alonso}},
  \bibinfo{author}{\bibfnamefont{E.~E.} \bibnamefont{Jenkins}},
  \bibinfo{author}{\bibfnamefont{A.~V.} \bibnamefont{Manohar}},  and
  \bibinfo{author}{\bibfnamefont{M.}\,\bibnamefont{Trott}},
  }\bibfield{title}{\emph {\bibinfo{title}{{Renormalization Group Evolution of
  the Standard Model Dimension Six Operators III: Gauge Coupling Dependence and
  Phenomenology}}}, }\href {\doibase 10.1007/JHEP04(2014)159}
  {\bibfield{journal}{\bibinfo{journal}{JHEP}\,}\textbf{\bibinfo{volume}{04}}\,(\bibinfo{year}{2014})\,\bibinfo{pages}{159}},
  \Eprint {http://arxiv.org/abs/1312.2014}{arXiv:1312.2014
  [hep-ph]}\BibitemShut {NoStop}%
\bibitem [{\citenamefont{Yu} and \citenamefont{Zhou}(2022)}]{Yu:2022ttm}%
  \BibitemOpen
  \bibfield{author}{\bibinfo{author}{\bibfnamefont{B.}\,\bibnamefont{Yu}} and
  \bibinfo{author}{\bibfnamefont{S.}\,\bibnamefont{Zhou}},
  }\bibfield{title}{\emph {\bibinfo{title}{{CP violation and flavor invariants
  in the seesaw effective field theory}}}, }\href@noop {}
  {\,(\bibinfo{year}{2022})}, \Eprint
  {http://arxiv.org/abs/2203.10121}{arXiv:2203.10121 [hep-ph]}\BibitemShut
  {NoStop}%
\bibitem [{\citenamefont{Bilal}(2008)}]{Bilal:2008qx}%
  \BibitemOpen
  \bibfield{author}{\bibinfo{author}{\bibfnamefont{A.}\,\bibnamefont{Bilal}},
  }\bibfield{title}{\emph {\bibinfo{title}{{Lectures on Anomalies}}},
  }\href@noop {} {\,(\bibinfo{year}{2008})}, \Eprint
  {http://arxiv.org/abs/0802.0634}{arXiv:0802.0634 [hep-th]}\BibitemShut
  {NoStop}%
\end{thebibliography}%

\end{document}